\documentclass[twocolumn,12pt]{aastex62}
\pdfoutput=1 
\usepackage{amsmath,amstext}
\usepackage[T1]{fontenc}
\usepackage{apjfonts,verbatim} 
\usepackage[figure,figure*]{hypcap}
\usepackage{color}
\usepackage[utf8,latin1]{inputenc}

\shortauthors{Levy et al.}

\begin{document}
\renewcommand{\comment}{}
\newcommand{\per}{$^{-1}$}
\newcommand{\pers}{$^{-2}$}
\newcommand{\D}			{$^\circ$}
\newcommand{\vrot}		{V$_{\rm rot}$}
\newcommand{\vrad}		{V_{\rm rad}}
\newcommand{\vsys}		{V_{\rm sys}}
\newcommand{\vflat}		{V_{\rm flat}}
\newcommand{\vbar}		{\bar{V}}
\newcommand{\rflat}		{R_{\rm flat}}
\newcommand{\rbulge}    {${\rm R}_{\rm e,bulge}$}
\newcommand{\ha}		{\mbox{\rm{H}$\alpha$}}
\newcommand{\hb}		{\mbox{\rm{H}$\beta$}}
\newcommand{\hg}		{\mbox{\rm{H}$\gamma$}}
\newcommand{\hd}		{\mbox{\rm{H}$\delta$}}
\newcommand{\ttco}		{$^{13}$CO}
\newcommand{\kms}		{\mbox{km\,s$^{-1}$}}
\newcommand{\kmskpc}		{\mbox{km\,s$^{-1}$\,kpc$^{-1}$}}
\newcommand{\hi}        {\mbox{\rm H{\small I}}}
\newcommand{\HI}        {\hi}
\newcommand{\HII}       {\mbox{\rm H{\small II}}}
\newcommand{\SII}		{\mbox{\rm [S{\small II}]}}
\newcommand{\NII}		{\mbox{\rm [N{\small II}]}}
\newcommand{\OIII}		{\mbox{\rm [O{\small III}]}}
\newcommand{\OII}		{\mbox{\rm [O{\small II}]}}
\newcommand{\OI}		{\mbox{\rm [O{\small I}]}}
\newcommand{\CII}		{\mbox{\rm [C{\small II}]}}
\newcommand{\HeI}		{\mbox{\rm He{\small I}}}
\newcommand{\htwo}      {\mbox{H$_{2}$}}
\newcommand{\Dv}		{$\Delta$V}
\newcommand{\eDv}		{$\sigma_{\Delta$V}}
\newcommand{\convol}	{{\fontfamily{cmtt}\selectfont convol}}
\newcommand{\regrid}	{{\fontfamily{cmtt}\selectfont regrid}}
\newcommand{\moment}	{{\fontfamily{cmtt}\selectfont moment}}
\newcommand{\miriad}	{{\fontfamily{cmtt}\selectfont Miriad}}
\newcommand{\ccdmom}	{{\fontfamily{cmtt}\selectfont ccdmom}}
\newcommand{\nmom}	{{\fontfamily{cmtt}\selectfont mom=32}}
\newcommand{\pipetd}	{\mbox{\rm{\small Pipe3D}}}
\newcommand{\chisq}		{$\chi_r^2$}
\renewcommand{\~}		{$\sim$}
\newcommand{\sigmaADC} {$\sigma_{\rm ADC}$}
\newcommand{\sigmaHg} {$\sigma_{\rm H\gamma}$}
\newcommand{\DvDz}		{$\Delta{\rm V}/\Delta z$}
\newcommand{\DlagDr}		{$\Delta{\rm Lag}/\Delta {\rm r}$}
\newcommand{\DAvDz}		{$\Delta{\rm A_{V}}/\Delta z$}
\newcommand{\hha}{$h$(\ha)}
\newcommand{\defgal}{IC\,480} 

\title{The EDGE-CALIFA Survey: Evidence for Pervasive Extraplanar Diffuse Ionized Gas in Nearby Edge-On Galaxies}

\author{Rebecca C. Levy}
\affiliation{Department of Astronomy, University of Maryland, College Park, MD 20742, USA}
\author{Alberto D. Bolatto}
\affiliation{Department of Astronomy, University of Maryland, College Park, MD 20742, USA}
\author{Sebasti\'{a}n F. S\'{a}nchez}
\affiliation{Instituto de Astronom\'{i}a, Universidad Nacional Aut\'{o}noma de M\'{e}xico, A.P. 70-264, 04510 M\'{e}xico, D.F.,  Mexico}
\author{Leo Blitz}
\affiliation{Department of Astronomy, University of California, Berkeley, CA 94720, USA}
\author{Dario Colombo}
\affiliation{Max-Planck-Institut f\"{u}r Radioastronomie, D-53121, Bonn, Germany}
\author{Veselina Kalinova}
\affiliation{Max-Planck-Institut f\"{u}r Radioastronomie, D-53121, Bonn, Germany}
\author{Carlos L\'{o}pez-Cob\'{a}}
\affiliation{Instituto de Astronom\'{i}a, Universidad Nacional Aut\'{o}noma de M\'{e}xico, A.P. 70-264, 04510 M\'{e}xico, D.F.,  Mexico}
\author{Eve C. Ostriker}
\affiliation{Department of Astrophysical Sciences, Princeton University, Princeton, NJ 08544, USA}
\author{Peter Teuben}
\affiliation{Department of Astronomy, University of Maryland, College Park, MD 20742, USA}
\author{Dyas Utomo}
\affiliation{Department of Astronomy, The Ohio State University, Columbus, OH 43210, USA}
\author{Stuart N. Vogel}
\affiliation{Department of Astronomy, University of Maryland, College Park, MD 20742, USA}
\author{Tony Wong}
\affiliation{Department of Astronomy, University of Illinois, Urbana, IL 61801, USA}

\correspondingauthor{Rebecca C. Levy}
\email{rlevy@astro.umd.edu}

\begin{abstract}
We investigate the prevalence, properties, and kinematics of extraplanar diffuse ionized gas (eDIG) in a sample of 25 edge-on galaxies selected from the CALIFA survey. We measure ionized gas scale heights from \ha\ and find that {\comment 90\%} have measurable scale heights with a {\comment median of  $0.8^{+0.7}_{-0.4}$\,kpc}. From the \ha\ kinematics, we find that 60\% of galaxies show a decrease in the rotation velocity as a function of height above the midplane. This lag is characteristic of eDIG, and we measure a median lag of 21\,\kmskpc\ which is comparable to lags measured in the literature. We also investigate variations in the lag with radius. \HI\ lags have been reported to systematically decrease with galactocentric radius. We find both increasing and decreasing ionized gas lags with radius, as well as a large number of galaxies consistent with no radial lag variation, {\comment and} investigate these results in the context of internal and external origins for the lagging ionized gas. We confirm that the \SII/\ha\ and \NII/\ha\ line ratios increase with height above the midplane as is characteristic of eDIG. The ionization of the eDIG is dominated by star-forming complexes (leaky \HII\ regions). We conclude that the lagging ionized gas is turbulent ejected gas likely resulting from star formation activity in the disk as opposed to gas in the stellar thick disk or bulge. This is further evidence for the eDIG being a product of stellar feedback and for the pervasiveness of this WIM-like phase in many local star-forming galaxies.
\end{abstract}


\section{Introduction}
\label{sec:intro}
The diffuse gas phases of the interstellar medium (ISM) are, by their very nature, harder to study than the denser phases. 
Nonetheless, they provide important windows into the evolution of galaxies.
The diffuse ionized gas phase (frequently referred to as Warm Ionized Medium or WIM in the Milky Way and Diffuse Ionized Gas or DIG in other galaxies), in particular, has important connections to star formation activity and accretion history. A large fraction (perhaps the majority) of ionized gas in galaxies is found in this diffuse phase \citep[e.g.][]{haffner09}. The existence of extraplanar diffuse ionized gas (eDIG) raises further questions about the formation and ionization of the DIG \citep[e.g.][]{dettmar90,rand90}. The origin of the eDIG is debated; possible formation {\comment and ionization} mechanisms include leaky \HII\ regions {\comment\citep[e.g.][]{haffner09,weber19}}, star formation feedback in the form of galactic fountains \citep[e.g.][]{shapiro76,bregman80}, post-asymptotic giant branch (AGB) stars in the stellar thick disk \citep{floresfajardo11,lacerda18}, outflows \citep[e.g.][]{lopezcoba19}, accretion from the intergalactic medium (IGM) \citep[e.g.][]{binney05}, or some combination of these processes. Properties of the eDIG, such as relation to the star formation rate (SFR), diagnostic line ratios tracing temperature and density, connections to extraplanar \HI, and kinematics can give insights into the formation {\comment and ionization} of the eDIG and to the formation history of the galaxy itself. 

Studies of eDIG in other galaxies are usually limited to edge-on systems, where the eDIG can be photometrically detected and separated from the disk. While there are large photometric studies of the eDIG \citep[][]{miller03I,rossa03a,rossa03b}, obtaining high resolution optical spectra of an entire disk has been observationally expensive. The kinematics of the eDIG, however, have proven to be very interesting:  the rotation velocity decreases with increasing height above the midplane of the galaxy \citep[e.g.][]{rand00,miller03II,fraternali04,heald06b,bizyaev17}. This vertical gradient in the rotation velocity (referred to as ``lag'') is also seen in extraplanar \HI\ \citep[e.g.][]{swaters97,fraternali02,zschaechner15a,zschaechner15b}. The advent of integral field unit (IFU) spectroscopy has revolutionized our ability to obtain high spatial and spectral resolution spectra of many galaxies. IFU galaxy surveys --- such as CALIFA \citep{sanchez12a,sanchez16}, MaNGA \citep{bundy15}, and SAMI {\comment\citep{croom12,bryant15}} --- enable the study of the kinematics of the eDIG in much larger samples. 

Using intermediate inclination star-forming disk galaxies drawn from the EDGE-CALIFA survey, \citet{levy18} find that the ionized gas (traced by H$\alpha$) rotates slower than the molecular gas for 75\% of their subsample. They attribute this difference in rotation velocity to a significant contribution from eDIG to the H$\alpha$ emission. In the midplane, the molecular and ionized gas should have the same rotation velocity. But if the eDIG rotates more slowly at greater heights above the midplane, the line-of-sight ionized gas rotation velocity will be consequently lowered. Indirect support for this hypothesis includes measured star formation rate surface densities above the empirical threshold for existence of the eDIG \citep{rossa03a}, \SII/\ha\ and \NII/\ha\ ratios higher than observed in galaxy midplanes \citep{haffner09}, and inferred ionized gas velocity dispersions large enough to support a thick ionized gas disk \citep{burkert10}. The results of \citet{levy18} reinforce the idea that eDIG is ubiquitous in star-forming galaxies.

With intermediate inclination galaxies, however, eDIG scale heights and the decrease in rotation velocity as a function of height cannot be directly measured. We, therefore, extend the work of \citet{levy18} using a sample of edge-on CALIFA galaxies to directly measure these eDIG properties. We show below that the measured ionized gas scale heights that are consistent with previous measurements. We find too that the ionized gas rotation velocity decreases with height for $\sim$75\% of the galaxies, and that the magnitude of this decrease (the lag) is also consistent with previous ionized gas lag measurements. We discuss how our results fit into the various eDIG formation scenarios. We also verify that the lags are indeed due to eDIG through analysis of the ionization properties, uniquely possible due to the large wavelength coverage of the CALIFA IFU survey. It appears that the eDIG is indeed a prominent component in star-forming galaxies, affecting the morphology and kinematics of these systems. {\comment This also complements studies of eDIG in edge-on galaxies done with  MaNGA \citep{jones17,bizyaev17} and SAMI \citep{ho16}, as well as studies of outflows in CALIFA \citep{lopezcoba17,lopezcoba19}.}

We summarize the observations and sample selection in Section \ref{sec:obs}. The method and results of fitting the ionized gas scale height are reported in Section \ref{sec:EOGSscaleheight}. The kinematic analysis and results are presented in Section \ref{sec:kinematics}. Section \ref{sec:origin} discusses how the eDIG could be formed and constraints placed on its origin from this analysis. Constraints on the source of the lagging extraplanar ionized gas are discussed in the context of the ionization in Section \ref{sec:ionization}. The results of this work are summarized in Section \ref{sec:summary}. 

\section{Observations and Data Reduction}
\label{sec:obs}
\subsection{The CALIFA Survey}
\label{ssec:califasurvey}
The CALIFA survey \citep{sanchez12a} observed 667 nearby (z = 0.005--0.03) galaxies as of the third data release (DR3). Full details of the CALIFA observations are presented in \citet{sanchez12a}, \citet{husemann13}, \citet{walcher14}, \citet{garciabenito15}, and \citet{sanchez16}, as well as in \citet{levy18}. A brief overview is presented here for context. All CALIFA galaxies are drawn from the Sloan Digital Sky Survey (SDSS). CALIFA used the PPAK IFU on the 3.5m Calar Alto observatory with two spectral gratings. The low resolution grating (V500) used here covered wavelengths from 3745--7500\,\AA\ with 6.0\,\AA\ (FWHM) spectral resolution, corresponding to a FWHM velocity resolution of 275\,\kms\ at \ha. The typical spatial resolution of the CALIFA PSF is 2.5", corresponding to \~0.8\,kpc at the mean distance of the galaxies. We note that the CALIFA PSF is a Moffat profile, not a Gaussian. The data used for this study come from the DR3 main sample as well as an additional 147 galaxies from the extension sample \citep[][]{sanchez16}\footnote{The CALIFA data cubes are publicly available at \url{http://califa.caha.es}.}. These data are gridded with 1" pixels. As in \citet{levy18}, the ``flux\_elines" data products are used \citep{pipe3DI,pipe3DII}. Additional masking was also applied to the CALIFA velocity fields using a SNR cut based on the integrated flux and error maps. Pixels with \ha\ SNR < 3.5 were blanked. Data products, such as line intensity and velocity maps, come from \pipetd\ version 2.2 \citep{pipe3DI,pipe3DII} provided in the final form by the CALIFA Collaboration.

The \ha\ fluxes were corrected for extinction by applying the \citet{calzetti00} extinction correction using the provided dust attenuation maps (A$_V$) from CALIFA. However, because the galaxies used here are edge-on, this correction is insufficient in the midplane where the extinction is much higher. We discuss the impact of extinction on our results in Appendix \ref{app:extinction}.

\subsection{The EDGE-CALIFA Survey}
\label{ssec:edgesurvey}
The EDGE-CALIFA survey \citep{bolatto17} measured CO in 126 nearby galaxies with CARMA in the D and E configurations. Full details of the survey, data reduction, and masking techniques are discussed in \citet{bolatto17}, and we present a brief overview here. The EDGE galaxies were selected from the CALIFA sample based on their infrared (IR) brightness and are biased toward higher star formation rates (SFRs) \citep[see Figure 6 of][]{bolatto17}. The EDGE sample is the largest sample of galaxies with spatially resolved CO, with typical angular resolution of 4.5\arcsec\ (corresponding to \~1.5\,kpc at the mean distance of the sample). Data cubes were produced with 20 \kms\ velocity channels. Data products used in this analysis are as described in \citet{bolatto17}\footnote{The EDGE CO data cubes and moment maps for the main sample are publicly available and can be downloaded from \url{www.astro.umd.edu/EDGE}.}, except for the velocity maps. At high inclinations, the lines can become skewed so that a first moment or Gaussian fit to determine the velocity centroid will underestimate the velocity in general. The CO velocity maps used here are the velocity of the line peak (moment=-3 in \miriad)\footnote{We note that this does not affect the CALIFA velocity maps because of the large instrumental line width; the lines remain Gaussian for very highly inclined systems.}. 

When comparing the velocity fields from the EDGE and CALIFA surveys, it is important to note that the velocities are derived using different velocity conventions: EDGE follows the radio convention, and CALIFA follows the optical convention. Because velocities in both surveys are referenced to zero, all velocities are converted to the relativistic velocity convention. In both the optical and radio conventions, the velocity scale is increasingly compressed at higher redshifts. Typical systemic velocities in the EDGE-CALIFA sample are \~4500 \kms, so this compression is non-negligible. The relativistic convention does not suffer from this compression effect. Differences between these velocity conventions and conversions among them can be found in Appendix A of \citet{levy18}. All velocities presented here are in the relativistic convention, unless otherwise noted.

\subsection{Selecting Edge-on Galaxies}
\label{ssec:EOGS}

\begin{figure*}
\label{fig:SDSSpanel}
\centering
\includegraphics[width=\textwidth]{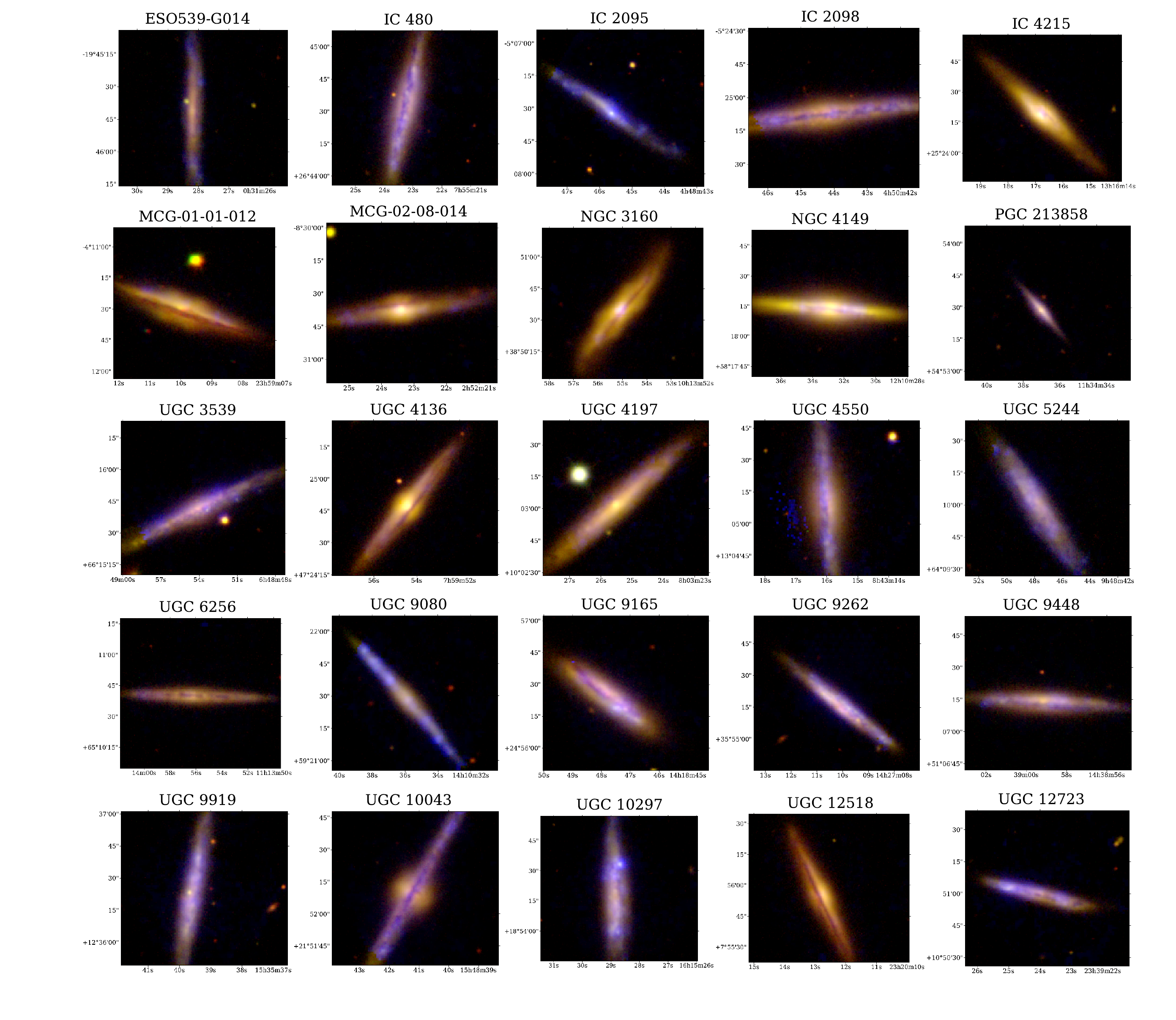}
\caption{Composite images for the 25 edge-on CALIFA galaxies showing the SDSS r- and g-bands (red, green) and (non-masked) \ha\ flux from CALIFA (blue). The horizontal axes show J2000 right ascension and the vertical axes show J2000 declination. Images are cropped to the CALIFA field-of-view \citep[74"$\times$64";][]{sanchez12a}.}
\end{figure*}

For this study, it is important to select the most edge-on systems for analysis to avoid interpreting a deviation in inclination away from edge-on as a detection of eDIG from either the photometry or kinematics (see Appendix \ref{app:inc} for a more detailed discussion of this effect). Starting with 814 galaxies from CALIFA DR3 and the extended sample \citep{sanchez16}, we first find those with inclinations of 90\D\ in HyperLEDA \citep{makarov14}. Inclinations in HyperLEDA are defined as
\begin{equation}
\label{eq:ledainc}
\sin^2i=\frac{1-10^{-2\log r_{25}}}{1-10^{-2\log r_o}}
\end{equation}
where $i$ is the inclination, $r_{25}$ is the axis ratio of the B-band 25th mag\,arcsec$^{-2}$ isophote, and
\begin{equation}
\label{eq:ro}
\log r_o =
\begin{cases}
0.43+0.0053t, & {\rm for \ } -5 \le t \le 7 \\
0.38, & {\rm for \ } t > 7
\end{cases}
\end{equation}
which accounts for intrinsic disk thickness based on the morphological type ($t$). This results in 156 galaxies with $i=90$\D. Because morphological types are highly uncertain in edge-on systems, we follow this step with a visual inspection of the SDSS images (such as those shown in Figure \ref{fig:FS}a), confirming the edge-on nature of each galaxy \citep{gunn98,gunn06,doi10,eisenstein11,alam15}. Dust lanes were used (if present) to visually confirm the edge-on nature of the galaxies; galaxies with dust lanes that were not centered in the midplane were excluded. If no dust lane was present, very thin systems were selected. Galaxies with visible spiral arms, bars, or other features that hinted at them not being perfectly edge-on were discarded. The inclination classifications from the Morphological Galaxy Catalog \citep[MGC;][]{mcg} were also used to confirm the edge-on nature of the galaxies; all galaxies are classified as edge-on in the MGC. This restricted the sample to 54 galaxies. From there, galaxies with robust \ha\ maps and clear rotation were selected. From these criteria, we construct a sample of 25 high-fidelity edge-on CALIFA galaxies. Composite SDSS and CALIFA images are shown in Figure \ref{fig:SDSSpanel}; the \ha\ flux maps shown in this image are not masked based on the \ha\ SNR. Composite SDSS images and \ha\ flux and velocity maps can be found in Figure \ref{fig:FS}a,b,c in Appendix \ref{app:addtlim}. The subsample of galaxies used here and relevant physical parameters are listed in Table \ref{tab:EOGSparams}.  Throughout, \defgal\ will be used as an example. 

Four of these galaxies have known ionized gas outflows \citep[Notes O in Table \ref{tab:EOGSparams};][]{lopezcoba17,lopezcoba19}. We do not exclude these galaxies from further analysis and will discuss the impacts of outflows in Section \ref{ssec:outflows}. Five of the galaxies have extraplanar ionized gas but do not meet the criteria to have an outflow according to \citet{lopezcoba19} (Notes E in Table \ref{tab:EOGSparams}). The selection criteria used here and by \citet{lopezcoba19} differ due to the goals of each study, so that our samples of galaxies with eDIG are not identical. We use a more stringent inclination cut than \citet{lopezcoba19}, who select galaxies with $i>70$\D. Whereas our primary selection criteria are edge-on systems with robust, clearly rotating \ha, \citet{lopezcoba19} further select only galaxies whose ionized gas line ratios increase with height off the midplane. To be selected as an outflow candidate (labeled O in Table \ref{tab:EOGSparams}), the \ha\ equivalent width must be greater than 3\,\AA\ and there must be some biconical morphology \citep{lopezcoba19}. Galaxies that do not meet the additional outflow requirements are labeled as having eDIG, but not an outflow \citep[labeled E in Table \ref{tab:EOGSparams};][]{lopezcoba19}. Three of the edge-on galaxies have robust CO measurements (IC\,480, UGC\,3539, and UGC\,10043), and comparisons between the molecular and ionized gas for these galaxies will be discussed in Section \ref{sssec:COcomp}.

\rotate
\begin{deluxetable*}{cccccccccccccccc}
\tabletypesize\footnotesize
\tablecaption{Parameters for the Edge-On CALIFA Galaxies \label{tab:EOGSparams}}
\tablehead{
\colhead{Name} & \colhead{ID} & \colhead{R.A.} & \colhead{Decl.} & \colhead{Type} & \colhead{Notes} & \colhead{Distance} & \colhead{D$_{25}$} & \colhead{PA} & \colhead{V$_{\rm sys}$} & \colhead{W4} & \colhead{SFR} & \colhead{$h$(H$\alpha$)} & \colhead{V(z=0)} & \colhead{Lag}  & \colhead{$\Delta$Lag/$\Delta$r} \\
 & & \colhead{(J2000 $^\circ$)} & \colhead{(J2000 $^\circ$)} & & & \colhead{(Mpc)} & \colhead{(kpc)} & \colhead{($^\circ$)} & \colhead{(\kms)} & \colhead{(mag)} & \colhead{(M$_\odot$ yr$^{-1}$)} & \colhead{(kpc)} & \colhead{(\kms)} & \colhead{(\kms\,kpc$^{-1}$)}  &\colhead{(\kms\,kpc$^{-2}$)}}
\startdata
ESO539-G014 & 15 & 7.86771 & -19.76127 & Scd & ... & 102.21 & 52.87$\pm$3.14 & 88.9 & 6984 & $6.94\pm0.10$ & $0.71^{+0.50}_{-0.12}$ & 1.10$\pm$0.80 & 144.5$^{+15.6}_{-15.7}$ & 18.9$^{+8.1}_{-8.0}$ & -3.4$^{+0.2}_{-0.2}$\\
IC 480 & 159 & 118.84634 & 26.74287 & Sbc & O & 66.02 & 33.60$\pm$2.07 & 256.7 & 4545 & $3.97\pm0.02$ & $4.58^{+3.21}_{-0.62}$ & 1.13$\pm$0.35 & 138.5$^{+15.6}_{-15.7}$ & 18.6$^{+8.4}_{-8.4}$ & -1.7$^{+1.5}_{-1.5}$\\
IC 2095 & 141 & 72.19066 & -5.12475 & Sc & ... & 40.71 & 17.27$\pm$1.29 & 327.7 & 2820 & ... & ... & 0.60$\pm$0.26 & 54.9$^{+18.7}_{-18.7}$ & 2.7$^{+17.7}_{-17.5}$ & 4.9$^{+1.3}_{-1.3}$\\
IC 2098 & 142 & 72.68451 & -5.41857 & Sc & E & 39.77 & 28.33$\pm$1.23 & 187.0 & 2776 & $4.57\pm0.03$ & $0.95^{+0.66}_{-0.13}$ & 0.62$\pm$0.32 & 91.8$^{+16.1}_{-15.2}$ & -10.4$^{+18.5}_{-17.4}$ & 8.2$^{+1.7}_{-4.9}$\\
IC 4215 & 615 & 199.07117 & 25.40582 & Sab & ... & 55.46 & 25.04$\pm$1.75 & 131.2 & 3828 & $5.07\pm0.04$ & $1.17^{+0.82}_{-0.16}$ & 1.64$\pm$0.58 & 151.5$^{+19.7}_{-18.9}$ & 16.8$^{+31.4}_{-29.8}$ & 20.9$^{+1.0}_{-14.5}$\\
MCG-01-01-012 & 936 & 359.79867 & -4.19247 & Sb & E & 82.50 & 37.77$\pm$2.61 & 339.7 & 5657 & $5.59\pm0.06$ & $1.60^{+1.12}_{-0.23}$ & 2.94$\pm$1.57 & 158.8$^{+30.0}_{-30.1}$ & 14.9$^{+20.0}_{-19.9}$ & 1.3$^{+2.1}_{-2.0}$\\
MCG-02-08-014 & 111 & 43.09772 & -8.51057 & Sab & ... & 71.80 & 37.22$\pm$2.23 & 188.2 & 4936 & $4.22\pm0.02$ & $4.28^{+3.00}_{-0.58}$ & 0.92$\pm$0.35 & 124.3$^{+29.6}_{-29.7}$ & 29.6$^{+22.6}_{-22.6}$ & 0.5$^{+1.5}_{-1.5}$\\
NGC 3160 & 319 & 153.47951 & 38.84297 & Sab & B & 98.52 & 35.75$\pm$3.43 & 50.8 & 6689 & $4.72\pm0.03$ & $5.09^{+3.57}_{-0.70}$ & ... & 192.0$^{+47.3}_{-44.4}$ & -9.7$^{+71.8}_{-19.6}$ & 8.0$^{+0.1}_{-2.3}$\\
NGC 4149 & 502 & 182.63695 & 58.30408 & Sb & BE & 43.76 & 17.45$\pm$1.57 & 174.8 & 3030 & $6.00\pm0.09$ & $0.31^{+0.22}_{-0.05}$ & 0.83$\pm$0.25 & 199.4$^{+22.2}_{-27.0}$ & 66.0$^{+17.3}_{-21.1}$ & 9.5$^{+9.3}_{-3.8}$\\
PGC 213858 & 5033 & 173.65409 & 54.89121 & Sab & ... & 83.13 & 14.44$\pm$2.83 & 310.9 & 5699 & $8.27\pm0.14$ & $0.14^{+0.10}_{-0.03}$ & ... & 117.8$^{+13.7}_{-13.8}$ & 16.0$^{+23.7}_{-23.4}$ & 0.9$^{+7.0}_{-6.0}$\\
UGC 3539 & 148 & 102.22453 & 66.26050 & Sbc & BO & 46.70 & 24.89$\pm$1.49 & 32.6 & 3231 & $4.47\pm0.03$ & $1.45^{+1.01}_{-0.20}$ & 0.72$\pm$0.37 & 117.2$^{+13.1}_{-13.1}$ & 14.8$^{+6.7}_{-6.7}$ & 1.9$^{+2.4}_{-2.4}$\\
UGC 4136 & 168 & 119.97651 & 47.41340 & Sa & ... & 95.86 & 44.29$\pm$3.04 & 51.1 & 6551 & $6.08\pm0.07$ & $1.37^{+0.96}_{-0.20}$ & 1.22$\pm$1.05 & 234.2$^{+37.0}_{-37.5}$ & 29.6$^{+25.8}_{-26.0}$ & -2.7$^{+0.5}_{-0.4}$\\
UGC 4197 & 174 & 120.85633 & 10.05058 & Sb & ... & 65.56 & 32.84$\pm$2.23 & 41.8 & 4514 & $5.33\pm0.04$ & $1.29^{+0.90}_{-0.18}$ & 0.58$\pm$0.35 & 197.0$^{+23.7}_{-32.4}$ & 69.1$^{+17.0}_{-32.0}$ & -0.6$^{+1.6}_{-0.8}$\\
UGC 4550 & 218 & 130.81647 & 13.08536 & Sb & E & 29.36 & 18.01$\pm$0.92 & 94.6 & 2070 & $4.64\pm0.03$ & $0.49^{+0.34}_{-0.07}$ & 0.47$\pm$0.30 & 101.5$^{+11.5}_{-11.7}$ & 25.8$^{+24.7}_{-24.7}$ & -8.8$^{+6.4}_{-1.9}$\\
UGC 5244 & 297 & 147.20058 & 64.16847 & Sc & ... & 43.33 & 18.99$\pm$1.41 & 122.3 & 3000 & $6.24\pm0.07$ & $0.24^{+0.17}_{-0.04}$ & 0.68$\pm$0.38 & 99.9$^{+13.3}_{-13.5}$ & 18.5$^{+11.4}_{-11.4}$ & -3.1$^{+1.4}_{-1.4}$\\
UGC 6256 & 383 & 168.48397 & 65.17769 & Sc & ... & 47.80 & 17.07$\pm$1.59 & 357.7 & 3336 & $7.06\pm0.11$ & $0.14^{+0.10}_{-0.02}$ & 1.03$\pm$0.52 & 110.8$^{+32.1}_{-27.3}$ & -2.8$^{+62.3}_{-25.7}$ & 4.2$^{+2.6}_{-8.4}$\\
UGC 9080 & 713 & 212.64959 & 59.35802 & Sc & ... & 43.58 & 15.63$\pm$1.44 & 130.4 & 3048 & $6.40\pm0.08$ & $0.21^{+0.15}_{-0.03}$ & 0.69$\pm$0.30 & 78.1$^{+13.4}_{-13.5}$ & 16.6$^{+15.9}_{-15.9}$ & 8.1$^{+2.3}_{-2.6}$\\
UGC 9165 & 731 & 214.69920 & 24.94047 & S0a & O & 74.87 & 26.43$\pm$2.55 & 144.8 & 5199 & $3.47\pm0.02$ & $9.33^{+6.54}_{-1.26}$ & 1.08$\pm$0.34 & 160.4$^{+13.4}_{-13.5}$ & 14.1$^{+5.0}_{-5.0}$ & 2.2$^{+0.6}_{-0.6}$\\
UGC 9262 & 747 & 216.79358 & 35.92219 & Sbc & E & 122.50 & 41.39$\pm$4.45 & 141.0 & 8318 & $4.04\pm0.02$ & $14.70^{+10.30}_{-1.99}$ & 1.97$\pm$0.97 & 200.1$^{+24.2}_{-24.2}$ & 16.8$^{+5.0}_{-4.9}$ & -0.1$^{+0.4}_{-0.4}$\\
UGC 9448 & 765 & 219.74656 & 51.12068 & Sb & ... & 31.50 & 14.73$\pm$1.02 & 356.6 & 2188 & $6.12\pm0.07$ & $0.14^{+0.10}_{-0.02}$ & 0.39$\pm$0.16 & 96.2$^{+17.2}_{-17.1}$ & 19.8$^{+25.7}_{-25.4}$ & -4.7$^{+10.7}_{-1.4}$\\
UGC 9919 & 805 & 233.91577 & 12.60638 & Sc & ... & 45.81 & 19.66$\pm$1.48 & 80.0 & 3170 & $5.30\pm0.04$ & $0.64^{+0.45}_{-0.09}$ & 0.99$\pm$0.43 & 107.2$^{+12.2}_{-12.3}$ & 15.7$^{+14.2}_{-14.2}$ & 5.3$^{+1.5}_{-1.5}$\\
UGC 10043 & 811 & 237.17269 & 21.86980 & Sbc & O & 30.62 & 19.57$\pm$0.99 & 60.6 & 2128 & $4.39\pm0.03$ & $0.66^{+0.47}_{-0.09}$ & 0.48$\pm$0.13 & 104.5$^{+15.9}_{-16.0}$ & 22.6$^{+18.4}_{-18.4}$ & -12.1$^{+1.0}_{-0.5}$\\
UGC 10297 & 827 & 243.87017 & 18.90479 & Sc & B & 32.94 & 20.63$\pm$1.03 & 272.6 & 2287 & $5.60\pm0.06$ & $0.25^{+0.18}_{-0.04}$ & 0.48$\pm$0.22 & 91.3$^{+10.0}_{-10.2}$ & 24.6$^{+8.0}_{-7.9}$ & 3.5$^{+1.7}_{-1.8}$\\
UGC 12518 & 910 & 350.05353 & 7.93222 & Sb & ... & 55.31 & 22.68$\pm$1.78 & 295.1 & 3818 & $5.64\pm0.06$ & $0.69^{+0.48}_{-0.10}$ & ... & 151.2$^{+25.6}_{-25.7}$ & 24.5$^{+46.3}_{-45.6}$ & 7.4$^{+1.1}_{-13.4}$\\
UGC 12723 & 926 & 354.84994 & 10.84992 & Sd & ... & 78.07 & 26.43$\pm$2.52 & 167.2 & 5359 & $5.07\pm0.04$ & $2.32^{+1.63}_{-0.32}$ & 1.00$\pm$0.66 & 98.8$^{+31.1}_{-31.1}$ & 6.7$^{+17.9}_{-17.7}$ & 1.6$^{+1.1}_{-1.1}$\\
\enddata
\tablecomments{The table lists  edge-on CALIFA galaxies, their CALIFA ID number, and their R.A. and Decl. from HyperLEDA (with offsets as needed).Morphological types and the diameter of isophote corresponding to 25 mag arcsec$^{-2}$ (D$_{25}$) are from HyperLEDA. Notes denote whether this galaxy has a bar (B), ring (R), is part of a multiple (M) (all from HyperLEDA), has an outflow (O) \citep{lopezcoba19}, or has extraplanar ionized gas (E) \citep{lopezcoba19}. Distances, position angles (PA), and systemic velocities (V$_{\rm sys}$ are from CALIFA. W4 is the WISE W4 (22$\mu$m) Vega magnitude (see Section \ref{ssec:dvdztrends}). SFR is the star formation rate calculated from the WISE W4 magnitudes (see Section \ref{ssec:dvdztrends}). \hha\ is the measured \ha\ scale height (see Section \ref{sec:EOGSscaleheight}). V(z=0) is the fitted rotation velocity in the midplane (see Section \ref{ssec:vvg}). \DvDz\ is the vertical gradient in the rotation velocity (see Section \ref{ssec:vvg}). The ``lag' reported here is $-$\DvDz. $\Delta$Lag/$\Delta$r\ is the radial gradient in the lag (i.e. the change in lag as a function of radius; see Section \ref{sssec:radialvar}). Uncertainties on \DvDz, V(z=0), and $\Delta$Lag/$\Delta$r\ reflect the 68\% confidence interval determined from the posterior likelihood distributions.
}\end{deluxetable*}


\section{Ionized Gas Scale Height Measurements}
\label{sec:EOGSscaleheight}

\begin{figure}
\label{fig:height}
\centering
\gridline{\fig{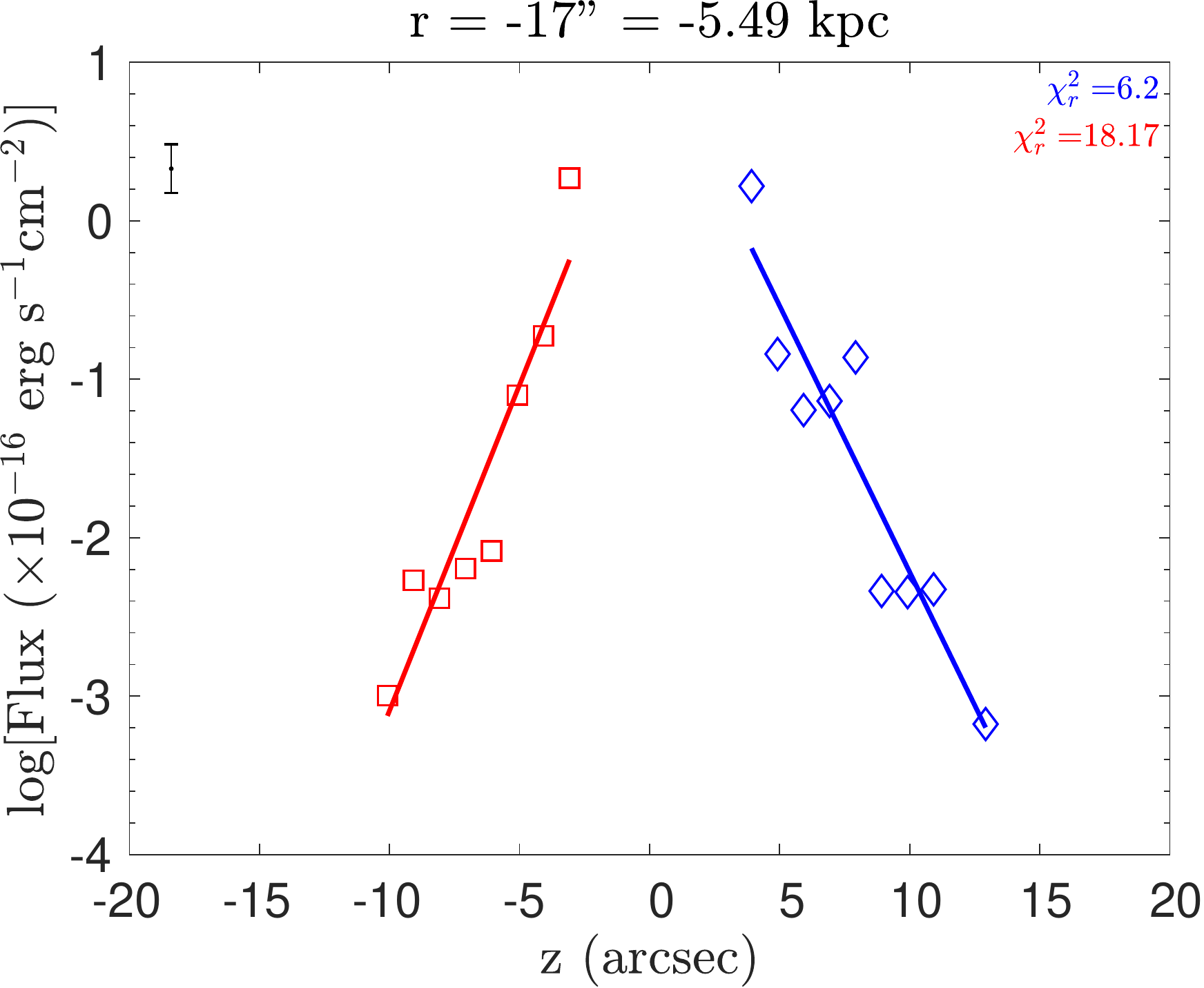}{0.5\columnwidth}{(a)}
\fig{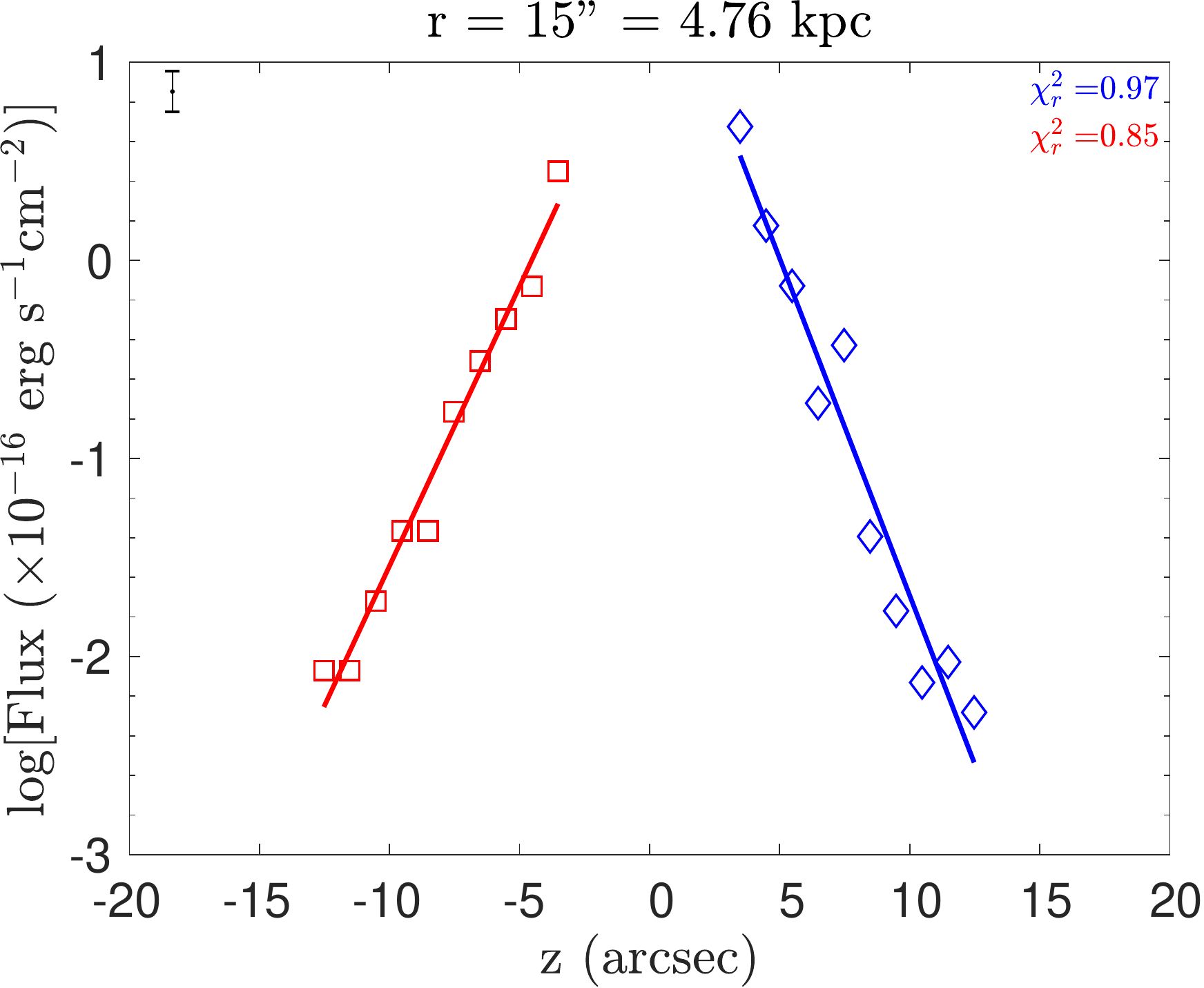}{0.5\columnwidth}{(b)}}
\vspace{-1em}
\gridline{\fig{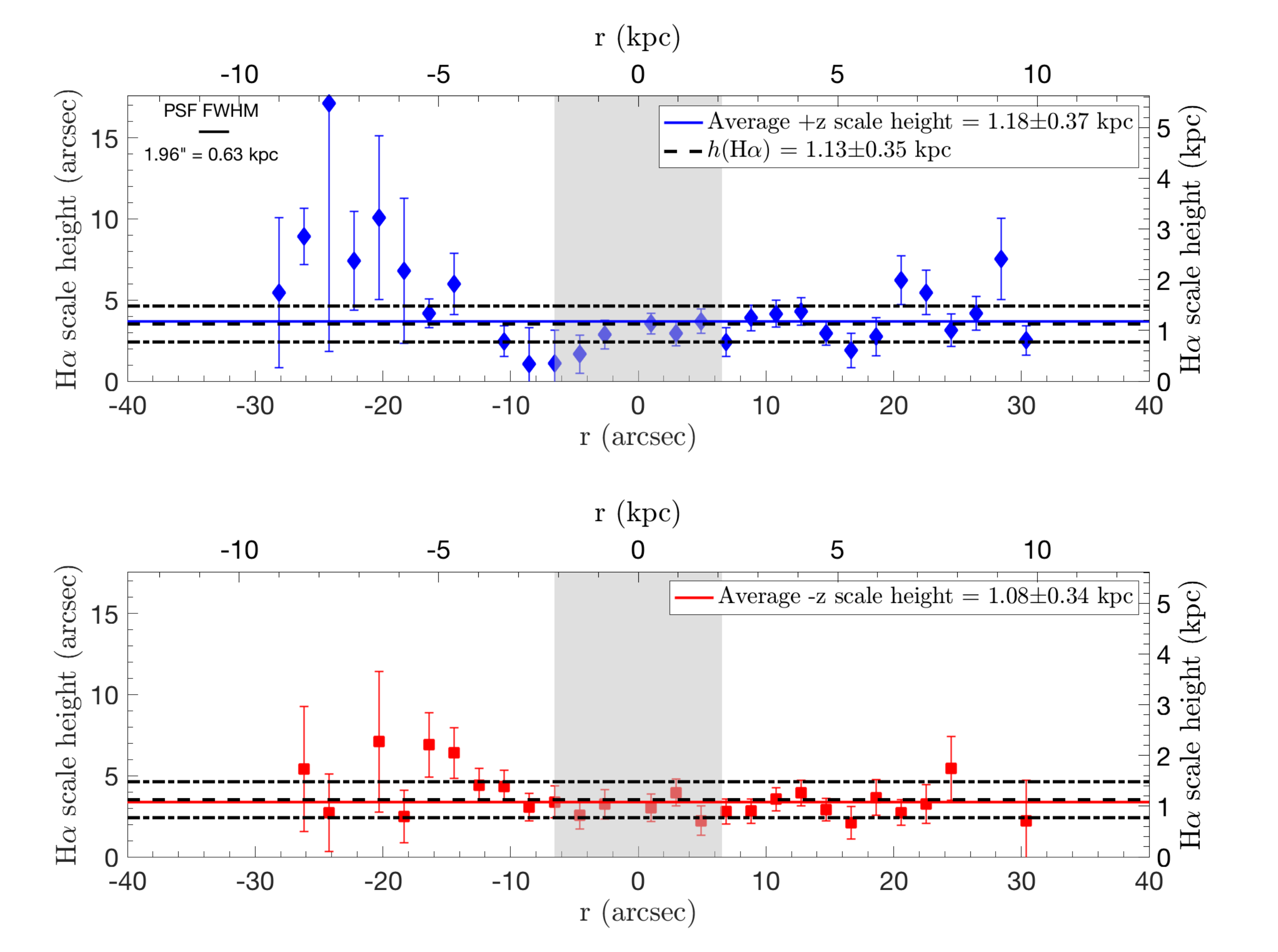}{0.5\textwidth}{(c)}}
\vspace{-1em}
\caption{\comment (a) The exponential fit to the \ha\ flux in \defgal\ as a function of distance from the midplane ($z$) in one radial bin ($r=-17"=-5.49$\,kpc). The open symbols show the flux measurements at each pixel above (blue diamonds) or below (red squares) the midplane. The solid lines show the exponential fits. The black error bars in the upper left corner show the typical uncertainty on the \ha\ flux measurements. (b) The same as (a) but for $r=15"=4.76$\,kpc. (c) The fitted exponential scale height of the \ha\ disk above (top; blue diamonds) and below (bottom; red squares) the midplane as a function of radius ($r$) for \defgal. The PSF FWHM has been removed in quadrature. The average scale height weighted by the uncertainties and corresponding weighted standard deviation are shown in the black dashed and dot-dashed lines. The gray shaded region shows the central radii excluded from the median (where $r<2$\,\rbulge). The solid black line in the upper left corner shows the FWHM of the PSF.} 
\end{figure}

{\comment We determine the ionized gas scale height by fitting an exponential function to the \ha\ intensity maps. The fitting of the \ha\ intensity as a function of distance from the midplane is performed in single pixel (1") increments along the major axis; we refer to these increments (which are parallel to the rotation axis of the galaxy) as ``radial bins''. Independent scale heights are fit to the emission above and below the midplane. Before finding the scale height, the location of the midplane at each pixel along the major axis is determined by fitting a Gaussian to the \ha\ intensity along the minor axis (above and below the midplane); the centroid pixel is taken to be the location of the midplane and is used to divide emission above and below the midplane. To avoid being biased by the (much brighter) emission from \HII\ regions in the midplane, we exclude the middle five pixels (the midplane and roughly one PSF FWHM above and below) from the fit. }
Fits for two radial bins in \defgal\ are shown in Figure \ref{fig:height}a,b. The reported \ha\ scale height is the exponential scale length. The PSF size is subtracted in quadrature from each fit to give an estimate of the scale height as a function of radius. {\comment We then interpolate the radial bins so that the spacing along the major axis is the PSF FWHM. The \ha\ scale height measurement at small galactocentric radii may be affected or dominated by a bulge or outflow as opposed to an eDIG disk. \citet{mendezabreu17} measure r-band bulge effective radii (\rbulge) for a subset of the CALIFA galaxies. Their photometric analysis excludes all highly inclined systems. Their sample of 404 galaxies is representative of the CALIFA sample as a whole, and we will assume their median \rbulge\ is representative of our sample as well. Using the redshift of each galaxy and H$_0=70$\,\kms\,Mpc$^{-1}$ \citep{mendezabreu17}, the median \rbulge\,$=1.05$\,kpc. We exclude radial bins with $r<2$\,\rbulge\,$=2.1$\,kpc (gray regions in Figure \ref{fig:height}c). We discuss the impact of bulge and outflow contamination on our results in Sections \ref{ssec:bulge} and \ref{ssec:outflows}. The average scale height (\hha) was found for each galaxy by averaging all radial bins above and below the midplane weighted inversely by the variance of each bin. The uncertainty ($\sigma_{h({\rm H}\alpha)}$) is the weighted standard deviation. These values are in Table \ref{tab:EOGSparams}. The \ha\ scale height as a function of radius is shown for \defgal\ in Figure \ref{fig:height}c. We find that $>$88\% of the subsample galaxies have a measurable eDIG scale height, where a measurable eDIG scale height has \hha$>\sigma_{h({\rm H}\alpha)}$.}

\begin{figure}
\label{fig:heightkde}
\centering
\includegraphics[width=\columnwidth]{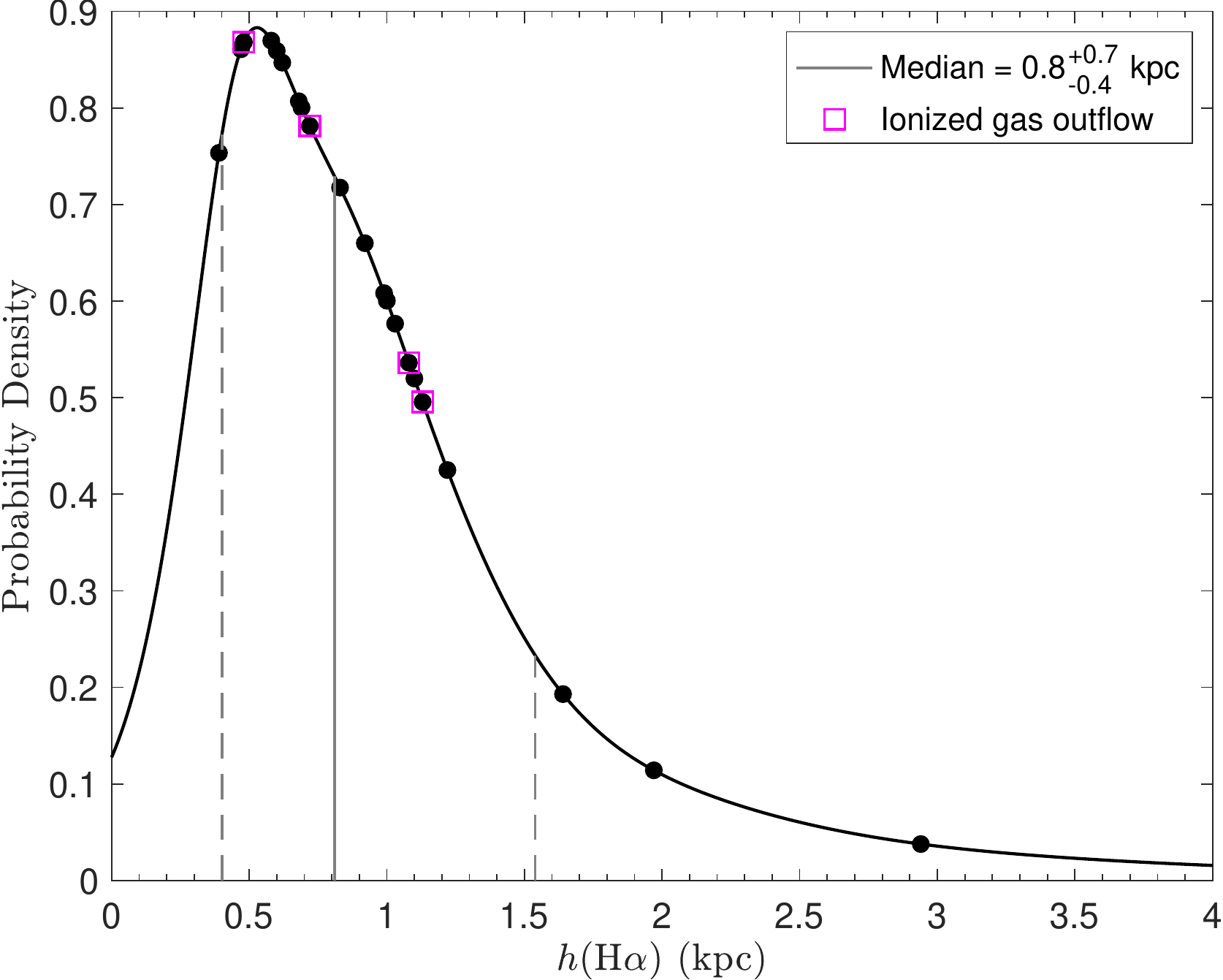}
\caption{The distribution of \hha\ all 25 galaxies. Individual galaxies are shown as the black points. Those galaxies with known ionized gas outflows are outlined with magenta squares \citep[][]{lopezcoba17,lopezcoba19}. {\comment The median \hha\ for this sample of edge-on galaxies is  $0.8^{+0.7}_{-0.4}$ kpc}, with values ranging from {\comment 0.3--2.9 kpc}.} 
\end{figure}

We investigate the distribution of \ha\ scale heights in this sample using a kernel density estimator (KDE). A KDE can be thought of as a histogram where the ``bin width'' is set by the uncertainty. Each \hha\ measurement is represented as a Gaussian, where the centroid is \hha\ and the width is set by the measurement uncertainty {\comment($\sigma_{h({\rm H}\alpha)}$). The individual Gaussians are summed and normalized to unit area to produce the distribution of \hha\, as shown in Figure \ref{fig:heightkde}.} Since the KDE is a probability distribution, the median of the distribution is where the cumulative distribution function is 0.5. We report this value as the median \hha\ for the sample and find the inner 68\% of the distribution in the same way. {\comment The median scale height of this sample is $0.8^{+0.7}_{-0.4}$ kpc}. Values range from {\comment 0.3--2.9\,kpc}, and the distribution is peaked around {\comment 0.5 kpc}. Galaxies with ionized gas outflows \citep[][]{lopezcoba17,lopezcoba19} are shown in magenta in Figure \ref{fig:heightkde}; those galaxies reside around the median scale height rather than populating the high \hha\ tail of the distribution. {\comment We investigate whether there are correlations between \hha\ and any global galaxy property (such as stellar mass, star formation rate, star formation rate surface density, etc.; see Section \ref{ssec:dvdztrends}) and find no trends with any parameter.}

{\comment
\subsection{Previous eDIG Scale Height Measurements}
\label{ssec:lithha}
We compare our eDIG scale heights to those previously measured in the literature. The most extensive sample of eDIG scale height measurements is from the MaNGA sample investigated by \citet{bizyaev17} who find a median \ha\ scale height of $1.2\pm0.5$ kpc over their sample of galaxies with detected eDIG. \citet{miller03I} also measured eDIG scale heights in a large sample of galaxies. The used both one- and two-component exponential fits, where the latter represents a brighter quiescent eDIG phase as well as a fainter disturbed phase. The median scale height for the more extended (disturbed) eDIG phase over their sample of 16 galaxies is $2.3\pm4.3$ kpc\footnote{\comment\citet{miller03I} report the average scale height of their sample (4.3 kpc). An unweighted mean is easily biased by a few galaxies with large scale heights. We consider the median or weighted mean to be more representative of the sample as a whole.}. If only a one-component exponential fit is used, the median scale height is $0.5\pm0.9$ kpc over the sample. We attempted to fit a two-component exponential model to our data, but the fits were poorly constrained because our data are not sensitive enough at large distances from the midplane. \citet{levy18} estimated the \ha\ scale height in their sample of intermediate-inclination EDGE-CALIFA galaxies using three methods: (1) from the \hg\ velocity dispersion; (2) using an asymmetric drift correction to estimate for the \ha\ velocity dispersion and the implied scale height; (3) using a suite of kinematic simulations. We refer the reader to Sections 5.5 and 5.7 of \citet{levy18} for the specifics of each method. For all three methods, possible \ha\ scale heights were limited to $\lesssim1.5$ kpc. Our observations of the edge-on galaxies in this subsample are well matched to this limit in general. We compiled an extensive list of eDIG scale heights from a variety of sources, observations, and fitting techniques \citep{rand97,wang97,hoopes99,collins00,collins01,miller03I,rosado13,bizyaev17}. The median eDIG scale height from these various techniques is $1.0\pm2.2$ kpc. We conclude that our \hha\ measurements are in good agreement with previous measurements of eDIG in nearby galaxies, although there is substantial scatter from galaxy to galaxy across all samples.}


\section{Ionized Gas Kinematics}
\label{sec:kinematics}

\subsection{Position-Velocity Diagrams}
\label{ssec:PVdiagrams}
\begin{figure}
\label{fig:PVheight}
\centering
\includegraphics[width=\columnwidth]{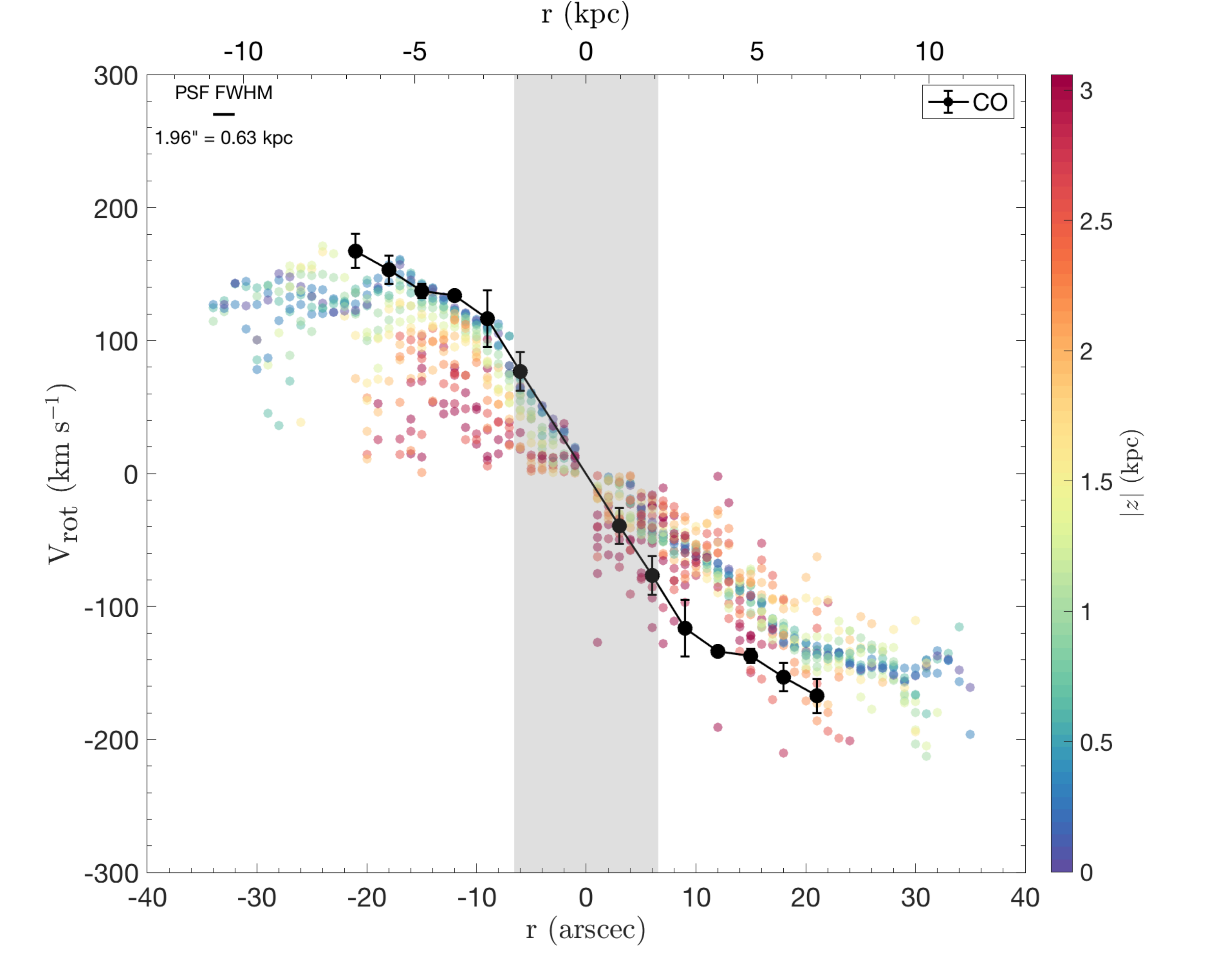}
\caption{PV diagram for \defgal. The colored dots show the \ha\ emission color-coded by distance from the midplane. The gray shaded region shows radii $<2$\,\rbulge. The \ha\ rotation velocity decreases with distance from the midplane. The black dots show the CO PV diagram (see Section \ref{sssec:COcomp}). The solid black line in the upper left corner shows the FWHM of the PSF. Similar figures for the other edge-on galaxies are shown in Figure \ref{fig:FS}d.} 
\end{figure}

In order to compare the kinematics as a function of distance from the midplane, we construct position-velocity (PV) diagrams by taking cuts parallel to the major axis in single-pixel increments along the minor axis. The systemic velocity is subtracted (listed in Table \ref{tab:EOGSparams}). {\comment Pixels with SNR\,$<5$ in either the \ha\ intensity or velocity maps are masked out.} Although we refer to this quantity as \vrot, it is likely that it includes contributions from non-rotational motions as well. None of the velocities have been corrected for inclination since we assume that all galaxies are perfectly edge on; we investigate the effects of extinction and inclination on our results in Appendices \ref{app:extinction} and \ref{app:inc} respectively. Figure \ref{fig:PVheight} shows the PV diagram for \defgal\ color-coded by distance from the midplane. Distances are converted from angular to physical units using the distance to each galaxy listed in Table \ref{tab:EOGSparams}.  

\subsection{Vertical Gradients in Ionized Gas Rotation Velocity}
\label{ssec:vvg}
\begin{figure}
\label{fig:vvg}
\centering
\includegraphics[width=\columnwidth]{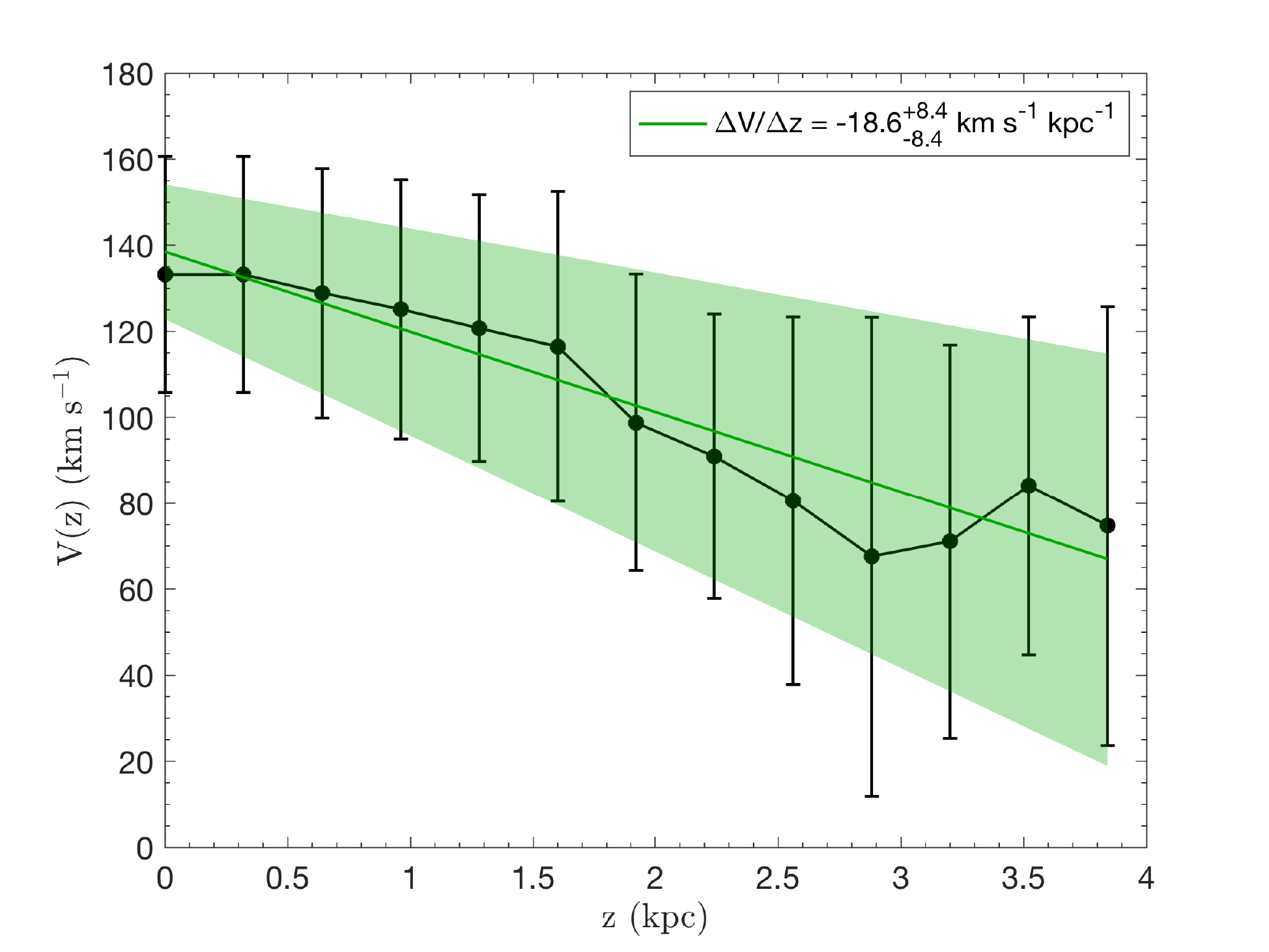}
\caption{The \ha\ rotation velocity as a function of height for \defgal. There is a clear, nearly linear decrease in \vrot\ with increasing distance from the midplane. The green line is the best fit. The green shaded region shows the 68\% confidence interval. Similar figures for the other edge-on galaxies are shown in Figure \ref{fig:FS}e.} 
\end{figure}

Both the \ha\ velocity fields and PV diagrams show evidence for a decrease in \vrot\ with distance from the midplane (Figures \ref{fig:PVheight} and \ref{fig:FS}c,d). We can further quantify this decrease in terms of the vertical gradient in the rotation velocity (\DvDz) with units of \kmskpc. The magnitude of \DvDz\ is often referred to as the ``lag'' in the literature. From the PV diagrams, the rotation velocity at each height is averaged, excluding radii less than 2\,\rbulge. This is shown in Figures \ref{fig:vvg} and \ref{fig:FS}e. We fit a line using a Bayesian method, where the slope is \DvDz. We marginalize over the intercept of the line and calculate the 68\% confidence interval from the posterior distribution of the slopes. The best fit line and confidence interval are also shown in Figures \ref{fig:vvg} and \ref{fig:FS}e. We argue that extinction does not significantly bias the observed \DvDz\ in Appendix \ref{app:extinction}.

\begin{figure*}
\label{fig:KDEvvg}
\centering
\gridline{\fig{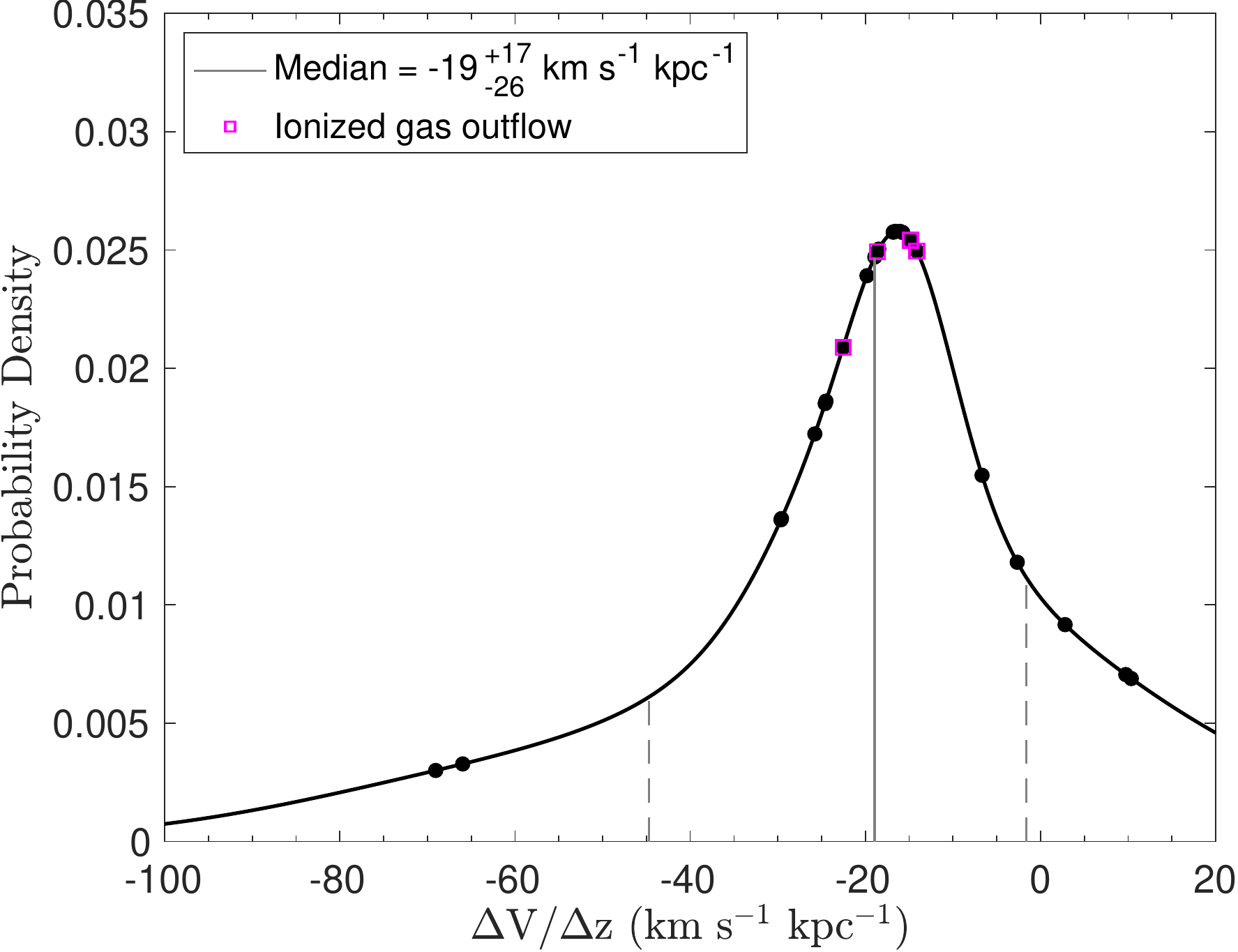}{0.455\textwidth}{(a)}
\fig{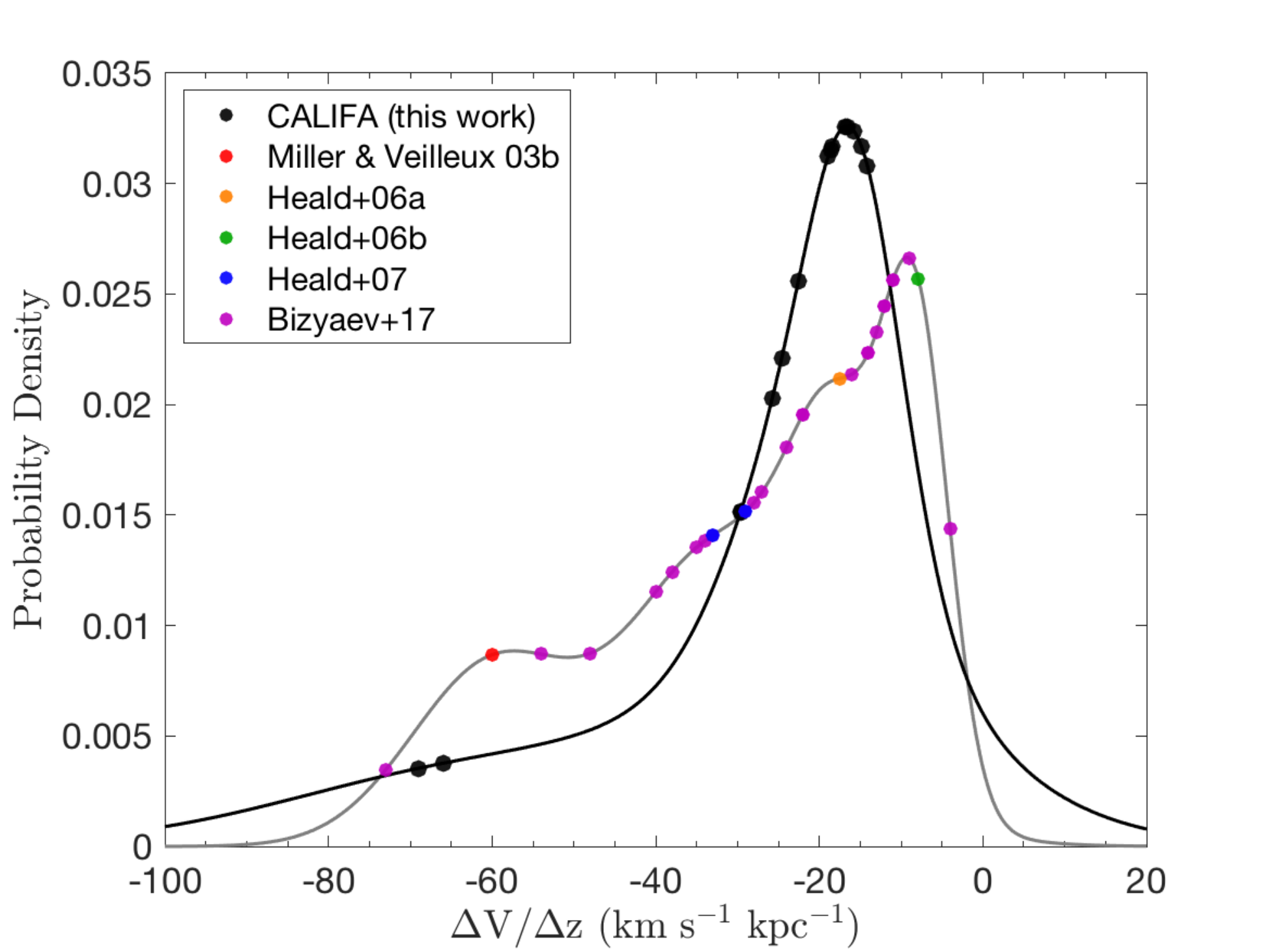}{0.5\textwidth}{(b)}}
\caption{(a) The KDE of \DvDz\ for the 25 edge-on CALIFA galaxies. The median \DvDz\ and inner 68\% of the distribution are marked by the solid and dashed gray lines. Galaxies with ionized gas outflows are marked with a magenta square \citep[][]{lopezcoba17,lopezcoba19}. (b) The KDE for the 17 CALIFA galaxies with significant lags (with respect to the measurement uncertainty, black) compared to the KDE of the literature values (gray) measured from \citet{miller03II}, \citet{heald06b,heald06a,heald07}, and \citet{bizyaev17}. The median \DvDz\ for the values pulled from the literature is $-25^{+16}_{-28}$ \kmskpc\ whereas the median for the CALIFA galaxies with lags is $-21^{+12}_{-27}$ \kmskpc.}
\end{figure*}

We investigate the distribution of \DvDz\ over the sample. We find that 60\% have measurable negative vertical gradients in the rotation velocity, where a measurable negative vertical gradient has \DvDz\,$+\sigma_{\rm\Delta V/\Delta z,+}<0$. The other 40\% are consistent with no gradient. There are no galaxies for which the \ha\ rotation velocity increases with height within the uncertainties (i.e. where \DvDz\,$-\sigma_{\rm\Delta V/\Delta z,-}>0$). We can further investigate the distribution of \DvDz\ using a KDE (Figure \ref{fig:KDEvvg}a). The distribution of \DvDz\ is strongly peaked around $-20$ \kmskpc\ with values ranging from $-70-10$ \kmskpc\ and a median of $-19^{+17}_{-26}$ \kmskpc. 

Extraplanar ionized gas which ``lags'' (i.e. rotates more slowly than) gas in the midplane has been observed in many systems, such as NGC\,891 \citep{heald06a}, NGC\,5775 \citep{heald06b}, NGC\,4302 \citep{heald07}, NGC\,2820 \citep{miller03II}, NGC\,4013 \citep{miller03II}, and 25 galaxies from the MaNGA survey \citep{bizyaev17}\footnote{Lagging extraplanar \HI\ has also been observed in several systems, but we limit the discussion here to the ionized gas properties.}. We compare the distribution and magnitudes of lags from this sample to those compiled from the literature listed above. Figure \ref{fig:KDEvvg}a shows the distribution for this sample of edge-on CALIFA galaxies. Because of our strict selection criteria (see Section \ref{ssec:EOGS}), we have sufficient detections along the minor axis to measure lags in all galaxies, and can measure lags which are consistent with zero within the uncertainties. To better compare with the literature values {\comment--- which only report lags where they are nonzero ---} we select the 17 galaxies in our subsample with significant nonzero lags given their uncertainties. Figure \ref{fig:KDEvvg}b shows the distribution of \DvDz\ for these galaxies with nonzero lags compared to the distribution of values from the literature listed above. Although our distribution is more peaked, the values from the literature also cluster between $\sim -40-0$ \kmskpc. The median \DvDz\ from the literature values is $-25^{+16}_{-28}$ \kmskpc, where the uncertainty is the inner 68\% of the distribution. The median \DvDz\ of the CALIFA galaxies with measurable lags is $-21^{+12}_{-27}$ \kmskpc. {\comment The full sample of 25 galaxies is used for the remainder of the analysis.}

In particular, we highlight the results of a systematic study of lagging eDIG using observations from the MaNGA survey \citep{bundy15}. \citet{bizyaev17} find lagging extraplanar ionized gas in at least 37\% of their sample of 67 edge-on galaxies (purple points in Figure \ref{fig:KDEvvg}b) compared to the 60\% of galaxies we find with lagging \ha\ in this study. The fraction of lagging eDIG they find is likely lower than in this study due to their selection criteria and their slightly coarser {\comment spatial} resolution. In their sample, edge-on galaxies have no evidence of spiral arms in the SDSS images, a dust lane projected near the galaxy midplane in the SDSS images (if present), and no sign of interaction \citep{bizyaev17}. They make no selection based on the SNR of the \ha\ velocity fields for their sample of 67 edge-on galaxies. As a consequence---as \citet{bizyaev17} point out---they cannot determine whether the other 63\% of the edge-on galaxies in their sample have lags or not. Because of our selection criteria (Section \ref{ssec:EOGS}), however, we can distinguish between galaxies with and without lags for our entire sample. In other words, the subsample of 67 galaxies used by \citet{bizyaev17} includes all edge-on, non-interacting MaNGA galaxies. Our subsample of 25 galaxies includes all edge-on, non-interacting CALIFA galaxies with \ha\ measurements which are robust enough to allow for a lag measurement to be made. The lower {\comment spatial} resolution of the MaNGA local sample ($\sim$1.5\,kpc) compared to CALIFA ($\sim0.8$\,kpc) also makes distinguishing extraplanar gas more difficult in the former. Therefore, the 37\% of galaxies with lags that \citet{bizyaev17} do find is a lower limit. 

We connect this result and those of \citet{levy18} and \citet{davis13} as being part of the same phenomenon. \citet{levy18} and \citet{davis13} found that the rotation velocity of the ionized gas is systematically lower than that of the molecular gas in intermediate inclination galaxies. \citet{levy18} argued that this was indeed due to eDIG with a vertical gradient in the rotation velocity. Viewed at an intermediate inclination, velocities from eDIG at different heights average along the line of sight, producing a smaller net ionized gas rotation velocity. A similar averaging is likely occurring in the early-type galaxies studied by \citet{davis13}, although the source of this more slowly-rotating ionized gas is likely due to the bulge rather than a layer of extraplanar gas. {\comment Neither \citet{davis13} nor \citet{levy18} find galaxies in which the ionized gas rotates faster than the molecular gas. Similarly, we find no galaxies in which the ionized gas rotation velocity increases with height above the midplane (given the uncertainties).} We note that the fractions of galaxies found to have some form of more slowly-rotating ionized gas not associated with the thin gas disk are similar among these three studies: 77\% of intermediate-inclination disk galaxies \citep{levy18}, 80\% of gas-rich early-type galaxies, and 60\% of edge-on disk galaxies show evidence of extraplanar ionized gas that rotates more slowly than the midplane. As {\comment there is evidence for lagging, extraplanar diffuse gas in} a large fraction of galaxies in surveys of the local universe, understanding the origin and properties of this diffuse ionized gas is important. {\comment Moreover, investigating galaxies without lagging eDIG or those with negative lags \citep[e.g.][]{peroux19} may give insights into the unique evolutionary and star formation histories of these systems.}

\subsubsection{Comparisons with the Molecular Gas}
\label{sssec:COcomp}
There are three galaxies in this subsample that have robust CO detections from the EDGE{\comment-CALIFA} survey \citep{bolatto17}: IC\,480, UGC\,3539, and UGC\,10043. For their galaxy and kinematic parameters, we refer the reader to the tables in \citet{bolatto17} and \citet{levy18}. All of these galaxies are candidates to host ionized gas outflows \citep[][]{lopezcoba17,lopezcoba19}. We repeat the analysis described in Sections \ref{ssec:PVdiagrams} and \ref{ssec:vvg} to derive PV diagrams and measure any lag in the CO velocity as a function of height. {\comment As shown in Table \ref{tab:colags},} the CO scale height can be measured only out to $\sim$0.3 kpc from the midplane {\comment(which is substantially smaller than the CO PSF)} and the uncertainties on each point and the gradient are large. {\comment The maximum heights probed by the CO are much smaller than the \ha\ scale heights listed in Table \ref{tab:EOGSparams}.} All three galaxies have CO lags consistent with zero, although this is limited by the small range of heights probed. {\comment Higher sensitivity and resolution data are needed for robust CO lag and scale height measurements.}

\begin{table}
\caption{CO Lags}
\label{tab:colags}
\centering
{\comment
\begin{tabular}{cccc}
\hline\hline
Galaxy & CO \DvDz & CO V(z=0) & Max. Height\tablenotemark{a}\\
 & (\kmskpc)  & (\kms) & (kpc) \\
\hline
IC\,480 & -1.5$^{+28.5}_{-46.1}$ & 136.3$^{+11.0}_{-10.0}$ & 0.32\\
UGC\,3539 & 1.4$^{+21.5}_{-23.1}$ & 134.3$^{+4.0}_{-4.0}$ & 0.23\\
UGC\,10043 & -24.3$^{+40.2}_{-41.0}$ & 142.3$^{+8.0}_{-10.0}$ & 0.30\\
\hline
\end{tabular}
\tablenotetext{a}{The is the maximum vertical extent probed by the CO data.}
}
\end{table}

\subsubsection{Implications for Galaxy Dynamical Mass Measurements}
Reiterating from \citet{levy18}, the dynamical mass inferred from the ionized gas rotation velocity of a galaxy with lagging ionized gas will be systematically underestimated if left uncorrected. {\comment Although the midplane \HII\ regions constitute the bulk of the ionized gas mass in a galaxy, more than half of the \ha\ luminosity comes from the (e)DIG \citep[e.g.][]{reynolds93,zurita00,poetrodjojo19}. This systematic underestimate of the dynamical mass will be worse in systems with smaller circular velocities and/or larger lags. For a galaxy with an eDIG scale height of 0.8\,kpc\ (the median for this sample), a circular velocity of 200\,\kms, and a lag of 21\,\kmskpc\ (the median for this sample), the dynamical mass will be underestimated by $\sim$13\%. This effect scales proportionally with the lag and eDIG scale height and inversely with the maximum rotation velocity.}

A measurement of the ionized gas velocity dispersion is needed to correctly determine the circular velocity and dynamical mass {\comment\citep[e.g.][]{ iorio17,aquinoortiz18,leung18}}. The low spectral resolution {\comment of CALIFA (275 \kms\ FWHM at $\lambda_{\rm H\alpha}$)} makes determining the ionized gas velocity dispersion difficult, {\comment if not impossible, for this sample} \citep[see][]{levy18}. Since the presence of eDIG is {\comment presumably} related to the star formation rate surface density \citep{rand96,rossa03a}, we would expect more eDIG in higher-redshift systems where SFRs are higher and galaxies are more compact on average. Moreover, since spatially-resolved galaxy studies at high redshift are difficult, inferring the presence of this lagging eDIG photometrically or kinematically will be difficult. Therefore, the potential to underestimate the dynamical masses of these high redshift systems using ionized gas rotation velocities is increased due to astrophysical and instrumental effects.


\section{The Origin of Extraplanar Gas}
\label{sec:origin}

\begin{figure*}
\label{fig:gasflowcartoon}
\centering
\includegraphics[width=0.9\textwidth]{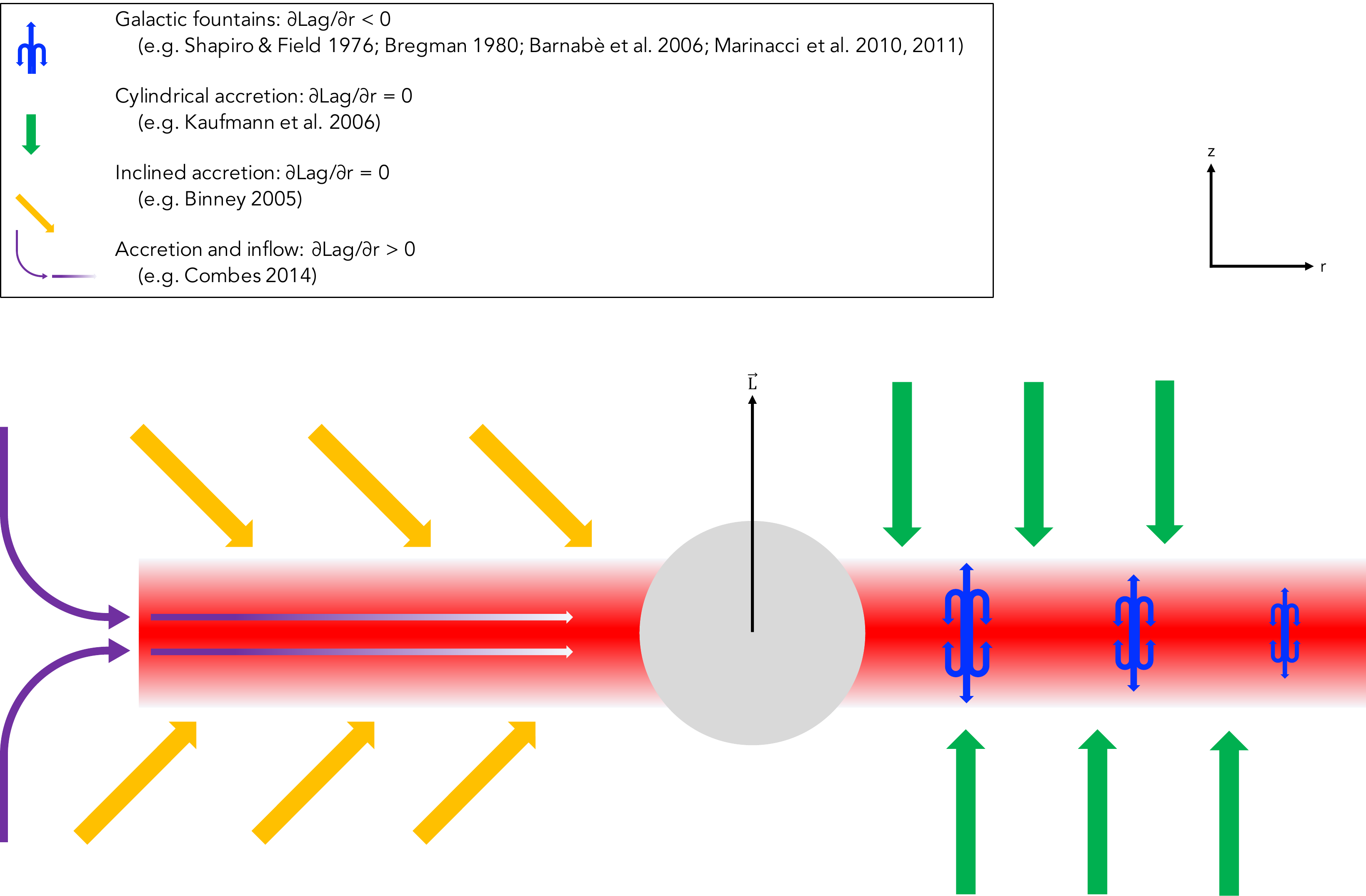}
\caption{A cartoon diagram of the various formation mechanisms for extraplanar gas discussed in Section \ref{sec:origin}. A schematic edge-on disk galaxy is shown, where the midplane is solid red, the bulge is the gray circle, and the lagging extraplanar gas is shown as the red-to-white gradient. Galactic fountains are shown in the blue curved arrow clusters, where the size of the {\comment symbol} is related to the {\comment magnitude} of the lag produced \citep{shapiro76,bregman80,barnabe06,marinacci10,marinacci11}. The green vertical arrows show cylindrical accretion from the corona \citep{kaufmann06}. Accretion which is inclined with respect to the angular momentum axis (vector labeled $\vec{\rm L}$) is shown in the gold slanted arrows \citep{binney05}. Accretion at the outskirts and the resulting inflow are shown in the purple arrows, where the purple-to-white gradient shows the decreasing inflow velocity from the outskirts to center \citep{combes14}. The legend gives the expected behavior of the radial variation of the lag ($\partial$Lag$/\partial r$) for each process (see Section \ref{sssec:radialvar} for details). {\comment As we discuss in Section \ref{sssec:radialvar}, it is not as straightforward to infer the origin of the eDIG from the data using the sign of the radial gradient in the lag as implied in this figure.}}
\end{figure*}

Both internal and external origins of lagging extraplanar gas have been suggested and substantiated. For internal origins, stellar feedback from overpressured superbubbles driven by star formation activity can eject material through galactic fountains. The material does not have enough momentum to escape the galaxy, but settles into a thick ionized disk  \citep[e.g.][]{shapiro76,bregman80}. Models of galactic fountain-produced extraplanar gas can reproduce the observed vertical gradients in the rotation velocity of extraplanar \HI\ and \ha\ \citep{barnabe06,marinacci10,marinacci11}. For external origins, material is accreted from the IGM or corona, \citep[e.g.][]{oort70,binney05,fraternali05,kaufmann06}. In simulations of accretion from the hot corona, gas is accreted cylindrically along the angular momentum axis, and vertical gradients in the rotation velocity agree with observed values \citep{kaufmann06}. In reality it is likely that both internal and external processes contribute \citep[e.g.][]{kaufmann06,haffner09,combes14}. These various formation scenarios are summarized schematically in Figure \ref{fig:gasflowcartoon}.

\subsection{Radial Variations in the Lag}
\label{sssec:radialvar}
Inferring the origin of the extraplanar gas directly from the data is difficult. The formation scenarios make different predictions for radial variations in the lag which can possibly be used to differentiate among them. If the extraplanar gas origin is internal to the galaxy, caused for example by galactic fountains, a decrease (shallowing) of the lag with radius might be expected. Physically, for a centrally concentrated potential, gas at smaller radii will overcome a larger change in the potential to be ejected to the same height as gas at larger radii. As a result, gas at smaller radii will have a larger lag than gas at larger radii \citep[for gas ejected to the same height at all radii; see][]{zschaechner15a}. If the extraplanar gas is due to accretion instead of galactic fountains, simulations of cylindrical accretion parallel to the angular momentum axis do not show evidence for radial variations in the lag \citep[see Figure 4 of][]{kaufmann06}. Even if gas is accreted with some inclination with respect to the angular momentum axis \citep{binney05}, no radial variation in the lag is expected. If, instead, cold material is accreted from streams in the outskirts of the galaxy \citep[e.g.][]{combes14}, radial inflows are expected. If the radial inflow velocity increases with radius (i.e. is largest at the outskirts of the galaxy where the gas is being accreted), then larger lags at larger radii might be expected (i.e. a steepening of the lag with radius). These scenarios for the origin of the eDIG gas are shown schematically in Figure \ref{fig:gasflowcartoon}. {\comment As we will show in this section, however, determining the origin of the extraplanar gas from measurements of the radial lag gradient is not as straightforward as laid out here.}

\subsubsection{\comment Radial Lag Gradients in the Edge-On CALIFA Galaxies}

\begin{figure}
\label{fig:radiallag}
\centering
\includegraphics[width=\columnwidth]{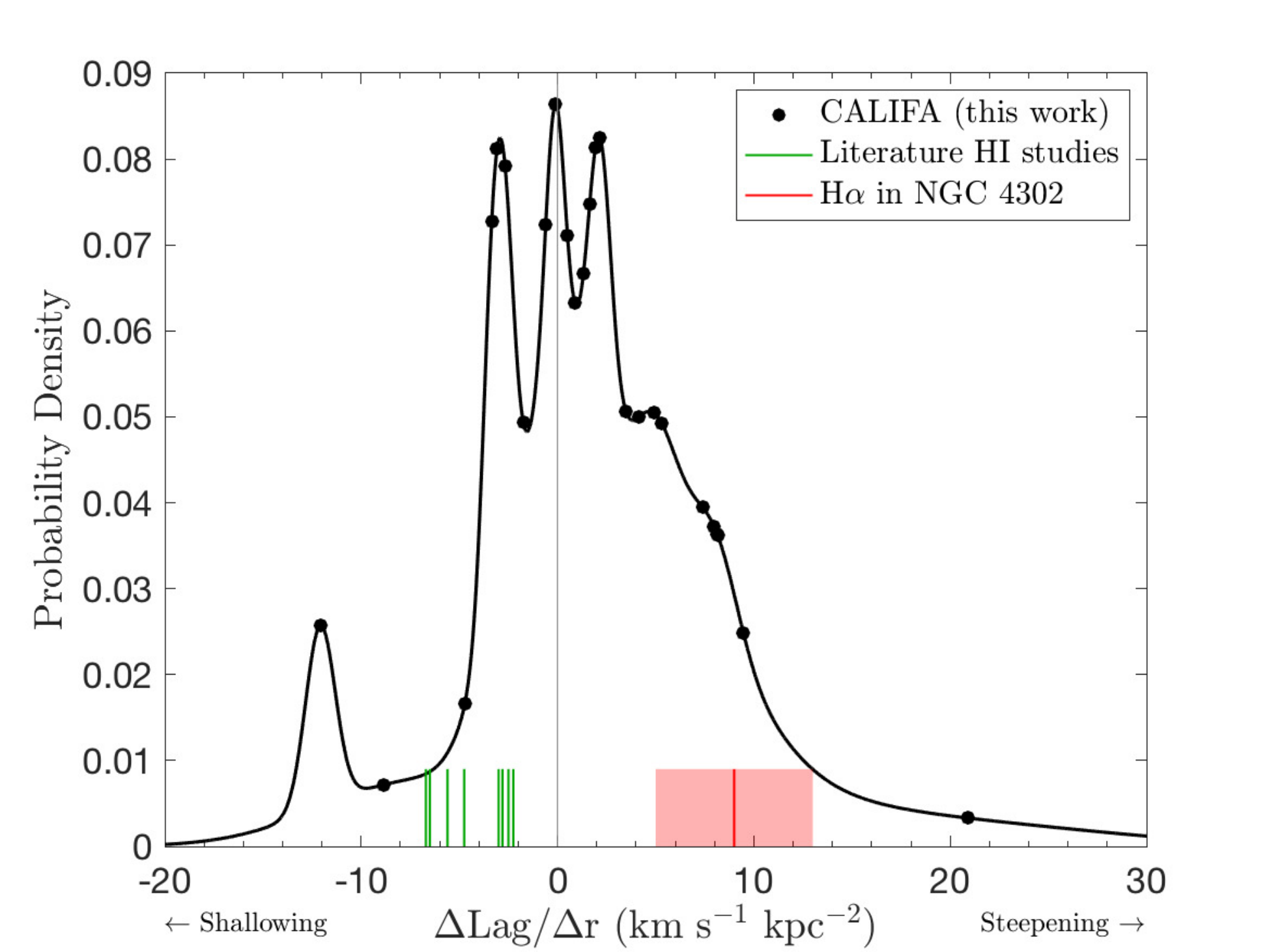}
\caption{The distribution of radial variations in the lag (\DlagDr) for the sample. We find no systematic shallowing (or steepening) of the lag with radius. The green line segments show radial lag variations from previous \HI\ studies (see values and references in Table \ref{tab:HIradiallag}). The red line and shaded region show the \ha\ radial lag variation and uncertainty in NGC\,4302 \citep{heald07}.}
\end{figure}

{\comment With the edge-on CALIFA sample, we can investigate radial trends in the measured eDIG lags. Following Section \ref{ssec:vvg}, we fit for the lag at each radial bin independently above and below the midplane. We find the median lag measurement for each radius and fit a line where the slope is the radial variation in the lag (\DlagDr). The distribution of \DlagDr\ over the sample is shown in Figure \ref{fig:radiallag} and values for individual galaxies are listed in Table \ref{tab:EOGSparams}. In our dataset, 36\% of the galaxies are consistent with no radial variation in the lag within the uncertainties. There are six galaxies (24\%) with lags that shallow with radius (significantly within the uncertainties) with a median \DlagDr\,$=-3.2\pm4.2$\,\kms\,kpc$^{-2}$. As shown in Figure \ref{fig:radiallag}, the shallowing lags we find for the eDIG are comparable to shallowing \HI\ lags (discussed in Section \ref{sssec:HIradiallags}). There are ten galaxies (40\%) with lags that steepen with radius (significantly given the uncertainties) with a median \DlagDr\,$=6.6\pm5.3$\,\kms\,kpc$^{-2}$. Unlike the systematically shallowing \HI\ lags, the \ha\ lags shallow, steepening, and remain constant with radius. If the sign of the radial variation in the lag reflects the origin of the extraplanar gas, these results suggest a mix of internal and external origins for the eDIG.}

\subsubsection{\comment Lags and Radial Lag Gradients Induced by a Thick-Disk Potential}
\label{sssec:potential}

{\comment The kinematics of the eDIG will be shaped by the potential, regardless of whether the extraplanar gas originated internally or externally to the system. We analytically investigate the anticipated lag and radial variation in the lag from a Miyamoto-Nagai potential which describes a three-dimensional axisymmetric potential of a disk galaxy with total mass $M$, radial disk scale length $r_0$, and vertical disk scale length $z_0$ \citep{miyamoto75}. From this potential, we derive V$_{\rm rot}(r,z)$, the lag ($-\partial $V$_{\rm rot}(r,z)/\partial z$), and the radial variation in the lag ($\partial{\rm Lag}/\partial r\equiv-\partial[\partial$V$_{\rm rot}(r,z)/\partial z]/\partial r$). The full derivation and corresponding equations can be found in Appendix \ref{app:derivation}. The analytic potential and resulting equations for the kinematics are agnostic to the origin of the extraplanar gas: the equations describe a disk with vertical scale height $z_0$ regardless of how the gas became extraplanar.}

\begin{figure*}
\label{fig:radiallaganalytic}
\centering
\gridline{\fig{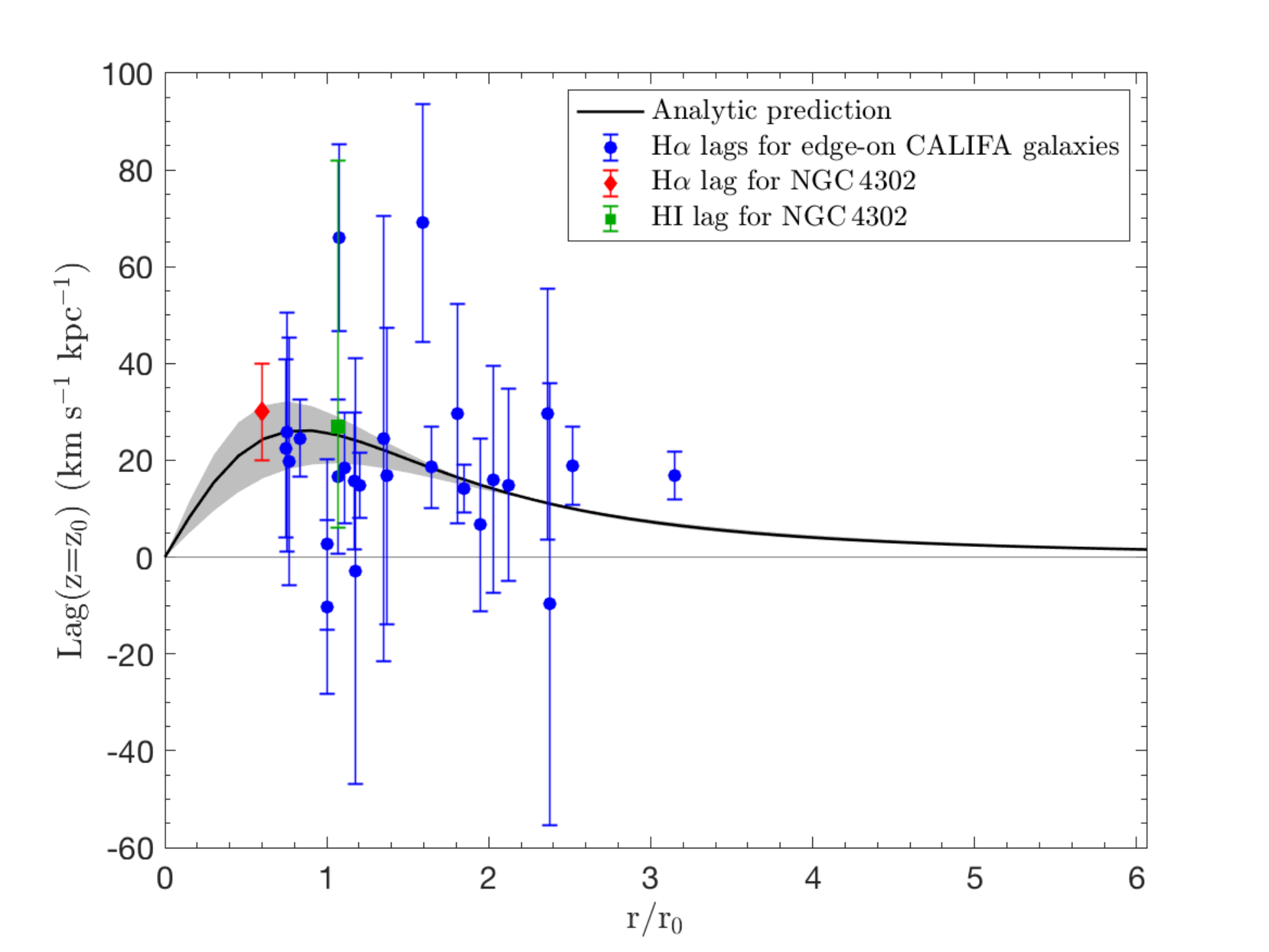}{0.5\textwidth}{(a)}
\fig{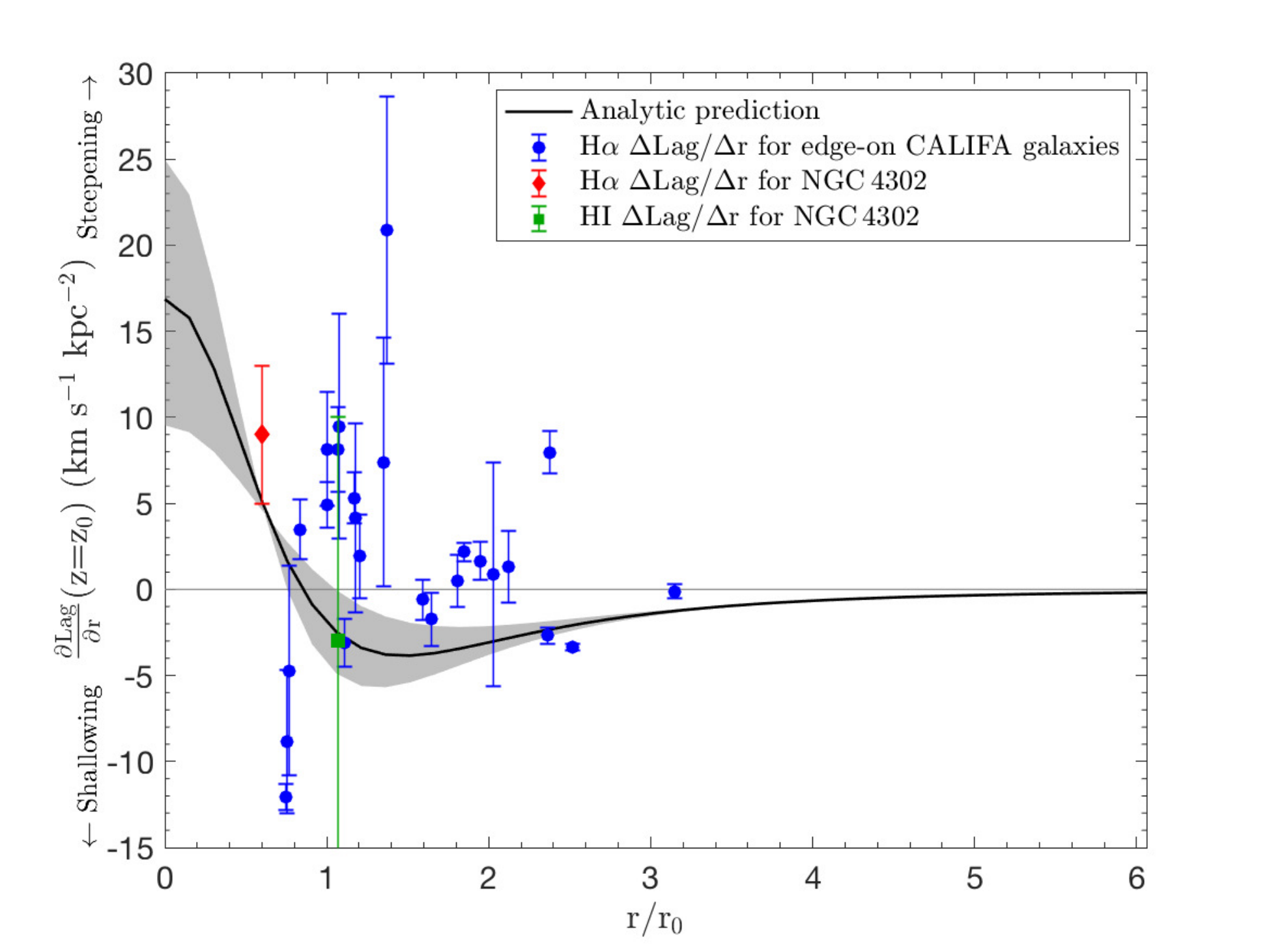}{0.5\textwidth}{(b)}}
\caption{The analytic (a) lag (Equation \ref{eq:dvdzz0}) and (b) $\partial$Lag$/\partial r$  (Equation \ref{eq:dlagdrz0}) as a function of $r/r_0$ at $z=z_0$ for a Miyamoto-Nagai potential. The lag shallows with radius for $r\gtrsim r_0$. In both panels, the black curve shows the analytic prediction for {\comment $z_0=0.8$\,kpc (the median \hha\ for this sample). The gray shaded region shows $z_0$ ranging from $0.3-1.5$\,kpc}, corresponding to the inner 68\% around the median \hha\ (Figure \ref{fig:heightkde}). The blue dots show the measurements of the (a) lag and (b) \DlagDr\ for each galaxy studied here. The red diamond shows the (a) lag ($30\pm10$\,\kmskpc) and (b) \DlagDr\ ($\sim9\pm4$\,\kms\,kpc$^{-2}$) for the ionized gas in NGC\,4302 \citep{heald07}. The green square shows the (a) lag ($27^{+55}_{-21}$\,\kmskpc) and (b) \DlagDr\ ($-3^{+31}_{-13}$\,\kms\,kpc$^{-2}$) for \HI\ in NGC\,4302 \citep{zschaechner15a}. See Section \ref{sssec:radialvar} for details.}
\end{figure*}

{\comment Figure \ref{fig:radiallaganalytic}a shows the lag at $z=z_0$ as a function of $r/r_0$ (from Equation \ref{eq:dvdzz0}) assuming $M=10^{11}\,{\rm M}_\odot$, $z_0=0.8$\,kpc (the median \hha\ for this sample), and $r_0=3.3$\,kpc \citep[the median stellar scale length for the EDGE-CALIFA sample;][]{bolatto17}. The lag increases sharply for $r\lesssim1\ r_0$, then decreases smoothly for $r\gtrsim1\ r_0$. Figure \ref{fig:radiallaganalytic}b shows $\partial$Lag$/\partial r$ as a function of $r/r_0$ (from Equation \ref{eq:dlagdrz0}). As in Figure \ref{fig:radiallaganalytic}a, the radial lag gradient steepens for  $r\lesssim1\ r_0$ and shallows for  $r\gtrsim1\ r_0$. Therefore, the sign of the radial lag gradient (i.e. whether the lag shallows or steepens with radius) induced by the potential depends on the radius at which the measurement is made within the galaxy. This kinematic signature will be imprinted on top of any kinematic imprint relating to the origin of the extraplanar gas.}

{\comment We can compare the analytic predictions from Equations \ref{eq:dvdzz0} and \ref{eq:dlagdrz0} to the measured lag and \DlagDr\ values. To find an average $r/r_0$ for each galaxy, we average the minimum and maximum radii used. The minimum radius is 2\,\rbulge; the maximum radius is assumed to be 30" (see Figure \ref{fig:FS}d). Angular distances are converted to kpc using the distance to each galaxy listed in Table \ref{tab:EOGSparams}. We assume $r_0=3.3$\,kpc as before, which is the median stellar scale length of the EDGE-CALIFA sample \citep{bolatto17}. The measured lag and \DlagDr\ values for each galaxy are shown as the blue dots in Figure \ref{fig:radiallaganalytic}. Given the assumptions, the ionized gas lags and \DlagDr\ we measure for this sample (Table \ref{tab:EOGSparams}) are broadly consistent with our analytic expectation, although there is significant scatter. As shown in Figure \ref{fig:radiallaganalytic}, this may be due to the radial regime probed by the \ha\ data. Radial lag gradients with magnitudes comparable to those we measure are induced by the potential (Figure \ref{fig:radiallaganalytic}b) regardless of the origin of the extraplanar gas. While large deviations from the analytic curve may reflect the origin of the extraplanar gas, the kinematic signature imprinted by the origin of the extraplanar gas is largely overwhelmed by the kinematics induced by the potential itself. Given the assumptions, we conclude that the majority of the radial lag gradient we measure is likely due to the potential and does not give much information about the origin of the eDIG. }

{\comment
\subsubsection{Reassessing Previous Conclusions about the Origin of Extraplanar Gas Based on Radial Lag Variations}
\label{sssec:HIradiallags}

\begin{table}
\caption{\HI\ Radial Lag Deviations from the Literature}
\label{tab:HIradiallag}
\centering
\begin{tabular}{ccc}
\hline\hline
Galaxy & \DlagDr\ & Ref \\
 & (\kms\,kpc$^{-2}$)  & \\
\hline
NGC\,891 & -2.5 & (1) \\
MW & -6.7 & (2) \\
NGC\,4244 & -2.25 & (3)  \\
NGC\,4665 & -5.6 & (4) \\
NGC\,5023 & -2.8 & (5) \\
NGC\,3044 & -4.75 & (6) \\
NGC\,4302 & -3$^{+31}_{-13}$ & (6) \\
NGC\,4013 & -6.5 & (7) \\
\hline
\end{tabular}
\tablerefs{(1) \citet{oosterloo07}, (2) \citet{marasco11}, (3) \citet{zschaechner11}, (4) \citet{zschaechner12}, (5) \citet{kamphuis13}, (6) \citet{zschaechner15a}, (7) \citet{zschaechner15b}}
\end{table}

Measured changes in the lag as a function of radius (\DlagDr) have been found for lagging \HI\ in several galaxies (Table \ref{tab:HIradiallag}). \citet{zschaechner15a} summarize many of these results, finding a systematic decrease in the lag with radius between $r\sim0.5-1.0\ R_{25}$. Assuming $R_{25}=3.2\ r_0$ \citep{persic91}, \HI\ lags tend to decrease between $r\sim1.6-3.2\ r_0$. For the galaxies studied by \citet{zschaechner15a}, many galaxies are probed out to $1.5\ R_{25}\sim4.8\ r_0$. This is in agreement with our analytic expectation (Figure \ref{fig:radiallaganalytic}), where lags induced by the potential tend to shallow with radius for $r \gtrsim 1\ r_0$.  In light of this modeling, the interpretation by \citet{zschaechner15a} that this shallowing of the lag with radius points toward an internal origin for the extraplanar \HI\ is, therefore, not a clear-cut as they claim. We suggest that the majority of the shallowing of the lag with radius may be a result of the thick-disk potential which overwhelms any kinematic signature of the origin of the extraplanar gas.}

{\comment Large kinematic studies of extraplanar ionized gas \citep[e.g.][]{bizyaev17,levy18} have lacked the ``radial resolution'' necessary to investigate radial trends in eDIG lags. To highlight one particular example comparing \HI\ and eDIG lags, we consider NGC\,4302, studied in \HI\ by \citet{zschaechner15a} and in \ha\ by \citet{heald07}. NGC\,4302 is the only galaxy in the literature with measured radial variations in the \ha\ lag. On the approaching side of the galaxy, both the \HI\ and \ha\ lags are constant with radius (although the \ha\ lag is much larger than the \HI\ lag). On the receding side, however, the \HI\ lag shallows slightly with radius whereas the \ha\ lag steepens by $\sim 36$\,\kmskpc\ at $r=4.25$\,kpc\,$=0.6\ r_0$ \citep[assuming $r_0=7$\,kpc;][]{heald07}, resulting in \DlagDr\,$\sim9\pm4$\,\kms\,kpc$^{-2}$. These measurements for the receding side of NGC\,4302 are overplotted in Figures \ref{fig:radiallag} and \ref{fig:radiallaganalytic}. At the radii probed by each measurement, the steepening \ha\ lag and shallowing \HI\ lags are completely reproduced by our analytic models (Figure \ref{fig:radiallaganalytic}). On the receding side of NGC\,4302, at least, tension between internal and external origins of the extraplanar \HI\ and \ha\ are alleviated by the potential dominating the kinematics. The difference in the kinematics between the approaching and receding sides of NGC\,4302 are, however, still unresolved.}

{\comment To conclude, measurements of radial variations in the lag cannot easily discriminate between an internal or external origin for the eDIG, as they are convolved with radial variations induced by the equilibrium potential. Perhaps large excursions from the analytic predictions could discriminate between an internal or external origin, but the uncertainties are large. }

\subsection{Trends with Galaxy Properties}
\label{ssec:dvdztrends}

\begin{figure*}
\label{fig:vvgtrends}
\centering
\includegraphics[width=\textwidth]{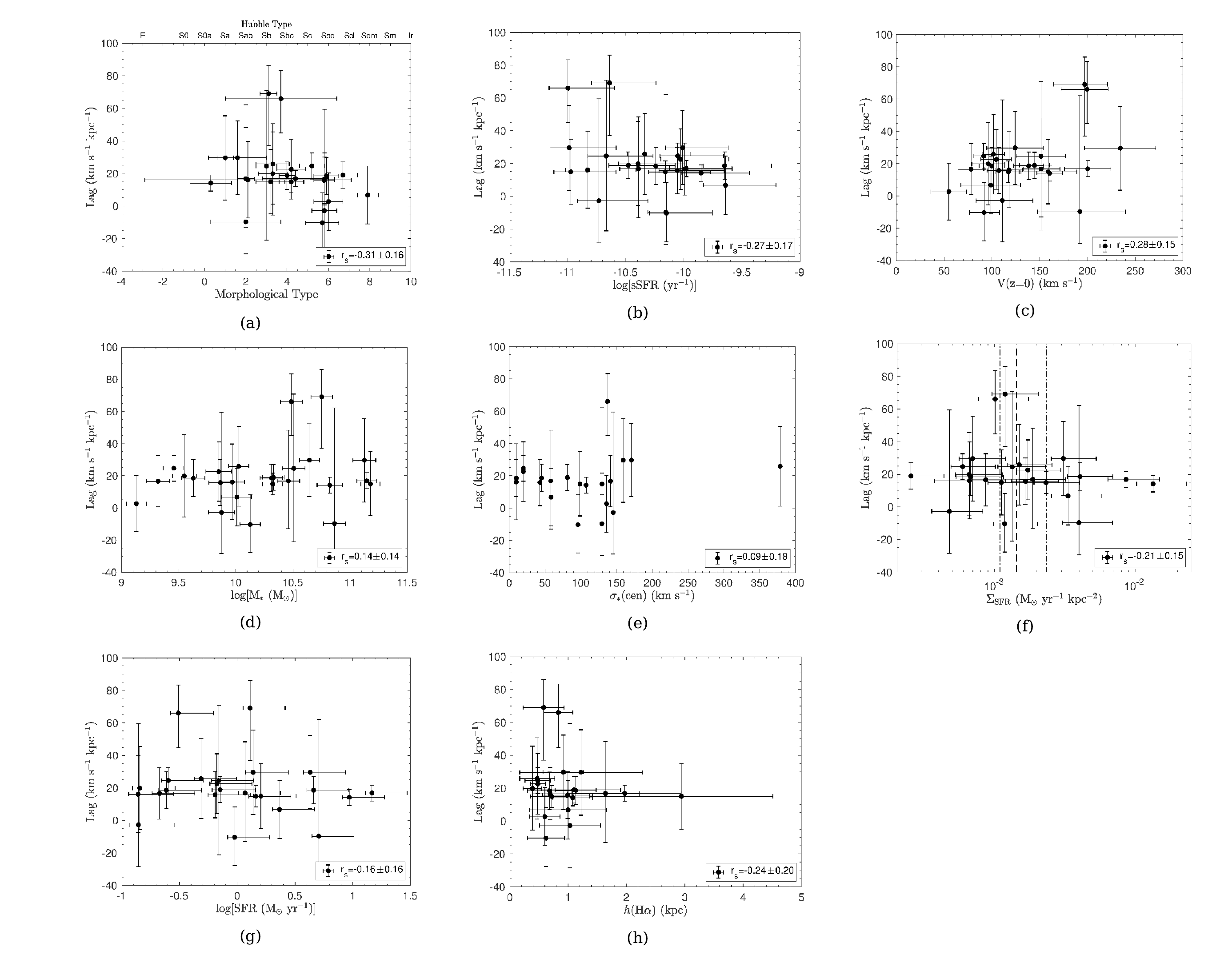}
\caption{\comment There are possible trends between the vertical gradient in the ionized gas rotation velocity (lag) and (a) the morphology.  We find no strong correlations between the measured ionized gas lag and the (b) sSFR, (c) rotation velocity in the midplane (V(z=0)), (d) stellar mass (M$_*$), (e) central stellar velocity dispersion ($\sigma_*$(cen)), (f)  SFR surface density ($\Sigma_{\rm SFR}\equiv$\,SFR/D$_{25}^2$), (g) SFR, or (h) measured \ha\ scale height (\hha). The vertical dashed and dot-dashed lines in (f) show the threshold $\Sigma_{\rm SFR}$ (and uncertainties) from \citet{rossa03a}.}
\end{figure*}

The relationship between the eDIG and global galaxy properties can give insights into the origin of the extraplanar gas. {\comment If the eDIG is a result of star formation activity in the disk (i.e. an internal origin), then trends between the eDIG scale height and/or the lag with some measure of the star formation activity might be expected \citep[e.g.][]{heald07}. To complicate matters, however, trends with the star formation activity may not be inconsistent with an external origin in a scenario where the accretion that creates the eDIG also powers the star formation. There is a relationship between the presence of eDIG and the SFR surface density ($\Sigma_{\rm SFR}$), where eDIG is ubiquitous for galaxies above a threshold $\Sigma_{\rm SFR}$ \citep{rand96,rossa03a}. From a sample of edge-on MaNGA galaxies, \citet{bizyaev17} find a trend between the \ha\ luminosity (a proxy for the SFR) and the eDIG scale height. Correlations between the magnitude of the lag and the star formation activity have yet to be found \citep[][and references therein]{zschaechner15a,bizyaev17,levy18}. \citet{levy18} did not find strong correlations between the difference between the molecular and ionized gas rotation velocity and any other global galaxy property, similar to previous photometric eDIG studies \citep[e.g.][]{rossa03b} as well as studies of extraplanar \HI\ \citep{zschaechner15b}. Previous evidence is, therefore, suggestive of an internal origin but inconclusive.}

We compare the measured lag to galaxy-wide properties, such as the \ha\ disk scale height (\hha), morphology, stellar mass (M$_*$), SFR, specific SFR (sSFR\,$\equiv$\,SFR/M$_*$), average SFR surface density ($\Sigma_{\rm SFR}\equiv$\,SFR/D$_{25}^2$), central stellar velocity dispersion ($\sigma_{*}({\rm cen})$), and V$_{\rm max}$. Morphologies and D$_{25}$ (the diameter of B-band isophote corresponding to 25 mag arcsec$^{-2}$) are from HyperLEDA; M$_*$ and $\sigma_{\rm *,cen}$ are from CALIFA; \hha\ measurements are described in Section \ref{sec:EOGSscaleheight}; {\comment V(z=0) is the intercept of the fitted vertical velocity gradient (Section \ref{ssec:vvg}, Table \ref{tab:EOGSparams}).} We derive the SFR (and hence sSFR and $\Sigma_{\rm SFR}$) from the WISE 22\,$\mu$m (W4) AB magnitudes reported by \citet{bitsakis19} which use the apertures from \citet{catalantorrecilla15}. {\comment We use the WISE W4 magnitudes to derive these quantities because extinction by dust in these edge-on systems renders the \ha\ extinction correction impossible, especially in the midplane.} We find inconsistencies in the magnitudes reported by \citet{bitsakis19} stemming from an improper application of a scaling factor to account for systematics (described in a few sentences). To convert from the AB magnitudes in \citet{bitsakis19} to the Vega magnitudes listed in Table \ref{tab:EOGSparams}, a value of 7.220 should be subtracted\footnote{This corrects for an older conversion between Vega and AB magnitudes used by \citet{bitsakis19} (6.604 instead of 6.620 \citep{jarrett11}) and an improper application of a scaling factor to account for systematics so that $7.220 = 6.604+(6.620-6.604/1.1)$}. It is known that star-forming galaxies measured with WISE W4 are systematically brighter by $\sim10$\% than inferred from \textit{Spitzer} IRS and 24\,$\mu$m observations \citep{wright10,jarrett13,brown14}, and we apply this scaling factor of 1/1.1 to the W4 Vega magnitudes. The W4 magnitudes in Table \ref{tab:EOGSparams} include this scaling factor. We then find the 22\,$\mu$m luminosity using L$(22\,\mu{\rm m}) = \frac{\nu_{\rm W4}}{\Delta\nu_{\rm W4}}F_{\rm W4}\times10^{-m_{\rm W4}/2.5}\times 4\pi d^2$, where $\nu_{\rm W4}$ is the W4 isophotal system frequency, $\Delta\nu_{\rm W4}$ is the W4 bandwidth, $F_{\rm W4}$ is the in-band flux \citep[all given in][]{jarrett11}, $m_{\rm W4}$ is the W4 Vega magnitude, and $d$ is the distance to the galaxy (both in Table \ref{tab:EOGSparams}). We then convert from L$(22\,\mu{\rm m})$ to SFR using the single-band linear calibrator without type-2 AGN developed by \citet{catalantorrecilla15}. We list our values of SFR inferred from the W4 magnitudes in Table \ref{tab:EOGSparams}\footnote{\citet{catalantorrecilla15} assume a Kroupa initial mass function (IMF) \citep{kroupa01} when deriving their SFR calibrators. A Salpeter IMF \citep{salpeter55} is assumed the EDGE and CALIFA surveys in general \citep{bolatto17,pipe3DII}. These different IMFs do not affect the results presented here, but this should be kept in mind when comparing the SFRs we derive here with other derivations of the SFR from the EDGE and CALIFA surveys.}. {\comment Finally, we use the values of M$_*$ and D$_{25}$ listed in Table \ref{tab:EOGSparams} to find sSFR$\equiv$SFR/M$_*$ and $\Sigma_{\rm SFR }\equiv{\rm SFR}$/D$_{25}^2$.}

We investigate whether there is a trend between the parameters by using the Spearman rank correlation coefficient (r$_{\rm s}$) which tests for any monotonic relationship between the two parameters. We estimate the uncertainty in r$_{\rm s}$ by using 1000 Monte Carlo iterations in which the values are allowed to randomly vary within the error bars, and r$_{\rm s}$ is computed for each iteration. The reported uncertainty on r$_{\rm s}$ is the standard deviation of the Monte Carlo iterations.

We find a weak inverse correlation between the lag and the morphology (r$_{\rm s}=-0.31\pm0.16$) (Figure \ref{fig:vvgtrends}a). This is similar to the trends with axis ratio and S\'ersic index found by \citet{bizyaev17} (with $r=0.48$ and $r=0.26$ respectively). We note, however, that determining the morphology, axis ratio, or S\'ersic index for an edge-on system is very difficult. {\comment We find no evidence for a trend with any of the other parameters (Figure \ref{fig:vvgtrends}b-h). The weak trends with sSFR, V(z=0), and $\Sigma_{\rm SFR}$ (classified by their non-zero correlation coefficients and uncertainties; Figure \ref{fig:vvgtrends}b,c,f) are ultimately not significant. If the two galaxies with the largest lags are removed, none of the weak trends in Figure \ref{fig:vvgtrends}b,c,f are significant. \citet{bizyaev17} found stronger trends between their lag measurements and V$_{\rm max}$ ($r=0.55$), M$_*$ ($r=0.5$), and $\sigma_{*}({\rm cen})$ ($r=0.55$). As pointed out by \citet{bizyaev17}, a trend with the maximum rotation velocity (V$_{\rm max}$ or V(z=0)) is likely an observational effect, since larger amplitude lags are easier to find in systems with large V$_{\rm max}$.}

If the extraplanar gas indeed has an internal origin, we might expect a correlation between the lag and $\Sigma_{\rm SFR}$. {\comment For gas ejected from the midplane through galactic fountains due to star formation activity, there should be some minimum level of widespread star formation (i.e. $\Sigma_{\rm SFR}$) needed to sustain a thick disk that covers the entire plane of the galaxy \citep{rand96}.} As in \citet{levy18}, we find no such trend (Figure \ref{fig:vvgtrends}f). {\comment From this idea of a minimum level of star formation activity,} \citet{rossa03a} define a threshold to have eDIG of L$_{\rm FIR}/$D$_{25}^2 = (3.2\pm0.5)\times 10^{40}$\,erg\,s$^{-1}$\,kpc$^{-2}$ \citep{rossa03a}, where L$_{\rm FIR}$ is the far infrared luminosity from 60 and 100\,$\mu$m emission. We convert L$_{\rm FIR}$ to L$_{\rm TIR}$ (total infrared, $8-1000\mu$m) using a factor of 1.6 \citep{sanders96} and then to a SFR using the single-band linear calibrator without type-2 AGN developed by \citet{catalantorrecilla15}. The threshold to have eDIG is $\Sigma_{\rm SFR} = 1.4^{+0.9}_{-0.3}\times10^{-3}$\,M$_{\odot}$\,yr$^{-1}$\,kpc$^{-2}$ (shown in Figure \ref{fig:vvgtrends}f). We find that $46^{+42}_{-21}$\% of galaxies in this subsample are above this threshold. We note that there are galaxies below this threshold with comparable lags to those above it (as indicated by the lack of trend).

The CALIFA data enable us to look for trends with properties of the stars---such as age, metallicity, and extinction---and to investigate the presence of vertical gradients in these parameters. Because these properties are degenerate, they are difficult to interpret. We present the results of this analysis in Appendix \ref{app:otherparams}, but the interpretation is beyond the scope of this paper.

{\comment As mentioned in Section \ref{sec:EOGSscaleheight}, we also test for trends between the eDIG scale height and the same global galaxy properties described here. We find no significant trends with any parameter.}

{\comment\subsection{Synthesis}}
\label{ssec:originsynthesis}
{\comment Determining the origin of the lagging extraplanar gas is difficult. Gradients in the lag with radius do not provide a straightforward diagnostic of the origin of the eDIG gas. The kinematic signatures imprinted by the equilibrium gravitational potential will likely overwhelm those relating to the origin of the eDIG material. We find no strong trends between the lag and galaxy properties, similar to previous studies of lagging \HI\ and eDIG \citep[e.g.][]{zschaechner15a,levy18}. We also do not find trends between the eDIG scale height and global galaxy properties. Therefore, although some evidence points toward an internal star formation-powered origin for the eDIG, it is possible that external inflows of gas may also contribute to the eDIG thickness.}

\section{The Ionization and Source of the Lagging Ionized Gas}
\label{sec:ionization}
 Extraplanar ionized gas can be found in thick gas disks from ejected material, galaxy bulges, the stellar thick disk, or outflows. {\comment These various locations and sources for the eDIG will have different ionization and other properties.} In the following subsections, we attempt to place limits on the source of the extraplanar ionized gas in these galaxies. 

It is generally thought that the eDIG is ionized primarily by leaky \HII\ regions \citep[see review by][and references therein]{haffner09}. Additional ionization sources are required, however, to explain the increase of ionized gas line ratios (\SII/\ha, \NII/\ha, \OIII/\hb) with height above the midplane {\comment \citep[e.g.][]{collins01,otte01,hoopes03}}. Such additional ionization mechanisms could include shocks \citep{rand98,collins01}, turbulent mixing layers \citep{rand98,binette09}, magnetic reconnection {\comment\citep{reynolds99,hoffmann12}}, cosmic rays \citep{wiener13}, photoelectric heating from small grains \citep{reynolds01a}, or hot, old, low-mass evolved stars {\comment\citep[HOLMES;][]{sokolowski91,floresfajardo11,weber19}.} {\comment For this paper, we will focus on ionization by HOLMES since we have the data in hand to constrain this mechanism, and many of the other processes are outside the scope of this paper. \citet{lopezcoba19} studied ionization by shocks in the context of galactic outflows using CALIFA.} HOLMES are commonly found in retired galaxies (galaxies which lie below the star-forming main sequence) and in the bulges and halos of disk galaxies and produce line ratios similar to low-ionization nuclear emission-line regions (LINERs) \citep{sarzi10,belfiore16,gomes16,lacerda18}. {\comment Recent ionization models of hot, young stars in the midplane of galaxies, however, can reproduce observed line ratios in the DIG indicating that additional ionization mechanisms may not be necessary \citep{weber19}.}

\begin{figure*}
\label{fig:BPT}
\centering
\includegraphics[width=\textwidth]{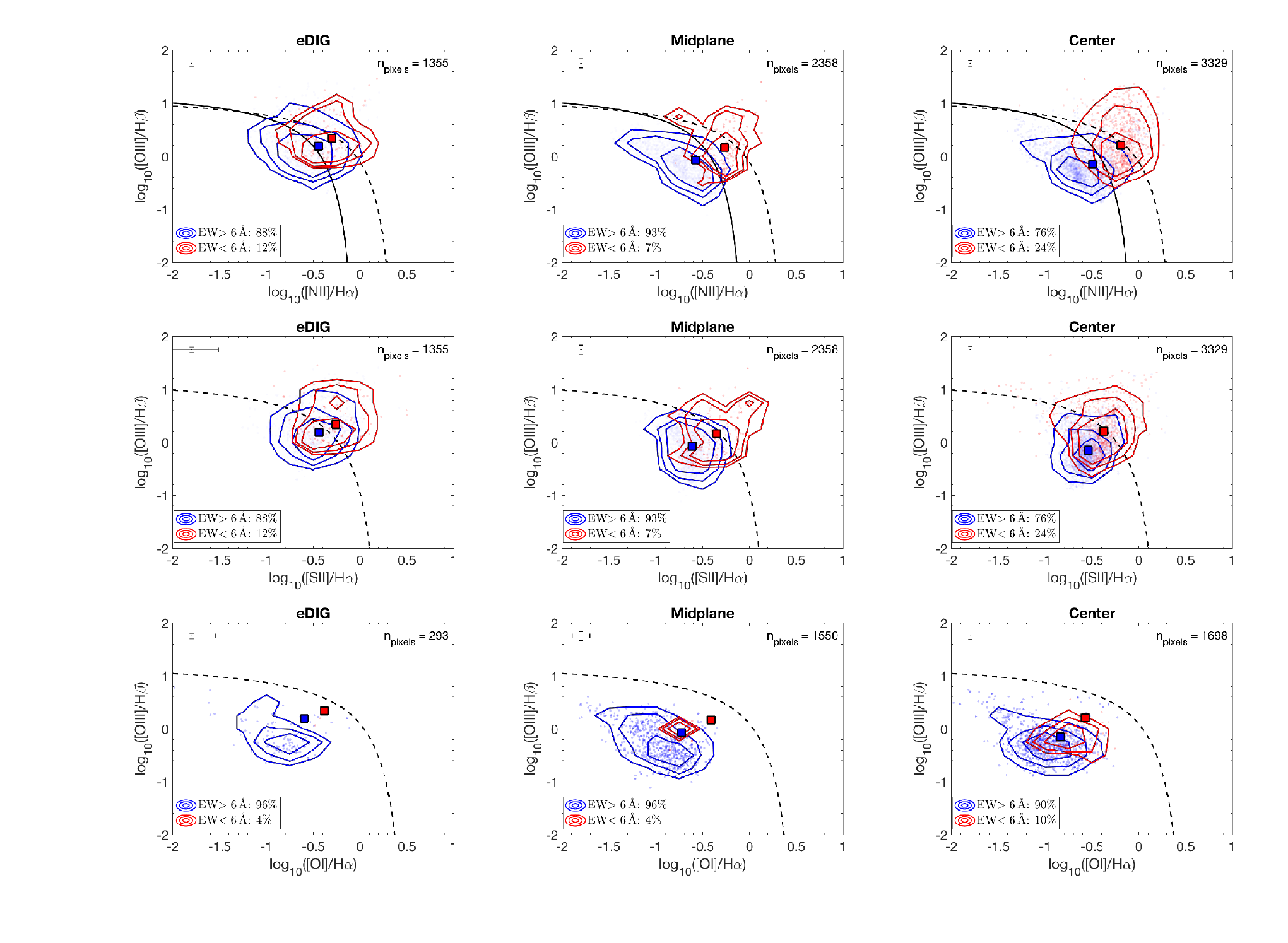}
\caption{Diagnostic diagrams for all 25 edge-on galaxies color-coded in bins of EW(\ha) \citep{sanchez14}. The left column plots pixels in the eDIG (i.e. excluding the center and midplane), the middle columns shows pixels in the midplane, and the right column shows pixels in the center. The top row uses \NII/\ha, the middle uses \SII/\ha, and the bottom uses \OI/\ha. Individual pixels are shown in the light points. Contours contain 50, 75, and 90\% of the pixels in each group. The large square symbols show the median of each group. Demarcations from \citet{kauffmann03} and \citet{kewley01} are shown (solid and dashed black curves). The percent of pixels in each group are in the legend, and the number of pixels in each region is in the upper right corner. Typical uncertainties for the individual pixels are shown in the upper left corner. {\comment The eDIG is primarily ionized by star-forming regions, rather than by the bulge, HOLMES, or some other process.} See Section \ref{sec:ionization} for details.} 
\end{figure*}

There are various diagnostic line-ratio diagrams used to investigate the ionization source of gas in galaxies. Here we will use \NII/\ha, \SII/\ha, and \OI/\ha\ versus \OIII/\hb\ and the \ha\ equivalent width (EW(\ha)) to infer the ionization properties of the eDIG in this sample. \citet{lacerda18} find that regions where the ionization is dominated by star formation have EW(\ha)\,$>$14\,\AA, and those with EW(\ha)\,$<$3\,\AA\ are dominated by HOLMES. These separations are fairly conservative; above 14\,\AA\ and below 3\,\AA, ionization will be dominated by star formation or HOLMES respectively. It is interesting, however, to include in the interpretation of intermediate EW(\ha). \citet{sanchez14} find a division between clumpy \HII\ regions and more diffuse gas at EW(\ha) = 6\,\AA. We adopt EW(\ha) = 6\,\AA\ as our nominal division between gas ionized by star forming \HII\ regions and more diffuse gas ionized by other means (HOLMES, shocks, etc). Maps of EW(\ha) are calculated in \pipetd\ for the CALIFA galaxies \citep{pipe3DI,pipe3DII}. We divide the galaxies in this sample into three spatial regions: eDIG, midplane, and center. We define the eDIG region as pixels with $|z|>$ \hha\ and $r>2$\,\rbulge, the midplane region as pixels with $|z|<$ \hha\ and $r>2$\,\rbulge, and the center region as pixels with $r<2$\,\rbulge. The diagnostic diagrams for each region color-coded by the EW(\ha) categories proposed by \citet{sanchez14} are shown in Figure \ref{fig:BPT}. {\comment Most of the eDIG is dominated by ionization from star-forming complexes ($\sim90$\%; Figure \ref{fig:BPT}). Only a small fraction is ionized by HOLMES ($\sim 10$\%). Although we do not analyze other ionization mechanisms (such as those mentioned in the introduction to this section), they must play a minor role since the eDIG ionization is dominated by star formation. A larger fraction of the midplane gas is ionized by star-forming complexes ($\sim95$\%), whereas a larger fraction of gas in the central regions is ionized by HOLMES ($\sim$20\%) and the distributions extend further above the \citet{kewley01} and \citet{kauffmann03} demarcation curves. If the more conservative EW(\ha) bins defined by \citet{lacerda18} are used instead, only 3\% of the eDIG is ionized by HOLMES, 66\% by star-forming regions, and 31\% by a mix of ionization processes \citep[these groups are not well separated; see][]{lacerda18}. In the center, 8\% is ionized by HOLMES and 45\% by a mix of ionization processes. We discuss these results in the context of the source of the ionized gas lags (stellar thick disk, bulge, outflows, and WIM-like gas) in the following subsections.}

\subsection{Is the Ionized Gas Related to a Stellar Thick Disk?}
\label{ssec:stellardisk}
It is possible that the lagging ionized gas at large heights off the midplane is associated with the stellar thick disk, rather than gas ejected from the midplane via star formation feedback. The stellar thick disk is known to rotate more slowly than the thin disk \citep[e.g.][]{pasetto12} and may have vertical gradients in the rotation velocity \citep[e.g.][]{spagna10}. {\comment The ionization of gas in the stellar thick disk is thought to be dominated by HOLMES \citep[][although see \citeauthor{weber19} \citeyear{weber19}]{floresfajardo11}. However, for eDIG regions---where we measure lags---ionization by HOLMES contributes only a small fraction ($\sim 10\%$; Figure \ref{fig:BPT}).} Therefore, we conclude that we are likely not measuring lags due to ionized gas associated with the stellar thick disk. 

\subsection{Contamination from a Bulge?}
\label{ssec:bulge}
It is also possible that we are measuring lags due to gas in the bulge which has larger velocity dispersions and lower rotation velocities. Although we exclude pixels with $r<2$\,\rbulge\ at all $z$, it is still possible that we are not masking all of the bulge emission. 

\begin{figure}
\label{fig:dvdzbulgeclass}
\centering
\includegraphics[width=\columnwidth]{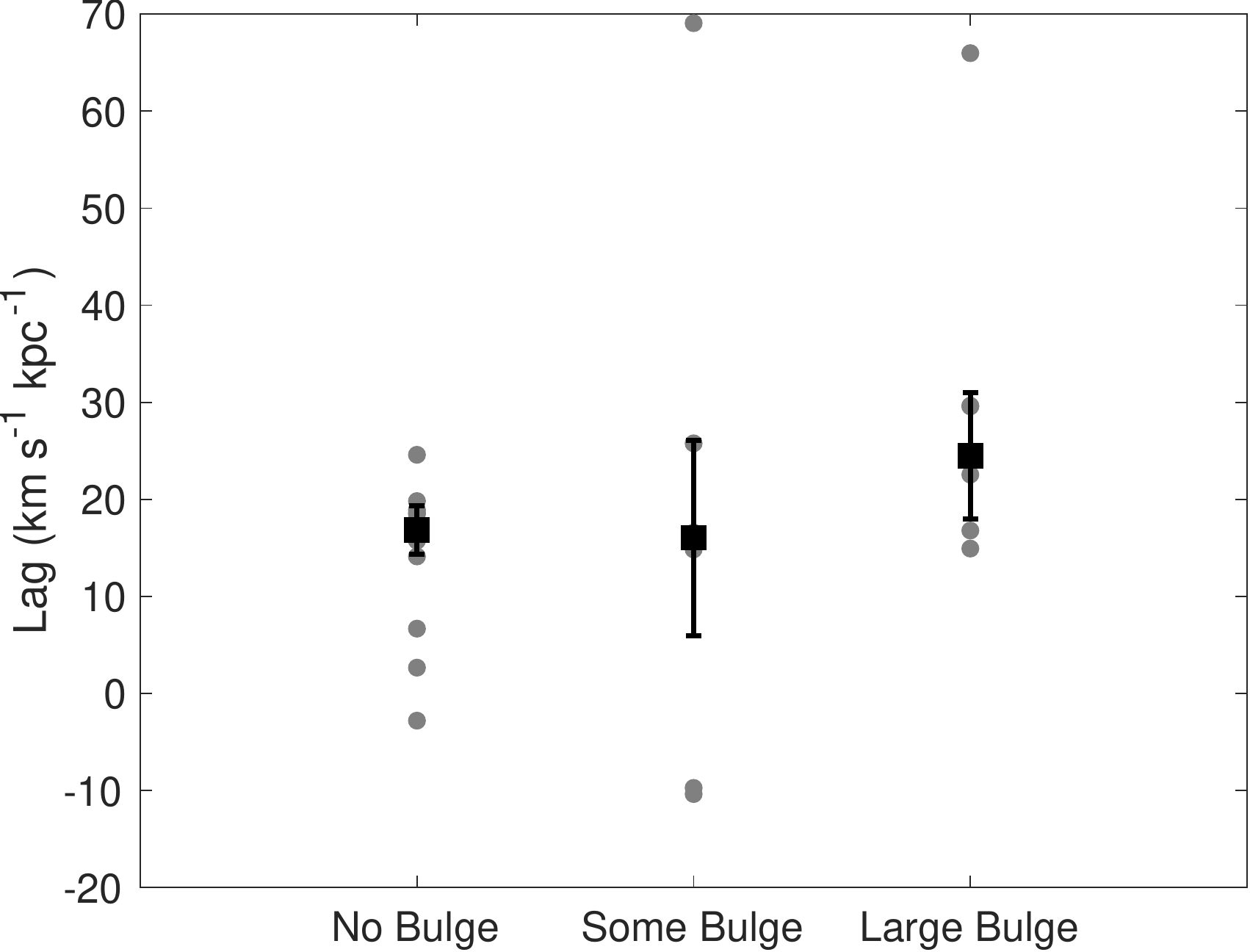}
\caption{There is no significant trend between the measured lag and the by-eye bulge classification. Points for individual galaxies are shown in gray (without error bars for clarity). Medians (with standard errors) for each bulge classification are shown in black squares and are consistent across bulge classifications.} 
\end{figure}

Because these galaxies are edge-on, there are no measurements of the bulge fractions or effective radii \citep[as was measured for the intermediately-inclined CALIFA galaxies by][]{mendezabreu17}. Making detailed decompositions of the bulges of these galaxies is out of the scope of this paper. Based on the SDSS images, we classified the galaxies into rough bulge groups by eye. As shown in Figure \ref{fig:dvdzbulgeclass}, there is no significant difference in lags measured for galaxies of different bulge classifications. 

We do find a weak inverse correlation between the amplitude of the lag and the Hubble type (Figure \ref{fig:vvgtrends}a). Earlier-type galaxies are more bulge-dominated and do seem to have larger lags, possibly indicating that the bulge is contributing to our measurement of the lags. We note, however, that the trend is not strong ($r_s=-0.31\pm0.16$) and that there are galaxies with lags $\sim20$\,\kmskpc\ across all Hubble types. We also note that morphological classifications for edge-on galaxies are highly uncertain. 

Finally, the ionization of gas in the bulge would be dominated by HOLMES \citep{sarzi10,belfiore16,gomes16}. From the line ratio diagrams (Figure \ref{fig:BPT}), we find that the fraction {\comment of the eDIG ionized by HOLMES} is small ($\sim10$\%). This fraction is higher and line ratios are more consistent with LINER-like emission in the center region where a bulge may dominate. We conclude that contamination from the bulge is not significantly affecting these results.

\subsection{Contamination from Ionized Gas Outflows?}
\label{ssec:outflows}
\citet{lopezcoba19} analyzed the prevalence of outflows in  edge-on CALIFA galaxies finding that 8\% are candidates to host outflows. A further 13\% present eDIG but do not meet their outflow criteria. There are four galaxies in our subsample which are outflow candidates and five which have eDIG according to \citet{lopezcoba19} (Notes O and E in Table \ref{tab:EOGSparams}). We note that our selection criteria are different so that our subsamples contain different galaxies. While we exclude the centers of galaxies from our analysis which should mitigate the effects of any outflows present, it is possible that this masking is insufficient as outflows can extend up to several kpcs towards the extraplanar regions. Therefore, the lags detected in these galaxies could be consequence of the presence of outflows.

To investigate the effect of outflows on our results, we mark outflow candidates in magenta in Figures \ref{fig:heightkde} and \ref{fig:KDEvvg}a. Neither those galaxies with the largest \ha\ scale heights (Figure \ref{fig:heightkde}) nor those with the largest lags (Figure \ref{fig:KDEvvg}a) are outflow candidates. We, therefore, conclude that outflows are not biasing our results in a significant way.

{\comment\subsection{The Lagging Ionized Gas is WIM-like}}
\label{ssec:eDIG}
Following the previous sections, we interpret the {\comment observed lagging extraplanar gas as a thick, turbulent eDIG layer, similar to the WIM in the MW, as opposed to gas associated with the bulge, stellar thick disk, or outflows. The majority of eDIG is ionized by star-forming regions (i.e. leaky \HII\ regions) as opposed to some other ionization source (Figure \ref{fig:BPT}). When two bins of EW(\ha) are used \citep[following][]{sanchez14}, $\sim$90\% the eDIG  is ionized by star formation (Figure \ref{fig:BPT}).}


\begin{figure*}
\label{fig:lineratios}
\centering   \gridline{\fig{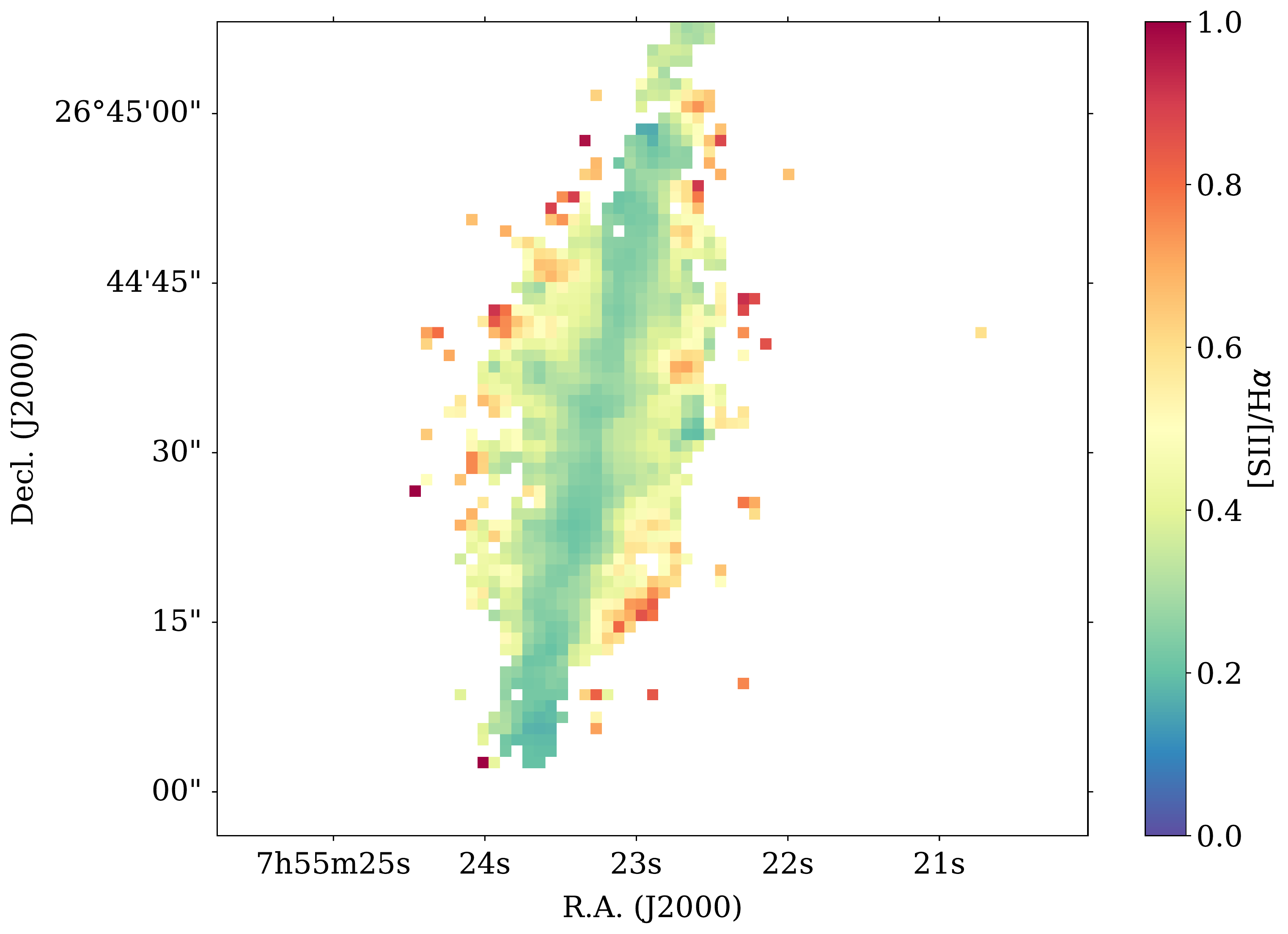}{0.33\textwidth}{(a)}
    \fig{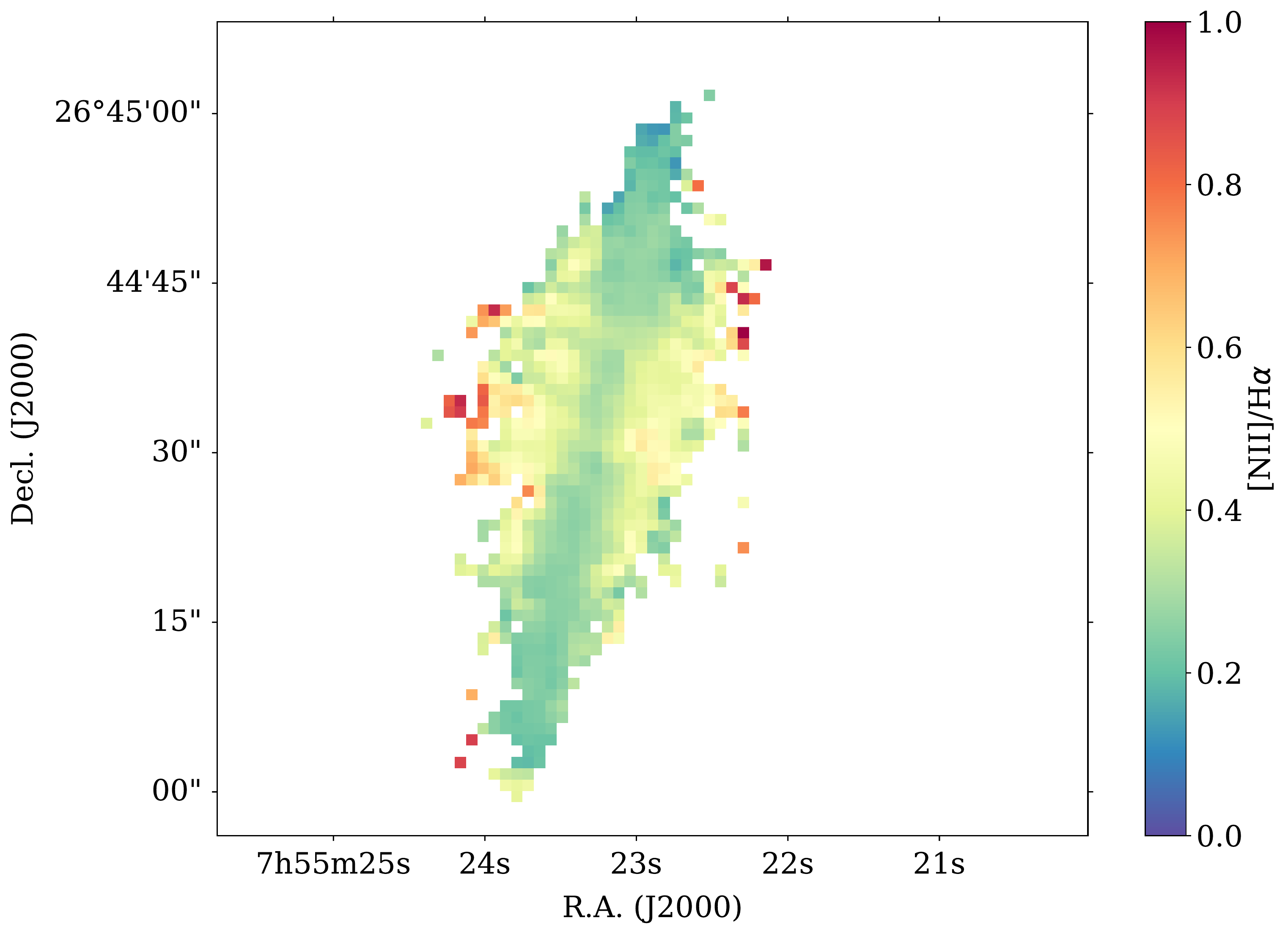}{0.33\textwidth}{(b)}}	\gridline{\fig{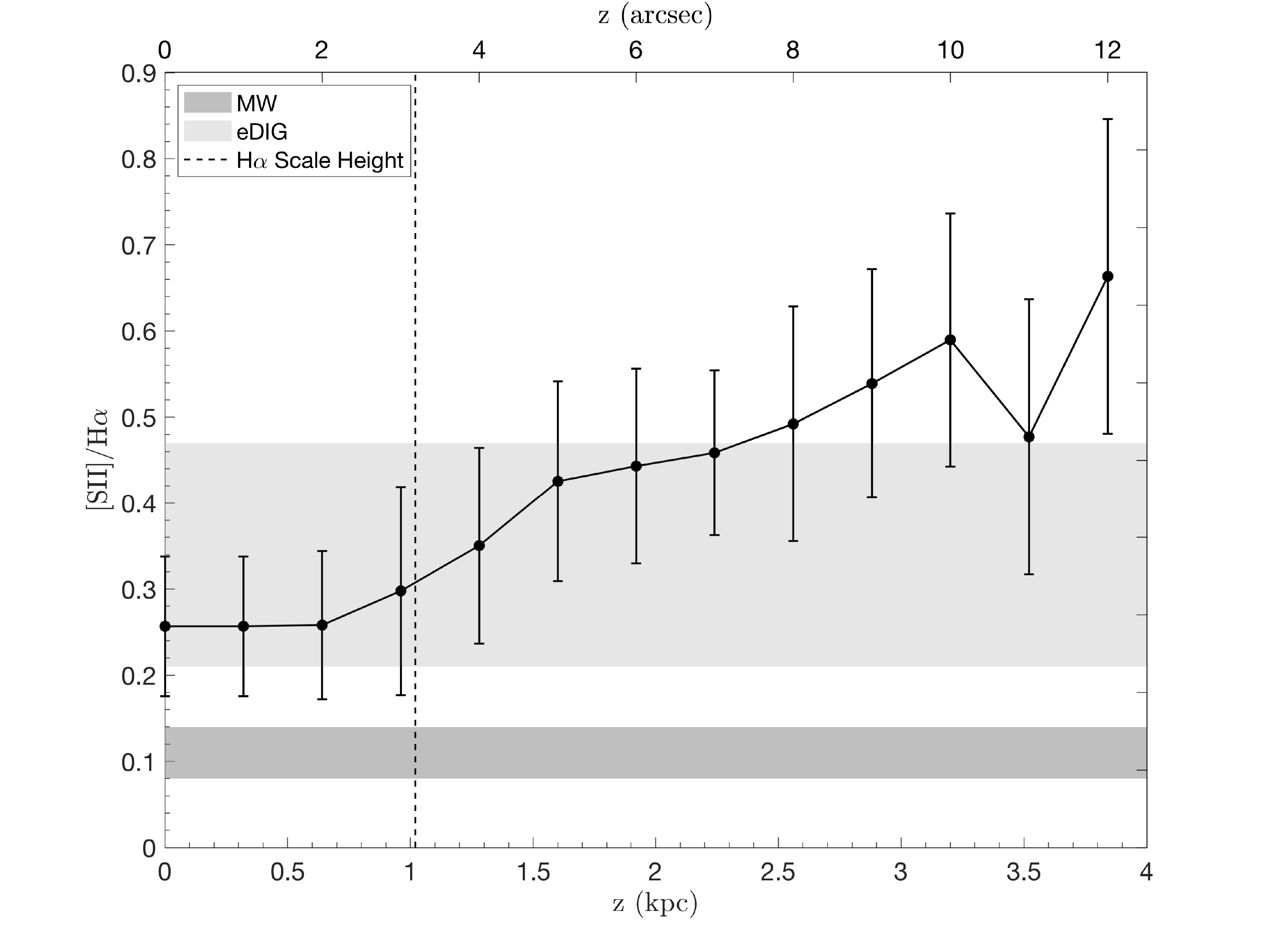}{0.42\textwidth}{(c)}
    \fig{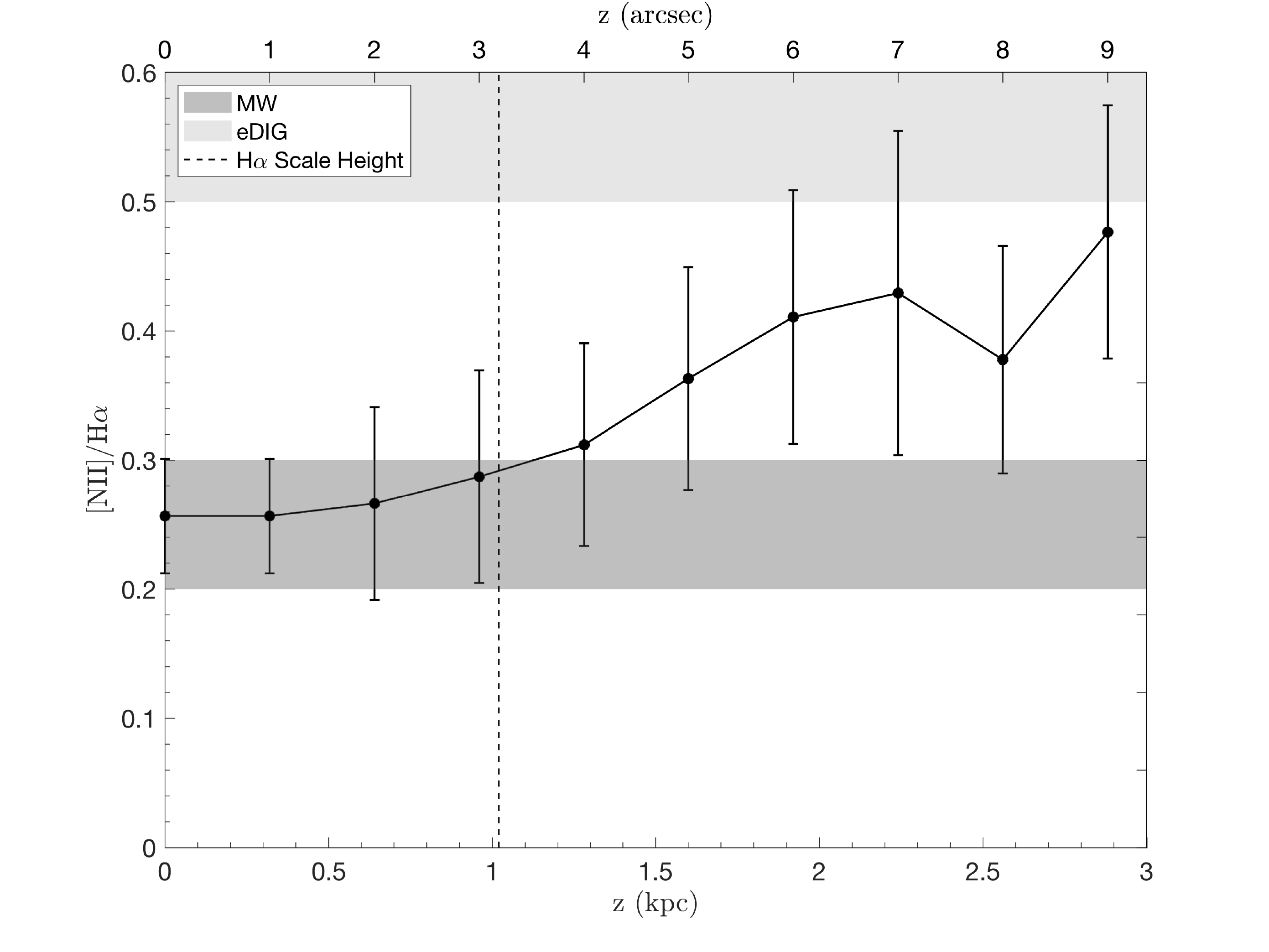}{0.42\textwidth}{(d)}}   \gridline{\fig{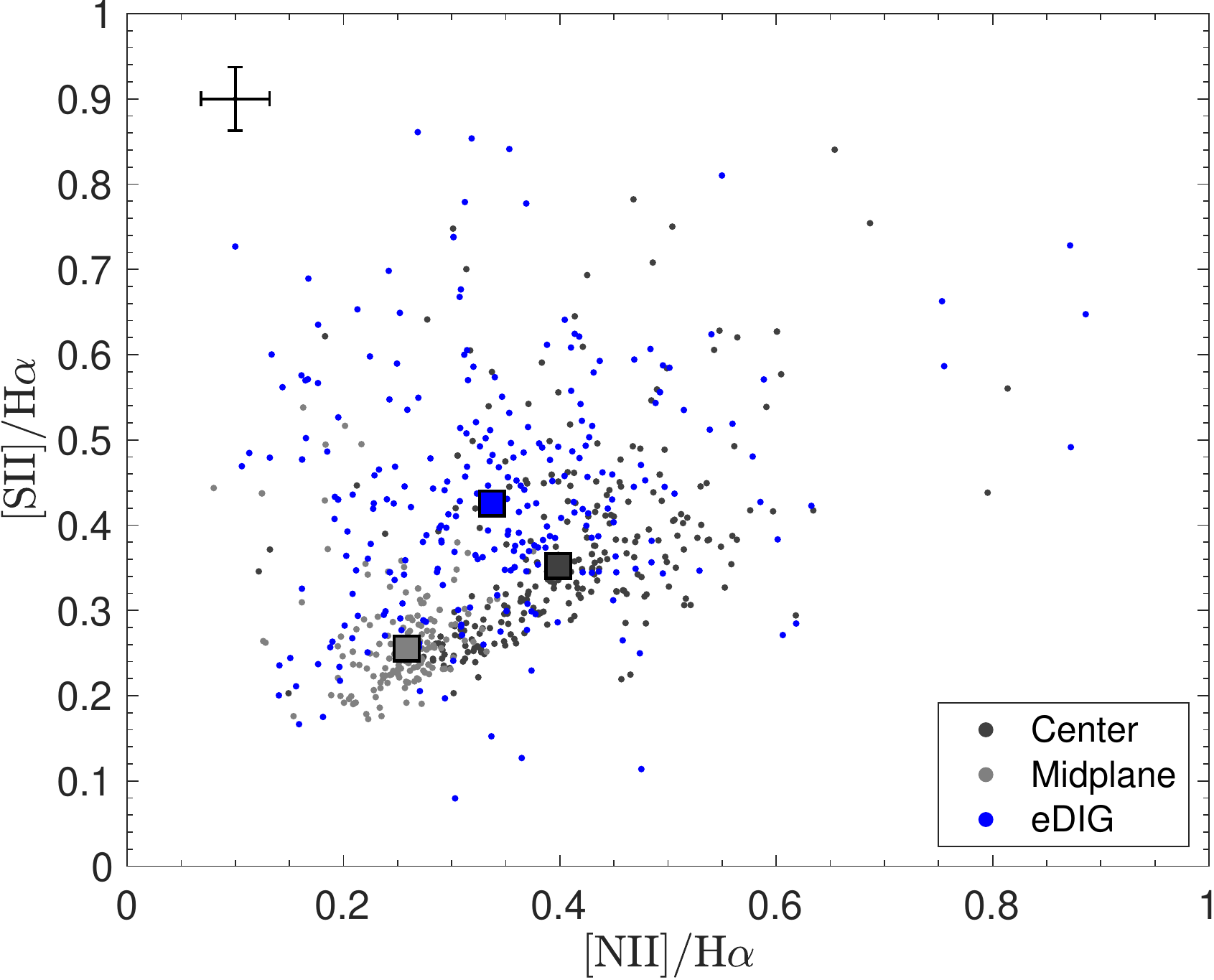}{0.42\textwidth}{(e)}
    \fig{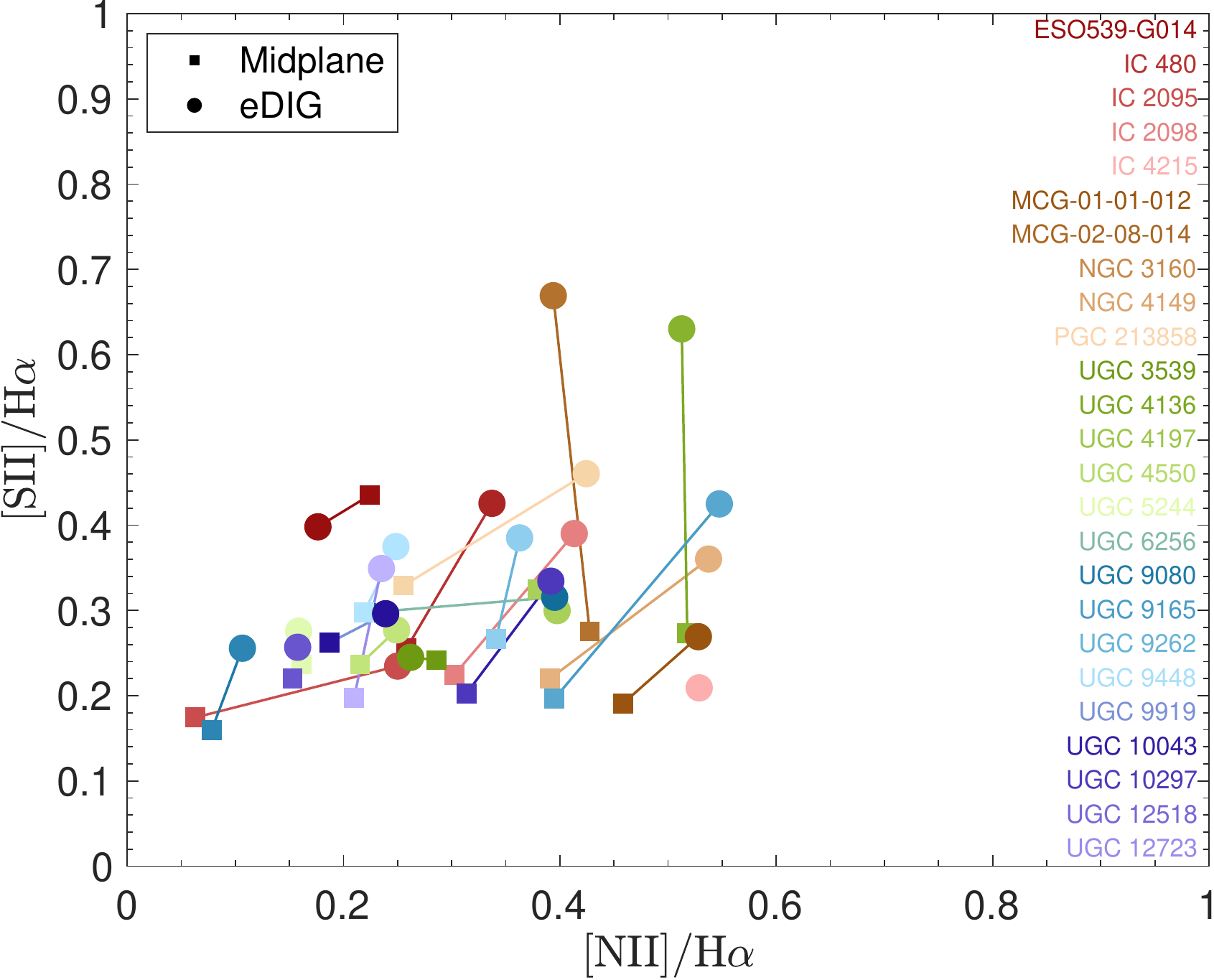}{0.42\textwidth}{(f)}}
\caption{For \defgal, the spatial distribution of extinction-corrected (a) \SII/\ha\ and (b) \NII/\ha. Median line ratios increase as a function of height above the midplane ($z$), as shown for (c) \SII/\ha\ and (d) \NII/\ha. (e) The \NII/\ha\ versus \SII/\ha\ for pixels with EW(\ha)\,$>6$\,\AA\ for \defgal. Points are separated by region in the galaxy. Representative error bars are shown in the upper left corner. Squares show the median of each region. As in the MW, the eDIG tends to have higher \SII/\ha\ than the midplane \citep[e.g.][]{haffner09}. (f) The median \NII/\ha\ versus \SII/\ha\ in the midplane (squares) and eDIG (circles) for all galaxies. Lines connect the midplane and eDIG points of the same galaxy (also shown in the colors). The eDIG of most galaxies has larger \SII/\ha\ than the midplane. Similar figures for (c), (d), and (e) are shown in Figure \ref{fig:FS} for the other edge-on galaxies.} 
\end{figure*}

Several line ratios are used as diagnostics of eDIG, as they trace variations in temperature and density. The ratios of \NII$\lambda$6583\AA\ and \SII$\lambda$6717\AA\ to \ha\ are strong functions of temperature and also trace variations in abundance and ionization \citep[see the review by][]{haffner09}. Previous studies show that \SII/\ha\ and \NII/\ha\ clearly increase with distance from the midplane in the eDIG, indicating a general increase in temperature with height and large variations in temperature and ionization fraction in the eDIG {\comment\citep[e.g.][]{madsen06,haffner09,ho16,jones17}}. In the midplane of the MW and a few other galaxies, \SII/\ha\,$=0.11\pm0.03$ and \NII/\ha \,$\sim0.25$ \citep{madsen04,madsen06}, whereas \SII/\ha\,$=0.34\pm0.13$ and \NII/\ha\,$\gtrsim 0.5$ in the eDIG \citep{blanc09,madsen04}. We note that some photoionization models of HOLMES have also been able to reproduce the observed line ratios \citep{floresfajardo11}{\comment, but other models of young, massive stars in the midplane can reproduce the observed line ratios and trends \citep[e.g.][]{barnes14,weber19}. In a stacked analysis of MaNGA galaxies, \citet{jones17} find a potential transition in the heating source from \HII\ regions to HOLMES, shocks, etc. at $\sim$6 kpc from the midplane based on the ratio of \OII/\ha. Our observations do not probe out to these distances (Figures \ref{fig:lineratios}c,d and \ref{fig:FS}f,g), which is consistent with midplane \HII\ regions dominating the ionization of these eDIG measurements.} 

CALIFA provides spatially-resolved maps of \SII\ and \NII\ at high SNR for the galaxies studied here. We mask pixels with \SII\ and \NII\ SNR\,$< 3$ and \ha\ SNR\,$<5$. We correct each line for extinction using the \citet{calzetti00} extinction curve. We note that this extinction correction will be insufficient in the midplane of these edge-on galaxies. We also apply a SNR cut of 3 in the line ratio maps (Figure \ref{fig:lineratios}a,b). To investigate trends in \SII/\ha\ and \NII/\ha\ with distance above the midplane, we take a radial average at each height (excluding radii\,$< 2$\,\rbulge). Indeed we find systematic increases in the line ratios with height above values found in the midplane (Figures \ref{fig:lineratios}c,d and \ref{fig:FS}f,g). 

Studies of the WIM in the MW reveal significant spatial variations in temperature and ionization fraction \citep[e.g.][]{haffner09}. These can be studied by comparing the \SII/\ha\ and \NII/\ha\ line ratios. Because the excitation temperatures of \SII\ and \NII\ are similar, \SII/\NII\ is nearly independent of the electron temperature (T$_{\rm e}$) \citep[see equation 1 in][]{haffner09}. Variations in \SII/\NII\ trace variations in S$^+$/N$^+$,  N/H,  and S/H. The \NII/\ha\ and \SII/\ha\ line ratios primarily trace variations in T$_{\rm e}$, although they are also sensitive to variations in the abundance of N or S \citep[see equation 2 in][]{haffner09}. In the MW, \citet{haffner09} find that the WIM has larger \SII/\ha\ ratios than regions near \HII\ regions (see their Figure 2). We plot the \SII/\ha\ line ratio against the \NII/\ha\ line ratio for the eDIG, midplane, and center regions for each galaxy (Figures \ref{fig:lineratios}e and \ref{fig:FS}h). Figure \ref{fig:lineratios}f shows the median line ratios for the eDIG and midplane for each galaxy in this sample. Indeed for many galaxies in this subsample, the eDIG {\comment region has} higher \SII/\ha\ ratios than in the midplane or center indicating variations in the S$^+$/S or S/H ratios as is seen in the MW. 

From this analysis of the line ratio and ionization diagnostics, we conclude that the eDIG in these galaxies is primarily ionized by photons escaping from \HII\ regions in the midplane, similar to the WIM in the Galaxy \citep[e.g.][]{haffner09}. {\comment We note that although the ionization is dominated by midplane \HII\ regions, this analysis of the ionization does not constrain whether the extraplanar gas originated internal or external to the galaxy (as was investigated in Section \ref{sec:origin}).}

\subsection{Synthesis}
To summarize this section, we investigate the source of the lagging ionized gas through the ionization properties. We consider four potential sources for the lagging ionized gas: gas in the stellar thick disk, gas in the bulge, outflows, and {\comment gas in a thick disk ionized by star-forming regions in the midplane}. From the diagnostic line ratio diagrams (Figure \ref{fig:BPT}), we show that the line ratios and EW(\ha) are consistent with ionization by star-forming regions (i.e. leaky \HII\ regions) dominating in the eDIG region. The fraction of ionization by HOLMES is small, unlike what is expected in the stellar thick disk \citep[e.g.][]{floresfajardo11,lacerda18} or bulge. We find no trend between the bulge fraction of these galaxies and the lag, again indicating that we are not being biased by incomplete masking of the bulge regions. We also rule out bias from outflows, as those galaxies with ionized gas outflows \citep{lopezcoba19} do not dominate the \ha\ scale height or lag distributions. Evidence for the lagging gas coming from an eDIG layer similar to the WIM in the MW is (1) ionization dominated by star-forming regions, (2) increasing \SII/\ha\ and \NII/\ha\ line ratios with height above the midplane which is characteristic of the eDIG, and (3) elevated \SII/\ha\ versus \NII/\ha\ in the eDIG region compared to the midplane.


\section{Summary}
\label{sec:summary}
We investigate the presence, properties, and kinematics of eDIG in a sample of 25 edge-on galaxies from the CALIFA survey. Much of the motivation for the study was an extension of the kinematic detection of eDIG in intermediate inclination EDGE-CALIFA galaxies by \citet{levy18}. We summarize our main results below, indicating relevant figures and/or tables. 
\begin{enumerate}
\itemsep0em
\item We measure exponential ionized gas scale heights ($h({\rm\ha}$)) in {\comment $\sim$90\% of this sample. We find $h({\rm\ha}) = 0.3-2.9$\,kpc, with a median $h({\rm\ha}) =0.8$\,kpc} over the sample (Table \ref{tab:EOGSparams}, Figure \ref{fig:heightkde}). These values are consistent with previous measurements of eDIG scale heights {\comment\citep[e.g.][]{rand97,wang97,hoopes99,collins00,collins01,miller03I,rosado13,bizyaev17,levy18}.} 
\item We investigate the rotation velocity (\vrot) as a function of height above the midplane ($z$) by constructing \ha\ PV diagrams in vertical bins. By radially averaging, we find a systematic decrease in \vrot$(z)$ in many of our sources (Figures \ref{fig:vvg} and \ref{fig:FS}e). We use a linear fit to find the vertical gradient in the rotation velocity (\DvDz, where the lag is -\DvDz) (Table \ref{tab:EOGSparams}). We find significant lags in 60\% of our galaxies with measurable lags ranging from 10--70\,\kmskpc. There are no galaxies for which \vrot\ increases with $z$. This is consistent with previous measurements of ionized gas lags in the literature (Figure \ref{fig:KDEvvg}).
\item We investigate radial variations in the lag, which may indicate an internal origin of the eDIG material. We find no indication of systematic shallowing (or steepening) of the lag with radius as is often seen in \HI\ (Figure \ref{fig:radiallag}, Table \ref{tab:EOGSparams}). Moreover, our analytic modeling indicates that radial variations in the lag are induced by the potential, regardless of the origin of the extraplanar gas (Figure \ref{fig:radiallaganalytic}). Disentangling this effect in order to use radial variations in the lag to deduce the origin of the extraplanar gas will be difficult.
\item There are no strong trends between \DvDz\ and global galaxy properties (Figure \ref{fig:vvgtrends}). There is potentially an inverse correlation between the lag and morphology (r$_{\rm s}=-0.31$; Figure \ref{fig:vvgtrends}a), but morphologies for edge-on galaxies are very uncertain. 
\item We use the ionization properties to discriminate among four potential sources for the lagging ionized gas: gas in the stellar thick disk, gas in the bulge, outflows, and extraplanar gas ionized by photons from star-forming regions. From the diagnostic line ratio diagrams, we show that the line ratios and EW(\ha) in the eDIG are consistent with ionization dominated by star-forming regions (Figure \ref{fig:BPT}). The fraction of {\comment the eDIG} ionized by HOLMES is small, unlike what is expected in the stellar thick disk or bulge. Evidence for an eDIG layer due to star formation is ionization dominated by star-forming regions (Figure \ref{fig:BPT}), increasing \SII/\ha\ and \NII/\ha\ line ratios with height above the midplane (Figures \ref{fig:lineratios}c,d and \ref{fig:FS}f,g), and elevated \SII/\ha\ versus \NII/\ha\ in the eDIG region compared to the midplane (Figures \ref{fig:lineratios}e,f and \ref{fig:FS}h). We can further rule out bulge contamination because there is no trend between the lag and bulge fraction (Figure \ref{fig:dvdzbulgeclass}). We also find no systematic influence from outflows on either $h(\rm\ha)$ or \DvDz\ (Figures \ref{fig:heightkde} and \ref{fig:KDEvvg}a). We, therefore, conclude that the lags are indeed due to a thick, eDIG layer similar to the WIM in the MW.
\end{enumerate}
To extend this work, it would be useful to measure the ionized gas velocity dispersions. It is thought that the increased velocity dispersion of the gas acts effectively as an additional pressure term so the gas can remain above or below the midplane \citep[e.g.][]{burkert10,marinacci10}. Due to the low spectral resolution of CALIFA, the ionized gas velocity dispersions cannot be measured. Higher spectral resolution observations of these galaxies are, therefore, necessary. It is also extremely useful to have spatially resolved observations of a dynamically cold tracer (such as CO) to compare with the ionized gas. This would allow us to determine whether the molecular and ionized gas rotation velocities indeed agree in the midplane, to measure the thin disk scale height, and to measure the velocity dispersion in the dynamically cold component. All of these measurements will help contextualize the ionized gas results. The EDGE Survey provides CO measurements for three of the galaxies studied here, but deeper CO observations in more galaxies (and with better spatial resolution if possible) are necessary. Finally, the neutral atomic gas (\hi) is also observed to lag the midplane in some systems \citep[e.g.][]{oosterloo07,kamphuis13,zschaechner11,zschaechner12,zschaechner15a,zschaechner15b}. More spatially resolved \hi\ measurements in systems with resolved ionized gas data will enable comparison between the neutral and ionized gas lags in a larger sample.

\acknowledgments
{\comment R.C.L. would like to thank Andrew Harris, Laura Lenki\'{c}, and Sylvain Veilleux for useful discussions and advice. The authors would like to thank the anonymous referee for their thorough reading and constructive comments.} R.C.L. and A.D.B. acknowledge support from the National Science Foundation (NSF) grants AST-1412419 and AST-1615960. A.D.B. also acknowledges visiting support by the Alexander von Humboldt Foundation. S.F.S. thanks CONACYT grant CB285080 and funding from the PAPIIT-DGAPA-IA101217 (UNAM) project. P.T. and S.N.V. acknowledge support from NSF AST-1615960. L.B. and D.U. are supported by the NSF under grants AST-1140063 and AST-1616924. D.C. acknowledges support by the Deutsche Forschungsgemeinschaft, DFG, through project number SFB956C. T.W. acknowledges support from the NSF through grants AST-1139950 and AST-1616199.  This study makes use of data from the EDGE (\url{www.astro.umd.edu/EDGE}) and CALIFA (\url{http://califa.caha.es}) surveys and numerical values from the HyperLeda database (\url{http://leda.univ-lyon1.fr}). Support for CARMA construction was derived from the Gordon and Betty Moore Foundation, the Kenneth T. and Eileen L. Norris Foundation, the James S. McDonnell Foundation, the Associates of the California Institute of Technology, the University of Chicago, the states of California, Illinois, and Maryland, and the NSF. CARMA development and operations were supported by the NSF under a cooperative agreement and by the CARMA partner universities. This research is based on observations collected at the Centro Astron\'{o}mico Hispano-Alem\'{a}n (CAHA) at Calar Alto, operated jointly by the Max-Planck Institut f\"{u}r Astronomie (MPA) and the Instituto de Astrofisica de Andalucia (CSIC). Funding for the Sloan Digital Sky Survey III has been provided by the Alfred P. Sloan Foundation, the U.S. Department of Energy Office of Science, and the Participating Institutions. The SDSS web site is \url{www.sdss3.org}. SDSS-III is managed by the Astrophysical Research Consortium for the Participating Institutions of the SDSS-III Collaboration including the University of Arizona, the Brazilian Participation Group, Brookhaven National Laboratory, Carnegie Mellon University, University of Florida, the French Participation Group, the German Participation Group, Harvard University, the Instituto de Astrof\'sica de Canarias, the Michigan State/Notre Dame/JINA Participation Group, Johns Hopkins University, Lawrence Berkeley National Laboratory, Max Planck Institute for Astrophysics, Max Planck Institute for Extraterrestrial Physics, New Mexico State University, New York University, Ohio State University, Pennsylvania State University, University of Portsmouth, Princeton University, the Spanish Participation Group, University of Tokyo, University of Utah, Vanderbilt University, University of Virginia, University of Washington, and Yale University. {\comment This research made use of APLpy\footnote{\label{foot:aplpy}\url{https://aplpy.github.io}}, an open-source plotting package for Python, Astropy\footnote{\label{foot:astropy}\url{www.astropy.org}}, a community-developed core Python package for Astronomy,  MatPlotLib\footnote{\label{foot:mpl}\url{www.matplotlib.org}}, NumPy\footnote{\label{foot:numpy}\url{www.numpy.org}}, and pandas\footnote{\label{foot:pandas}\url{www.pandas.pydata.org}}.}

\facilities{CAO:3.5, CARMA, Sloan, \comment{Spitzer}}
\software{\comment{APLpy\textsuperscript{\ref{foot:aplpy}} \citep{aplpy}, Astropy\textsuperscript{\ref{foot:astropy}} \citep{astropyI,astropyII},  MatPlotLib\textsuperscript{\ref{foot:mpl}} \citep{matplotlib}, Miriad \citep{miriad},  NumPy\textsuperscript{\ref{foot:numpy}} \citep{numpyI,numpyII}, pandas\textsuperscript{\ref{foot:pandas}} \citep{pandas}, \pipetd\ \citep{pipe3DI,pipe3DII}}}

\bibliographystyle{yahapj}

\appendix
\section{Effects of Extinction and Inclination}
\subsection{Extinction Effects}
\label{app:extinction}

\begin{figure}
\label{fig:extsim}
\centering
    \gridline{\fig{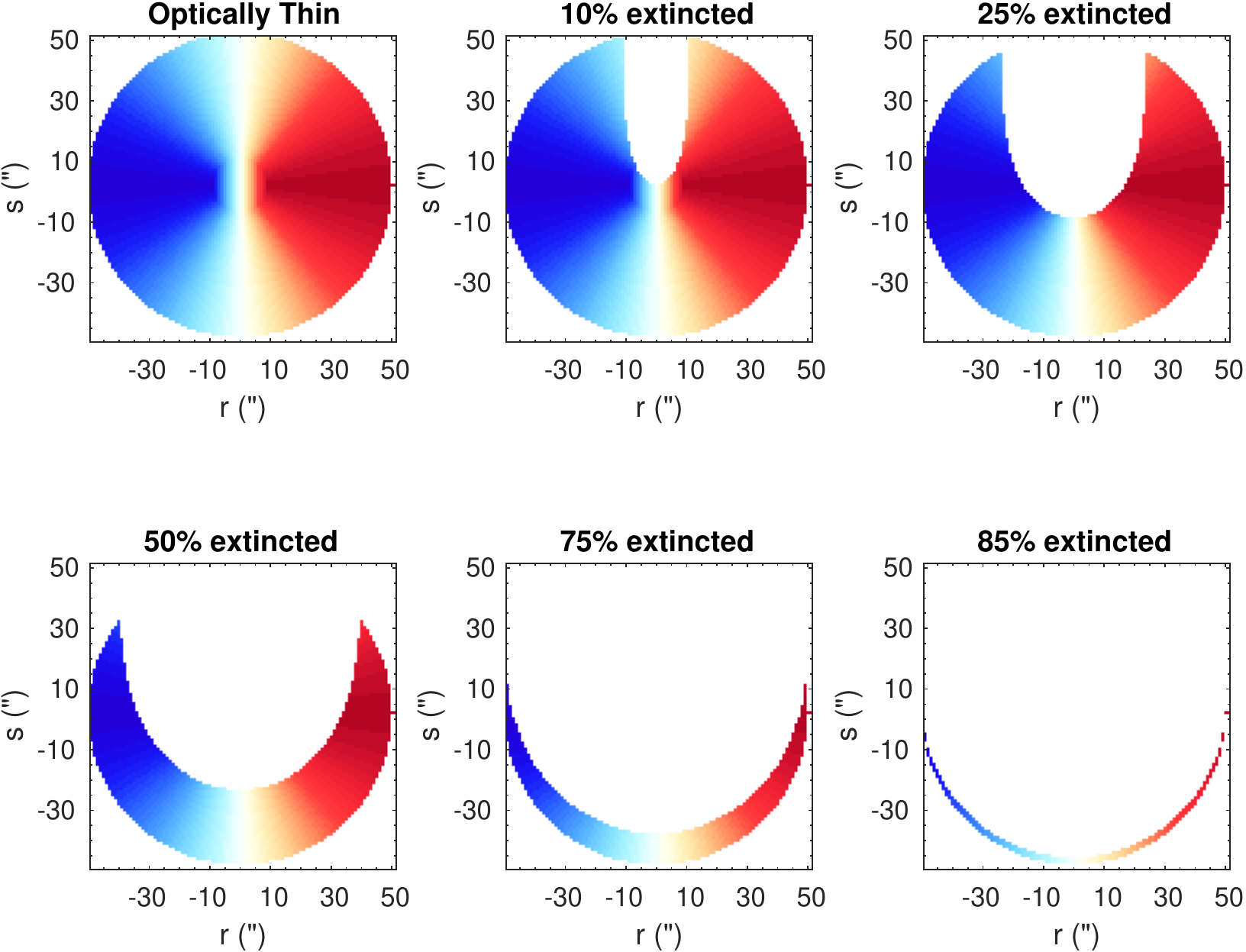}{0.5\columnwidth}{(a)}
    \fig{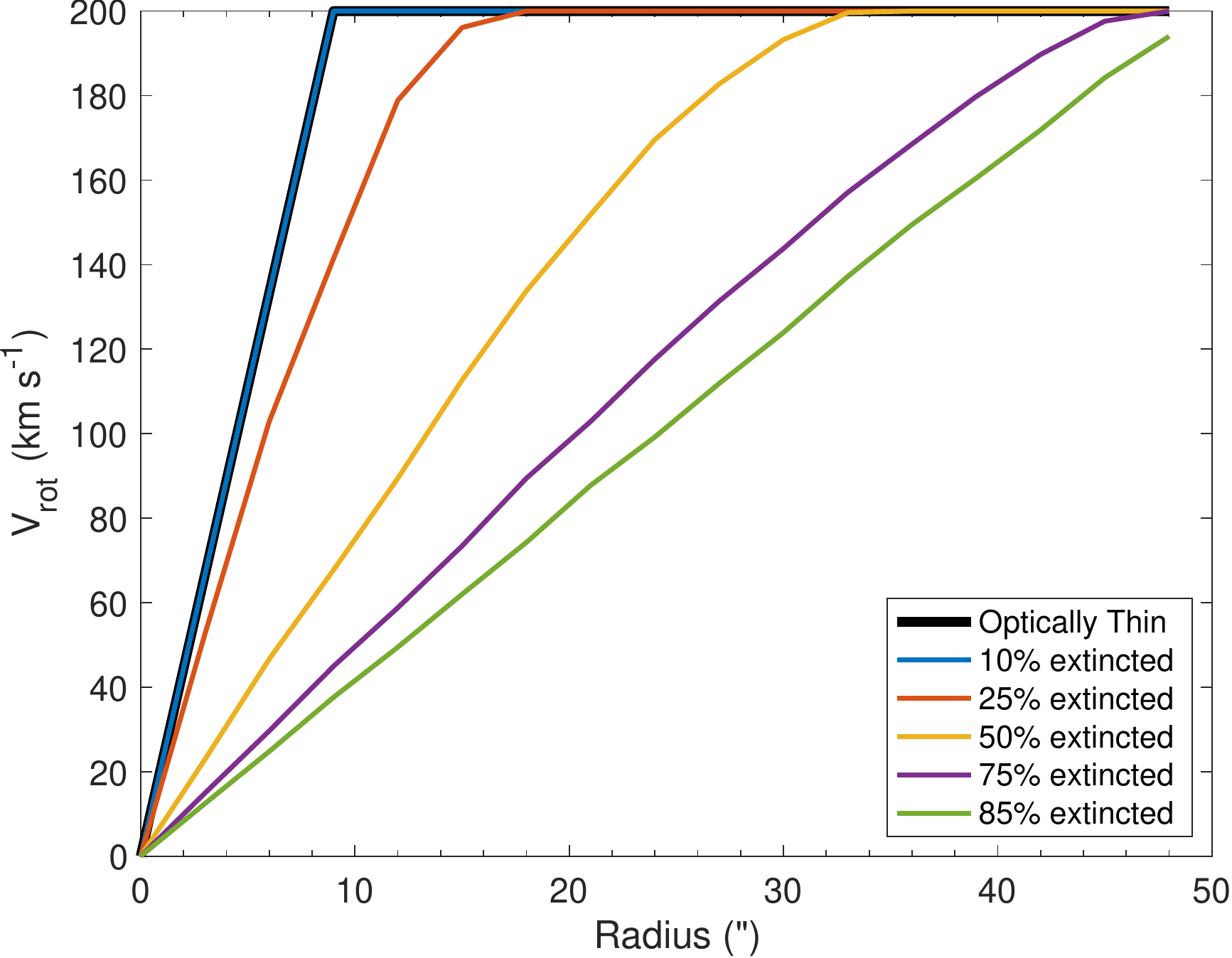}{0.5\columnwidth}{(b)}}
\caption{(a) Our simulated velocity fields with different amounts of dust extinction along the line of sight (quoted as the percent of the disk area extincted). The observer is in the plane of the page looking up. The color-scale ranges from -200--200\,\kms.  (b) The rotation curves resulting from the simulated velocity fields above. With increasing extinction, the rotation curves becomes more linear rather than flattening.} 
\end{figure}

While the \ha\ flux maps are extinction corrected, this correction is insufficient for the midplanes of the edge-on systems analyzed here. Extinction will be less severe away from the disk midplane, but it is nevertheless difficult to correct for. We investigate the effect of extinction on our results in two ways. First, we perform simulations to investigate the effects of extinction on the shape of the rotation curves. We create a thin disk with an input rotation curves which rises linearly and then flattens at some turnover radius (first panel in Figure \ref{fig:extsim}a and the black curve in Figure \ref{fig:extsim}b). Part of the disk is obscured above some level of extinction along the line of sight. This is shown in Figure \ref{fig:extsim}a, where the observer is in the plane of the page looking up into the disk, where the percent of the disk area extincted by dust is quoted above the panels. We then measure the rotation curve by taking a PV cut along the major axis. As shown in Figure \ref{fig:extsim}b, as the extinction increases, the rotation curve becomes more linear, approaching solid-body rotation. This result has been known for decades \citep[e.g.][]{goad81,bosma92}, but bears repeating here. In the context of this paper, ionized gas rotation velocities may be underestimated in the midplane from this effect. With increasing distance from the midplane, however, the extinction will be less and the true rotation velocity can be recovered. Since our measured vertical gradients in the rotation velocity extend a few kpc from the midplane and still show this linear decrease in the rotation velocity, we do not expect that extinction is greatly affecting our measured gradients.  

\begin{figure}
\label{fig:extgrad}
\centering
\gridline{\fig{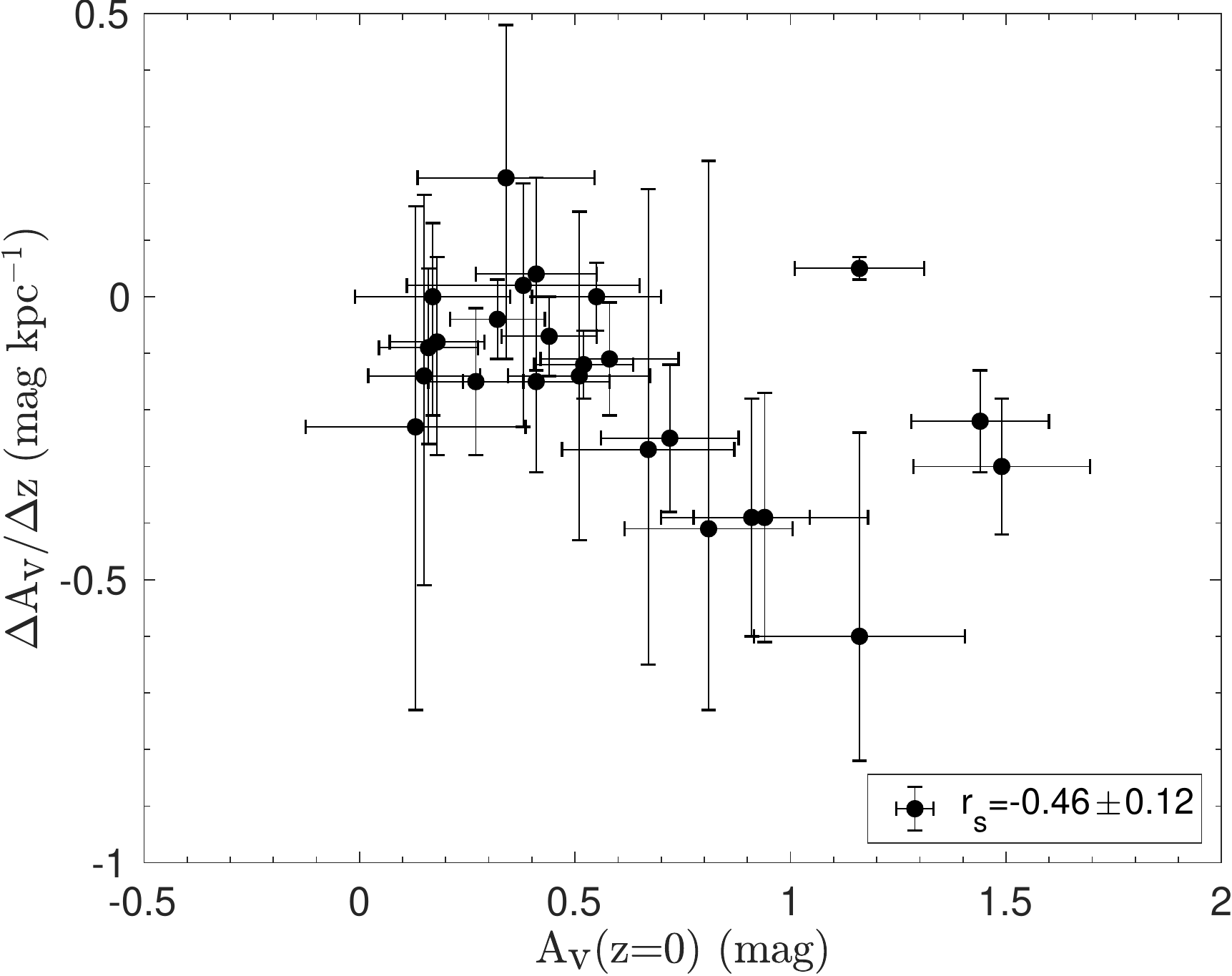}{0.3\columnwidth}{(a)}
\fig{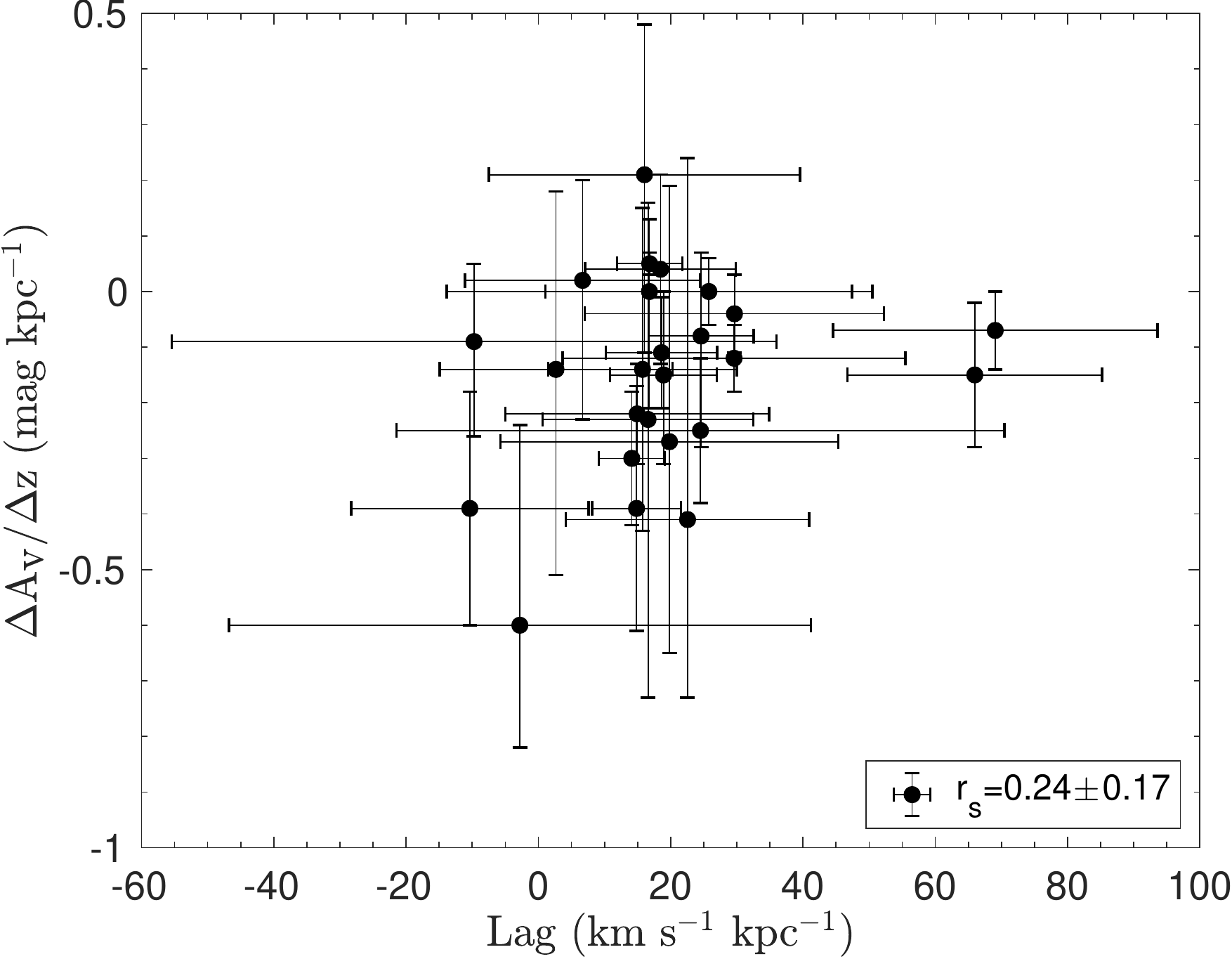}{0.3\columnwidth}{(b)}
    \fig{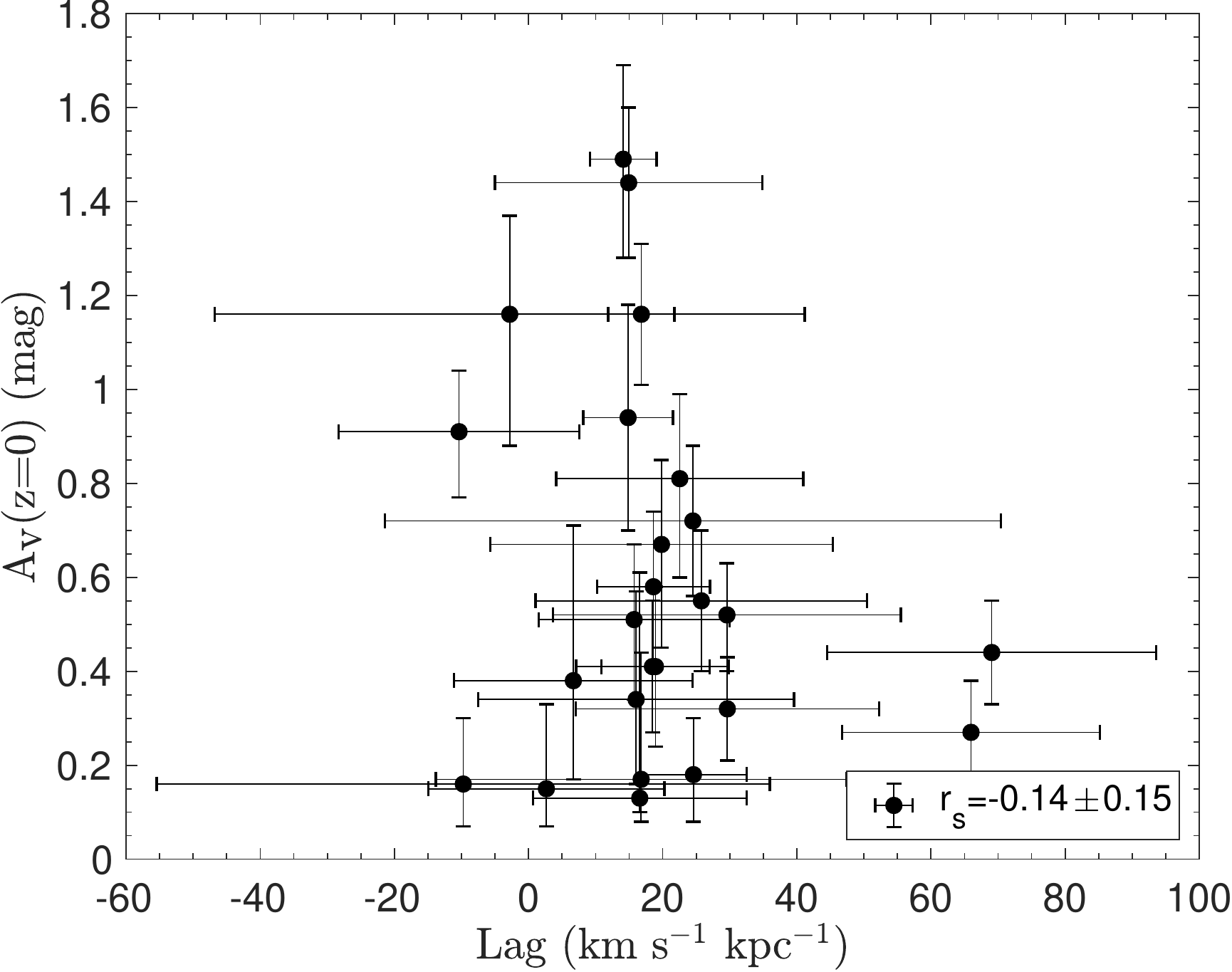}{0.3\columnwidth}{(c)}}
\gridline{\fig{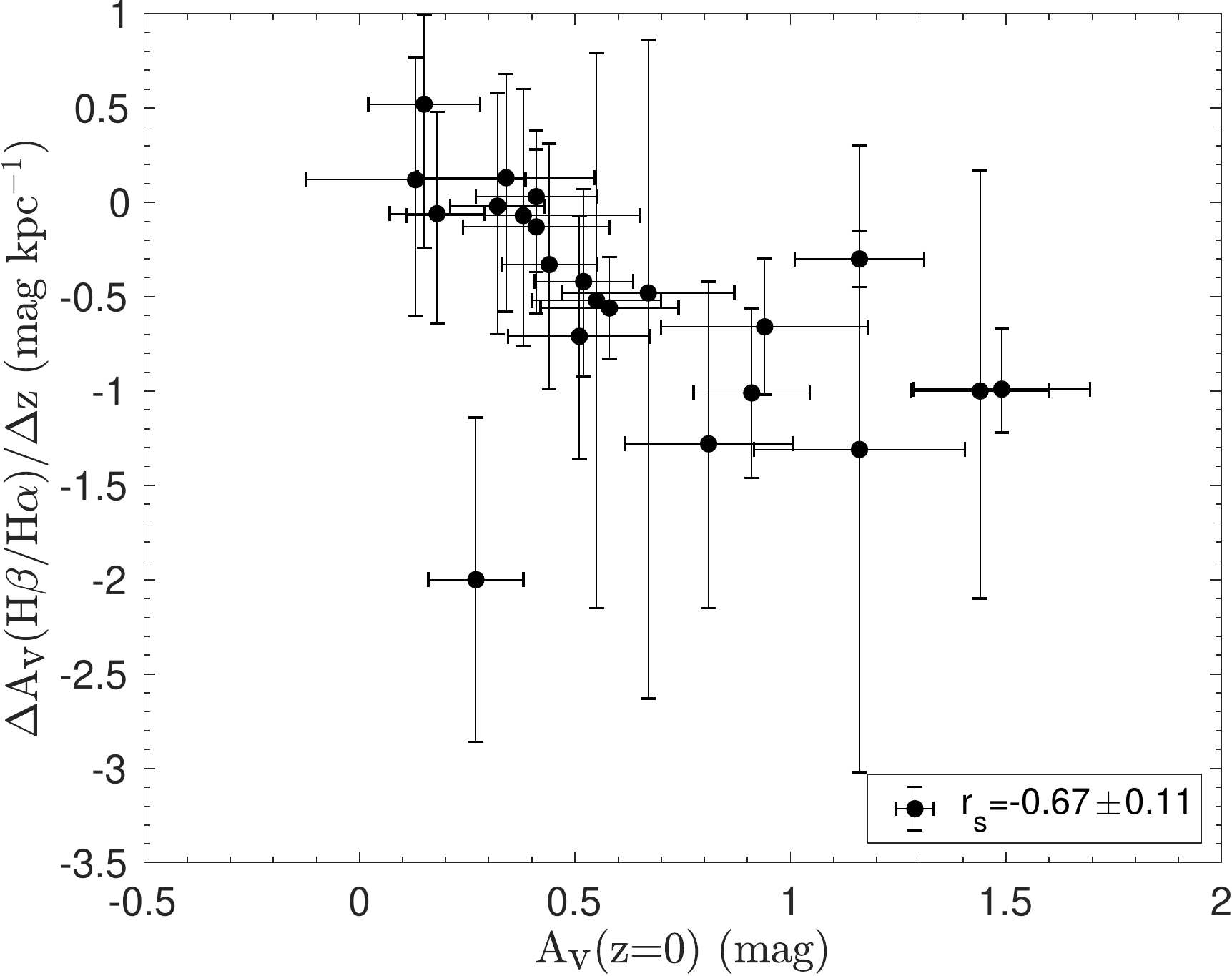}{0.3\columnwidth}{(d)}
\fig{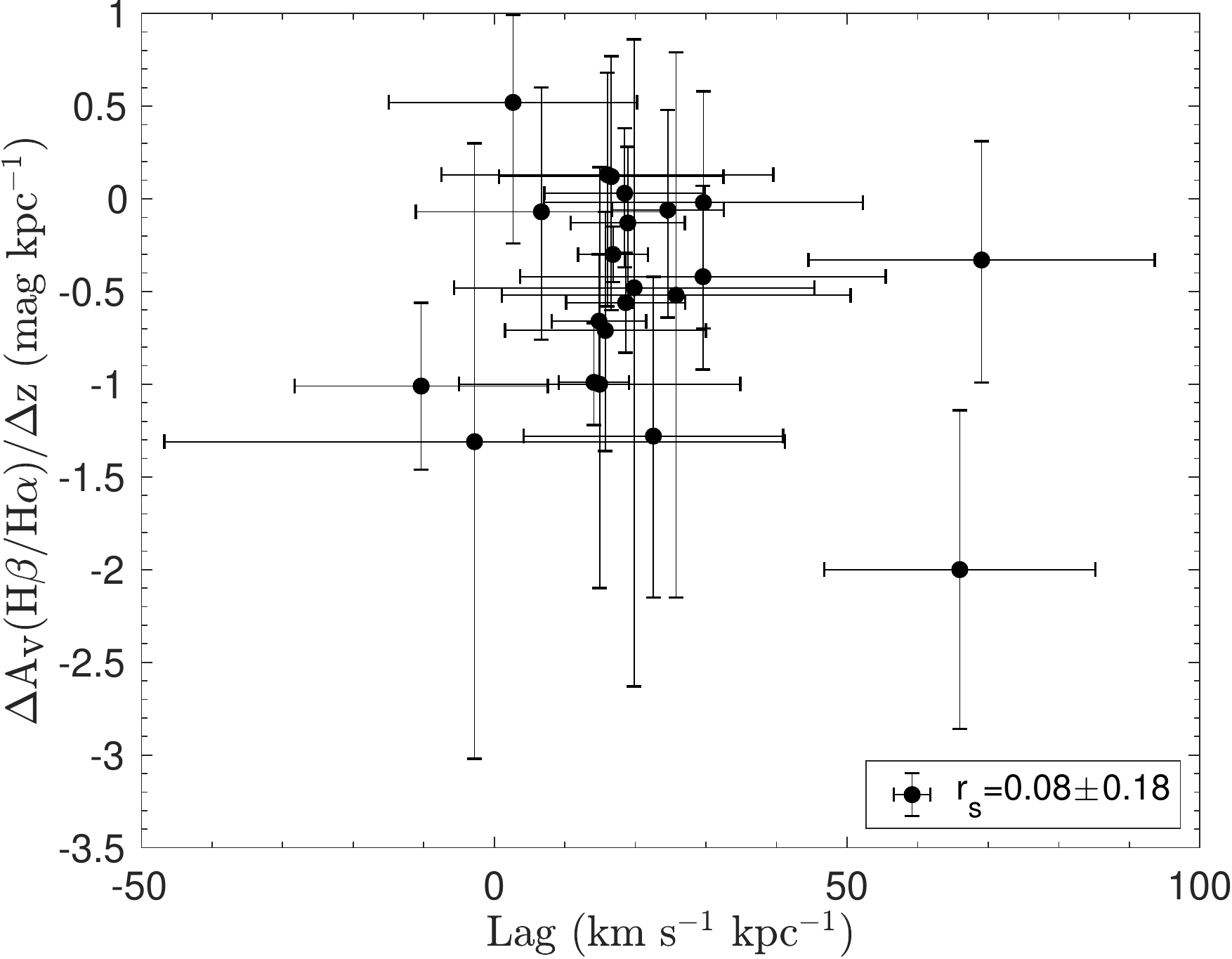}{0.3\columnwidth}{(e)}
    \fig{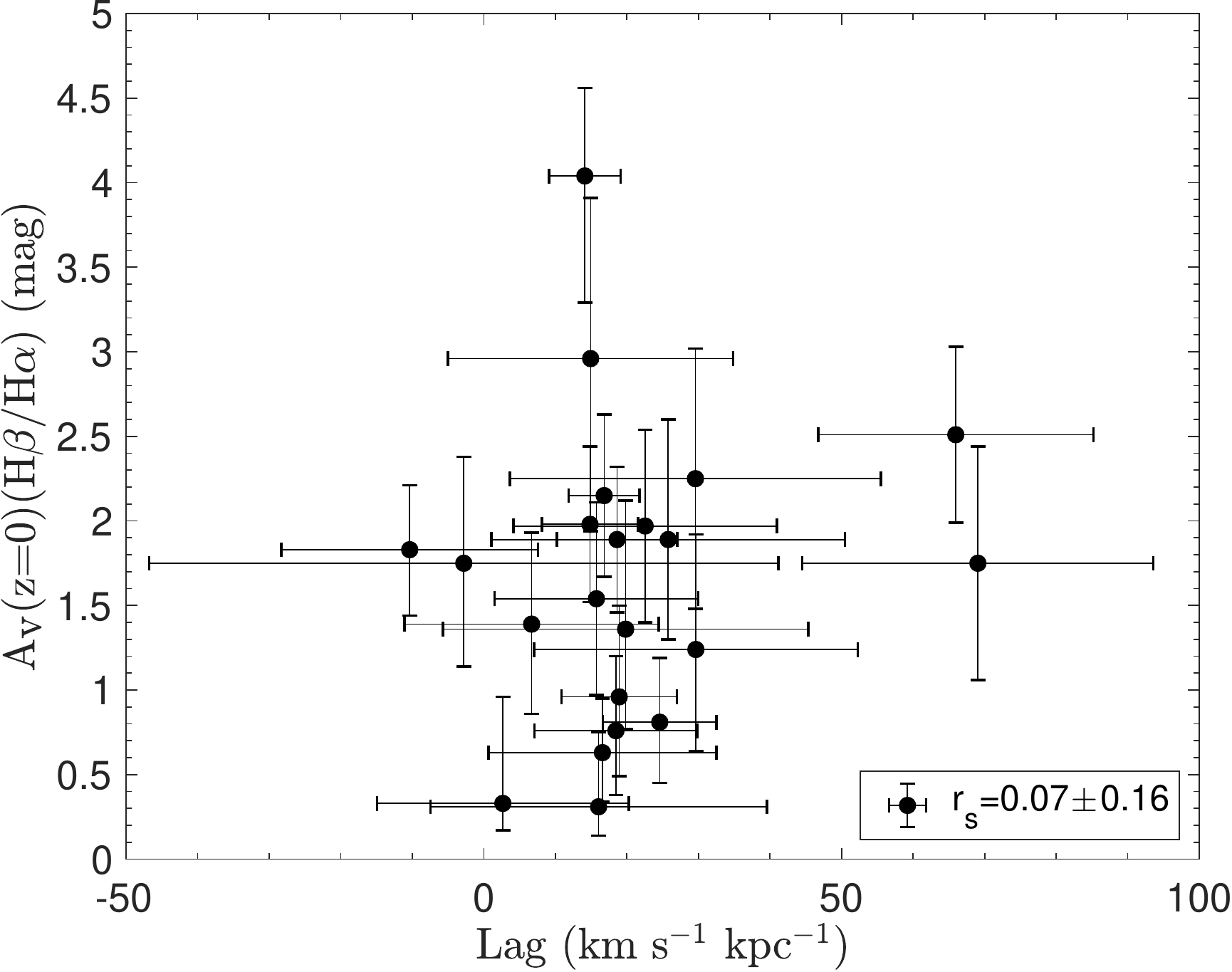}{0.3\columnwidth}{(f)}}
\caption{(a) \DAvDz\ and A$_{\rm V}(z=0)$ (extinctions measured from the stellar population synthesis modeling) are correlated, where the galaxies with the highest extinctions decrease the most as a function of height. Neither (b) \DAvDz\ nor (c) A$_{\rm V}(z=0)$ are correlated with the measured velocity lag, indicating that the measurement of the lag is not affected significantly by extinction. (d,e,f) The same as (a,b,c), but the extinctions are determined from the Balmer Decrement.} 
\end{figure}

We investigate our previous claim that the extinction drops with distance from the midplane directly. Using the same method as described in Section \ref{ssec:vvg}, we determine the vertical gradient in the extinction (\DAvDz). This is the extinction determined from the stellar population synthesis modeling \citep[see][]{pipe3DI,pipe3DII}, and it is degenerate with the stellar age and metallicity. We find that the extinction decreases with distance from the midplane for $76\%^{+12\%}_{-32\%}$ of galaxies.  We list the values of \DAvDz\ and A$_{\rm V}(z=0)$ in Table \ref{tab:othergrads}. A$_{\rm V}(z=0)$ and \DAvDz\ are inversely correlated (Figure \ref{fig:extgrad}a), such that the galaxies with the largest extinctions in the midplane decrease the most as a function of distance. By $z=1$, most galaxies have A$_{\rm V}<0.5$ mag. Away from the midplane, extinction will play a diminished role. The measured lag is not correlated with either \DAvDz\ or A$_{\rm V}(z=0)$ (Figures \ref{fig:extgrad}b,c), again indicating that extinction is not significantly biasing our lag measurements.

Because the extinction from the stellar population synthesis modeling is affected by stellar age and metallicity, we also investigate trends with the extinction derived from the Balmer decrement, assuming Case B recombination and an intrinsic ratio of I(\ha)/I(\hb)\,=\,2.86 \citep{osterbrock89}. We find vertical gradients in the extinction from the Balmer decrement (\DAvDz(\hb/\ha)) and the extinction in the midplane (A$_{\rm V}(z=0)$(\hb/\ha)) in the same way as before. Although A$_{\rm V}(z=0)(\hb/\ha)>{\rm A}_{\rm V}(z=0)$ in general, we find the same general trends as with the extinctions derived from the stellar population synthesis modeling (Figures \ref{fig:extgrad}d,e,f). The measured lag is not correlated with \DAvDz(\hb/\ha) or A$_{\rm V}(z=0)(\hb/\ha)$ (Figures \ref{fig:extgrad}d,e), confirming indicating that extinction is not significantly biasing our lag measurements.

\subsection{Inclination Effects}
\label{app:inc}

\begin{figure}
\label{fig:inc}
\centering
\includegraphics[width=0.5\columnwidth]{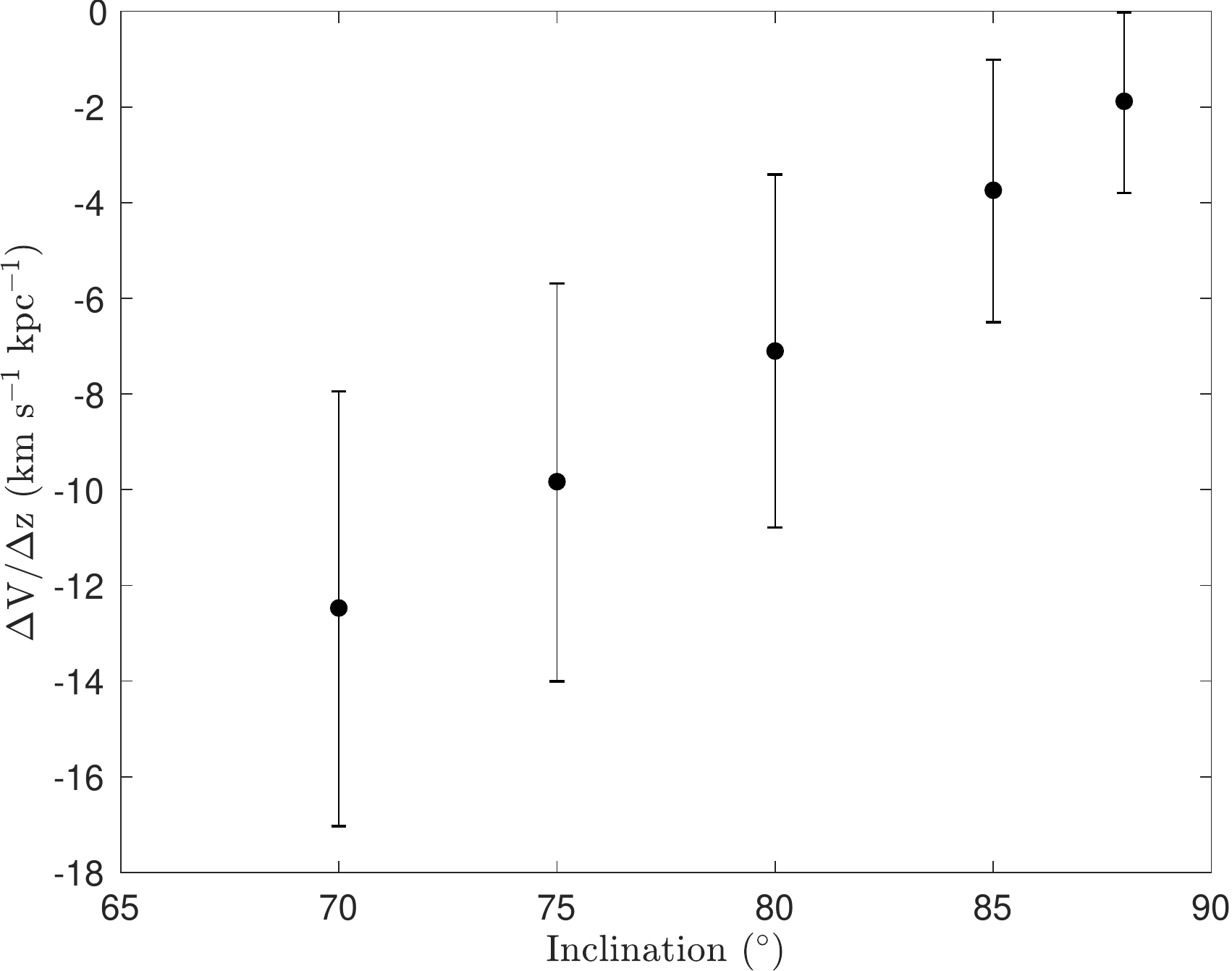}
\caption{For a simulated thin disk, we measure the lag induced by an inclination $<90$\D.} 
\end{figure}

If the galaxies used in this study are not perfectly edge-on, then the measured ionized gas scale height will be a combination of gas above the midplane and in the disk. Moreover, the lag measurements will also be a combination of more slowly rotation gas in the eDIG and gas at different radii in the disk. This is why our selection criteria for edge-on galaxies are fairly restrictive (Section \ref{ssec:EOGS}).
Previous analysis and modeling has shown that the photometric vertical scale height increases by 20\% if the inclination deviates from 90\D\ by 5\D\ \citep{degrijs97}. We investigate the effect of inclination on the measured lag. We simulate thin disks (that would not have an intrinsic lag) given an input rotation curve that rises linearly until 1.77\,kpc \citep[based on][]{levy18} and is then constant with \vrot\,=\,200\,\kms. We then incline the disk and measure the vertical gradient in the rotation velocity as before (see Section \ref{ssec:vvg}) assuming that the simulated galaxy is edge-on. We show the measured vertical gradients in the rotation velocity as a function of the true inclination in Figure \ref{fig:inc}. Because the simulated disks are thin and have no intrinsic lag, all of the resulting measured lag is due to measuring gas in the midplane due to the inclination. If the inclination deviates from 90\D\ by 5\D, the induced lag is \~4\,\kmskpc, but larger deviations can produce appreciable lags. Our selection criteria are fairly stringent to minimize this effect, and the resulting sample consist of galaxies that are very edge-on (see Section \ref{ssec:EOGS} and Figure \ref{fig:SDSSpanel}).

\section{Vertical Gradients in Other Galaxy Properties}
\label{app:otherparams}

\setcounter{table}{1}
\begin{deluxetable}{ccccccccc}
\tablecaption{Measured Gradients for Other Galaxy Properties \label{tab:othergrads}}
\tablehead{
\colhead{Name} & \colhead{$\Delta$Age/$\Delta$z} & \colhead{Age(z=0)} & \colhead{$\Delta$[Z/H]/$\Delta$z} & \colhead{[Z/H](z=0)} & \colhead{$\Delta$A$_{\rm v}$/$\Delta$z} & \colhead{A$_{\rm v}$(z=0)} & \colhead{$\Delta$A$_{\rm v}$(\hb/\ha)/$\Delta$z} & \colhead{A$_{\rm v}$(\hb/\ha)(z=0)} \\
 & \colhead{(dex\,kpc$^{-1}$)} & \colhead{(log[yr])} & \colhead{(dex\,kpc$^{-1}$)} & \colhead{(log[Z$_{\odot}$])} & \colhead{(dex\,kpc$^{-1}$)} & \colhead{(mag)}  & \colhead{(dex\,kpc$^{-1}$)} & \colhead{(mag)}}
\startdata
ESO539-G014 & -0.10$^{+0.09}_{-0.09}$ & 9.09$^{+0.22}_{-0.22}$& 0.011$^{+0.036}_{-0.036}$& -0.243$^{+0.088}_{-0.090}$& -0.15$^{+0.15}_{-0.16}$& 0.41$^{+0.17}_{-0.17}$& -0.13$^{+0.41}_{-0.46}$ & 0.96$^{+0.54}_{-0.47}$\\
IC 480 & -0.05$^{+0.07}_{-0.07}$ & 8.93$^{+0.10}_{-0.11}$& 0.065$^{+0.039}_{-0.039}$& -0.393$^{+0.062}_{-0.066}$& -0.11$^{+0.10}_{-0.10}$& 0.58$^{+0.16}_{-0.16}$& -0.56$^{+0.27}_{-0.27}$ & 1.89$^{+0.43}_{-0.43}$\\
IC 2095 & -0.01$^{+0.42}_{-0.42}$ & 8.31$^{+0.28}_{-0.29}$& 0.001$^{+0.114}_{-0.115}$& -0.199$^{+0.068}_{-0.068}$& -0.14$^{+0.32}_{-0.37}$& 0.15$^{+0.18}_{-0.08}$& 0.52$^{+0.47}_{-0.76}$ & 0.33$^{+0.63}_{-0.16}$\\
IC 2098 & 0.05$^{+0.12}_{-0.13}$ & 8.90$^{+0.11}_{-0.11}$& 0.087$^{+0.059}_{-0.077}$& -0.299$^{+0.082}_{-0.070}$& -0.39$^{+0.21}_{-0.21}$& 0.91$^{+0.13}_{-0.14}$& -1.01$^{+0.45}_{-0.45}$ & 1.83$^{+0.38}_{-0.39}$\\
IC 4215 & -0.20$^{+0.17}_{-0.12}$ & 9.72$^{+0.15}_{-0.25}$& 0.032$^{+0.053}_{-0.054}$& -0.175$^{+0.082}_{-0.084}$& 0.00$^{+0.13}_{-0.21}$& 0.17$^{+0.27}_{-0.09}$& ... & ...\\
MCG-01-01-012 & -0.02$^{+0.07}_{-0.07}$ & 9.18$^{+0.13}_{-0.13}$& -0.001$^{+0.042}_{-0.043}$& -0.103$^{+0.114}_{-0.116}$& -0.22$^{+0.09}_{-0.09}$& 1.44$^{+0.16}_{-0.16}$& -1.00$^{+1.17}_{-1.10}$ & 2.96$^{+0.95}_{-1.02}$\\
MCG-02-08-014 & -0.08$^{+0.08}_{-0.08}$ & 9.51$^{+0.10}_{-0.11}$& 0.054$^{+0.049}_{-0.050}$& -0.261$^{+0.078}_{-0.080}$& -0.04$^{+0.07}_{-0.07}$& 0.32$^{+0.11}_{-0.11}$& -0.02$^{+0.60}_{-0.68}$ & 1.24$^{+0.68}_{-0.60}$\\
NGC 3160 & -0.12$^{+0.04}_{-0.04}$ & 9.71$^{+0.10}_{-0.10}$& -0.005$^{+0.020}_{-0.020}$& -0.025$^{+0.052}_{-0.056}$& -0.09$^{+0.14}_{-0.17}$& 0.16$^{+0.14}_{-0.09}$& ... & ...\\
NGC 4149 & -0.05$^{+0.08}_{-0.08}$ & 9.32$^{+0.10}_{-0.11}$& -0.013$^{+0.046}_{-0.047}$& -0.123$^{+0.066}_{-0.070}$& -0.15$^{+0.13}_{-0.13}$& 0.27$^{+0.11}_{-0.11}$& -2.00$^{+0.86}_{-0.86}$ & 2.51$^{+0.52}_{-0.52}$\\
PGC 213858 & -0.05$^{+0.29}_{-0.29}$ & 8.82$^{+0.30}_{-0.30}$& 0.080$^{+0.065}_{-0.100}$& -0.169$^{+0.106}_{-0.084}$& 0.21$^{+0.27}_{-0.32}$& 0.34$^{+0.23}_{-0.18}$& 0.13$^{+0.55}_{-0.71}$ & 0.31$^{+0.44}_{-0.17}$\\
UGC 3539 & -0.10$^{+0.17}_{-0.18}$ & 8.93$^{+0.17}_{-0.18}$& 0.094$^{+0.054}_{-0.067}$& -0.385$^{+0.106}_{-0.088}$& -0.39$^{+0.22}_{-0.22}$& 0.94$^{+0.24}_{-0.24}$& -0.66$^{+0.36}_{-0.36}$ & 1.98$^{+0.46}_{-0.46}$\\
UGC 4136 & -0.05$^{+0.06}_{-0.06}$ & 9.67$^{+0.13}_{-0.14}$& -0.015$^{+0.034}_{-0.034}$& -0.093$^{+0.070}_{-0.072}$& -0.12$^{+0.06}_{-0.06}$& 0.52$^{+0.11}_{-0.12}$& -0.42$^{+0.49}_{-0.50}$ & 2.25$^{+0.77}_{-0.77}$\\
UGC 4197 & -0.18$^{+0.07}_{-0.07}$ & 9.50$^{+0.14}_{-0.15}$& 0.024$^{+0.024}_{-0.024}$& -0.151$^{+0.068}_{-0.068}$& -0.07$^{+0.07}_{-0.07}$& 0.44$^{+0.11}_{-0.11}$& -0.33$^{+0.64}_{-0.66}$ & 1.75$^{+0.69}_{-0.69}$\\
UGC 4550 & 0.21$^{+0.07}_{-0.07}$ & 8.58$^{+0.14}_{-0.14}$& -0.002$^{+0.033}_{-0.033}$& -0.269$^{+0.074}_{-0.076}$& -0.00$^{+0.06}_{-0.06}$& 0.55$^{+0.15}_{-0.15}$& -0.52$^{+1.31}_{-1.63}$ & 1.89$^{+0.71}_{-0.59}$\\
UGC 5244 & -0.26$^{+0.12}_{-0.13}$ & 8.77$^{+0.13}_{-0.13}$& 0.007$^{+0.073}_{-0.073}$& -0.373$^{+0.088}_{-0.092}$& 0.04$^{+0.17}_{-0.17}$& 0.41$^{+0.14}_{-0.14}$& 0.03$^{+0.35}_{-0.40}$ & 0.76$^{+0.44}_{-0.38}$\\
UGC 6256 & -0.06$^{+0.32}_{-0.32}$ & 8.93$^{+0.28}_{-0.28}$& 0.014$^{+0.089}_{-0.095}$& -0.269$^{+0.096}_{-0.094}$& -0.60$^{+0.36}_{-0.22}$& 1.16$^{+0.21}_{-0.28}$& -1.31$^{+1.61}_{-1.71}$ & 1.75$^{+0.63}_{-0.61}$\\
UGC 9080 & -0.08$^{+0.57}_{-0.50}$ & 8.24$^{+0.32}_{-0.35}$& 0.165$^{+0.008}_{-0.239}$& -0.309$^{+0.094}_{-0.086}$& -0.23$^{+0.39}_{-0.50}$& 0.13$^{+0.48}_{-0.03}$& 0.12$^{+0.65}_{-0.72}$ & 0.63$^{+0.32}_{-0.29}$\\
UGC 9165 & -0.11$^{+0.10}_{-0.10}$ & 9.14$^{+0.18}_{-0.18}$& 0.027$^{+0.034}_{-0.034}$& -0.293$^{+0.066}_{-0.066}$& -0.30$^{+0.12}_{-0.12}$& 1.49$^{+0.20}_{-0.21}$& -0.99$^{+0.32}_{-0.23}$ & 4.04$^{+0.52}_{-0.75}$\\
UGC 9262 & 0.04$^{+0.01}_{-0.01}$ & 8.96$^{+0.08}_{-0.08}$& 0.013$^{+0.009}_{-0.009}$& -0.209$^{+0.054}_{-0.054}$& 0.05$^{+0.02}_{-0.02}$& 1.16$^{+0.15}_{-0.15}$& -0.30$^{+0.15}_{-0.15}$ & 2.15$^{+0.48}_{-0.48}$\\
UGC 9448 & 0.21$^{+0.31}_{-0.32}$ & 8.83$^{+0.17}_{-0.17}$& -0.018$^{+0.121}_{-0.104}$& -0.331$^{+0.072}_{-0.078}$& -0.27$^{+0.46}_{-0.38}$& 0.67$^{+0.18}_{-0.22}$& -0.48$^{+1.34}_{-2.15}$ & 1.36$^{+0.76}_{-0.59}$\\
UGC 9919 & 0.12$^{+0.26}_{-0.26}$ & 8.44$^{+0.21}_{-0.21}$& 0.022$^{+0.094}_{-0.108}$& -0.365$^{+0.114}_{-0.106}$& -0.14$^{+0.29}_{-0.29}$& 0.51$^{+0.16}_{-0.17}$& -0.71$^{+0.64}_{-0.65}$ & 1.54$^{+0.57}_{-0.57}$\\
UGC 10043 & 0.04$^{+0.29}_{-0.29}$ & 8.65$^{+0.11}_{-0.12}$& 0.130$^{+0.031}_{-0.157}$& -0.375$^{+0.074}_{-0.060}$& -0.41$^{+0.65}_{-0.32}$& 0.81$^{+0.18}_{-0.21}$& -1.28$^{+0.86}_{-0.87}$ & 1.97$^{+0.57}_{-0.57}$\\
UGC 10297 & 0.08$^{+0.16}_{-0.16}$ & 8.79$^{+0.13}_{-0.13}$& 0.044$^{+0.089}_{-0.134}$& -0.349$^{+0.090}_{-0.080}$& -0.08$^{+0.15}_{-0.20}$& 0.18$^{+0.12}_{-0.10}$& -0.06$^{+0.54}_{-0.58}$ & 0.81$^{+0.38}_{-0.36}$\\
UGC 12518 & -0.21$^{+0.13}_{-0.13}$ & 9.70$^{+0.11}_{-0.12}$& -0.056$^{+0.057}_{-0.055}$& -0.037$^{+0.074}_{-0.076}$& -0.25$^{+0.13}_{-0.13}$& 0.72$^{+0.16}_{-0.16}$& ... & ...\\
UGC 12723 & -0.16$^{+0.19}_{-0.19}$ & 8.79$^{+0.30}_{-0.30}$& 0.061$^{+0.053}_{-0.054}$& -0.297$^{+0.092}_{-0.092}$& 0.02$^{+0.18}_{-0.25}$& 0.38$^{+0.33}_{-0.21}$& -0.07$^{+0.67}_{-0.69}$ & 1.39$^{+0.54}_{-0.53}$\\
\enddata
\tablecomments{Measured vertical gradients and midplane values for the stellar age, stellar metallicity ([Z/H]), nebular extinction (A$_{\rm v}$), and extinction from the Balmer decrement (A$_{\rm v}$(\hb/\ha)). See Appendix \ref{app:otherparams} for more details.}
\end{deluxetable}

In addition to vertical gradients in the rotation velocity, we also investigate vertical gradients in other spatially resolved galaxy properties from CALIFA. Such properties include the stellar age and metallicity, nebular extinction (A$_{\rm v}$), and extinction from the Balmer decrement (A$_{\rm v}$(\hb/\ha)). The stellar age and metallicity and A$_{\rm v}$ come from the stellar population synthesis done by the CALIFA collaboration \citep[for details on the derivation of these maps, see][]{pipe3DI,pipe3DII}. The extinction based on the Balmer decrement is calculated from the \ha\ and \hb\ emission line flux maps following \citet{dominguez13} (which assumes Case B recombination \citep{osterbrock89} and a \citet{calzetti00} extinction curve). These parameters are not independent from one another due to degeneracies between age, color, extinction, and metallicity when modeling the spectra. Attempting to disentangle these degeneracies is beyond the scope of this paper. We, nevertheless, fit vertical gradients in these quantities using the same method as described in Section \ref{ssec:vvg} and present the gradients and fitted midplane values in Table \ref{tab:othergrads}. We do discuss the results based on the nebular extinction in the previous section as they relate to the effect of extinction on our results.

\section{Analytic Derivation of the Lag and Radial Variation in the Lag}
\label{app:derivation}
In Section \ref{sssec:radialvar}, we investigate radial variations in the lag in the context of the origin of the eDIG. We begin our analytic derivation assuming a Miyamoto-Nagai potential:
\begin{equation}
\label{eq:potential}
\phi(r,z) = \frac{GM}{\bigg\{r^2+\Big[r_0+\big(z^2+z_0^2\big)^{1/2}\Big]^2\bigg\}^{1/2}}
\end{equation}
where $M$ is the total mass of the galaxy and $r_0$ and $z_0$ are the radial and vertical scale lengths \citep{miyamoto75}. This potential assumes that baryons in the disk dominate the potential (a maximal disk) and ignores the dark matter contribution. This assumption is likely valid at small radii (where $r \ll r_s$, where $r_s$ is the dark matter scale length), but may break down further out in the disk. Studies of the Milky Way's dark matter halo suggest $r_s\sim20$\,kpc \citep{hooper17}, whereas our \ha\ measurements probe out to $\lesssim 10$\,kpc. The rotation velocity for the potential in Equation \ref{eq:potential} is
\begin{equation}
\label{eq:vrot}
{\rm V}_{\rm rot}(r,z) = \frac{(GM)^{1/2}r}{\bigg\{r^2+\Big[r_0+\big(z^2+z_0^2\big)^{1/2}\Big]^2\bigg\}^{3/4}}
\end{equation}
where V$_{\rm rot}^2=r\frac{\partial\phi(r,z)}{\partial r}$ \citep[e.g.][]{binney08}. The vertical gradient in \vrot\ is then
\begin{equation}
\label{eq:dvdz}
{\rm Lag} \equiv \frac{-\partial{\rm V}_{\rm rot}}{\partial z} =\frac{3(GM)^{1/2}rz\Big[r_0+\big(z^2+z_0^2\big)^{1/2}\Big]}{2\big(z^2+z_0^2\big)^{1/2}\bigg\{r^2+\Big[r_0+\big(z^2+z_0^2\big)^{1/2}\Big]^2\bigg\}^{7/4}}.
\end{equation}
The radial variation in the lag ($\partial{\rm Lag}/\partial r$) is
\begin{equation}
\label{eq:dlagdr}
\frac{\partial{\rm Lag}}{\partial r} \equiv \frac{-\partial(\partial{\rm V_{rot}}/\partial z)}{\partial r} = \frac{3z(GM)^{1/2}\Big[r_0+\big(z^2+z_0^2\big)^{1/2}\Big]\Big[2r_0^2+4r_0\big(z^2+z_0^2\big)^{1/2}+2z_0-5r^2+2z^2\Big]}{4\big(z^2+z_0^2\big)^{1/2}\Big[r_0^2+2r_0\big(z^2+z_0^2\big)^{1/2}+z_0+r^2+z^2\Big]^{11/4}}.
\end{equation}
For simplicity, we evaluate Equations \ref{eq:dvdz} and \ref{eq:dlagdr} at $z=z_0$, so that
\begin{equation}
\label{eq:dvdzz0}
{\rm Lag}|_{z=z_0} = \frac{-\partial{\rm V}_{\rm rot}}{\partial z}\Big|_{z=z_0} = \frac{3\sqrt{GM}\big(r_0+\sqrt{2}z_0\big)r}{2\sqrt{2}\Big[r^2+\big(r_0+\sqrt{2}z_0\big)^2\Big]^{7/4}}
\end{equation}
and
\begin{equation}
\label{eq:dlagdrz0}
\frac{\partial{\rm Lag}}{\partial r}\Big|_{z=z_0} = \frac{-\partial(\partial{\rm V}_{\rm rot}/\partial z)}{\partial r}\Big|_{z=z_0} = \frac{3\sqrt{GM}\big(r_0+\sqrt{2}z_0\big)\big(2r_0+4\sqrt{2}r_0z_0+4z_0-5r^2\big)}{4\sqrt{2}\big(r^2+r_0^2+2\sqrt{2}r_0z_0+2z_0^2\big)^{11/4}}.
\end{equation}
Figure \ref{fig:radiallaganalytic} shows Equations \ref{eq:dvdzz0} and \ref{eq:dlagdrz0} as a function of $r/r_0$. 

\section{Additional Multipanel Images}
\label{app:addtlim}

We show multipanel images for all of the galaxies studied here in Figure \ref{fig:FS}. See the caption at the end of Figure \ref{fig:FS} for details.

\clearpage
\setcounter{table}{\value{figure}} 
\begin{longtable*}{c} 
\label{fig:FS}
\endfirsthead
\endhead
\endfoot
\endlastfoot
\includegraphics[width=\textwidth]{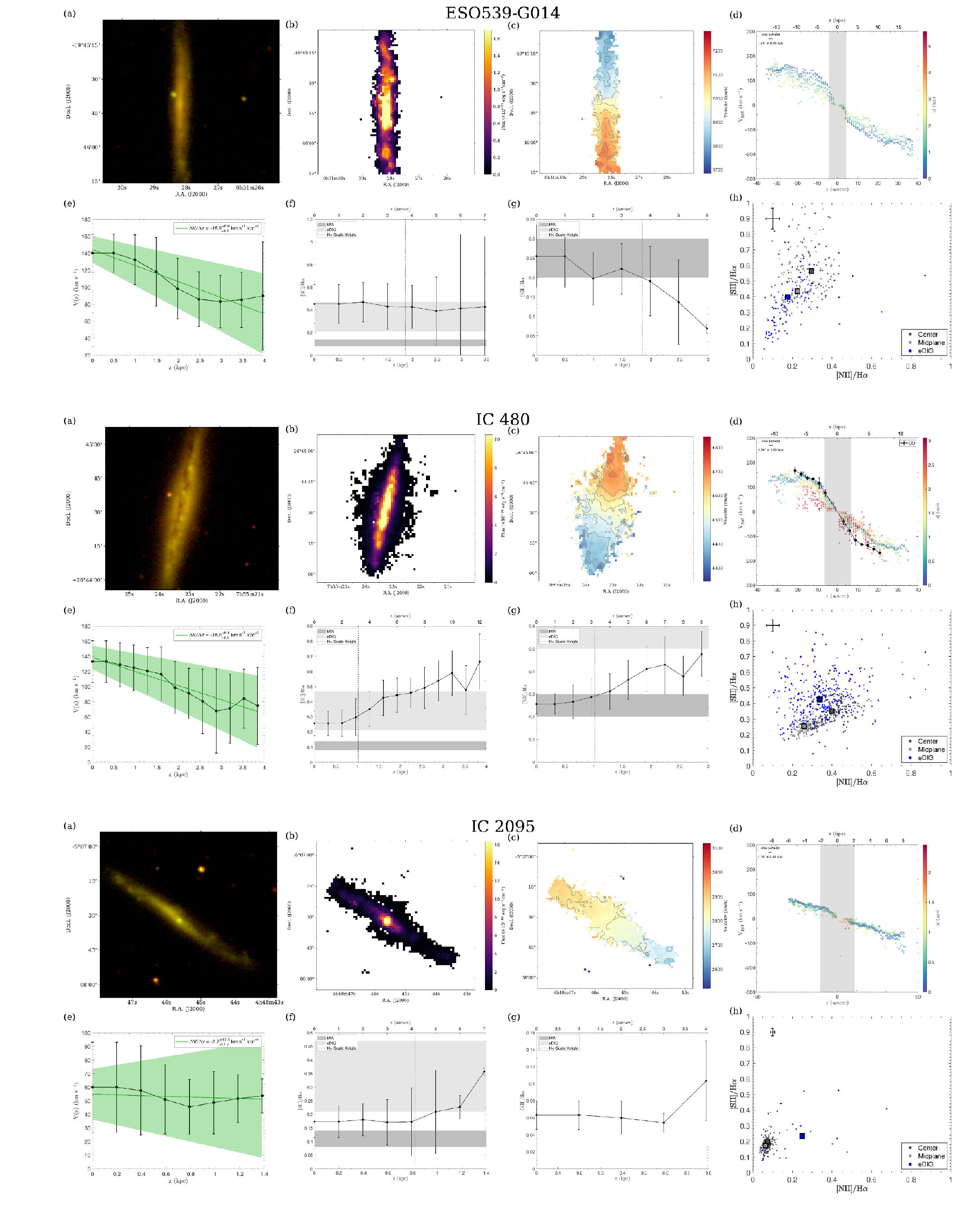}\\
\includegraphics[width=\textwidth]{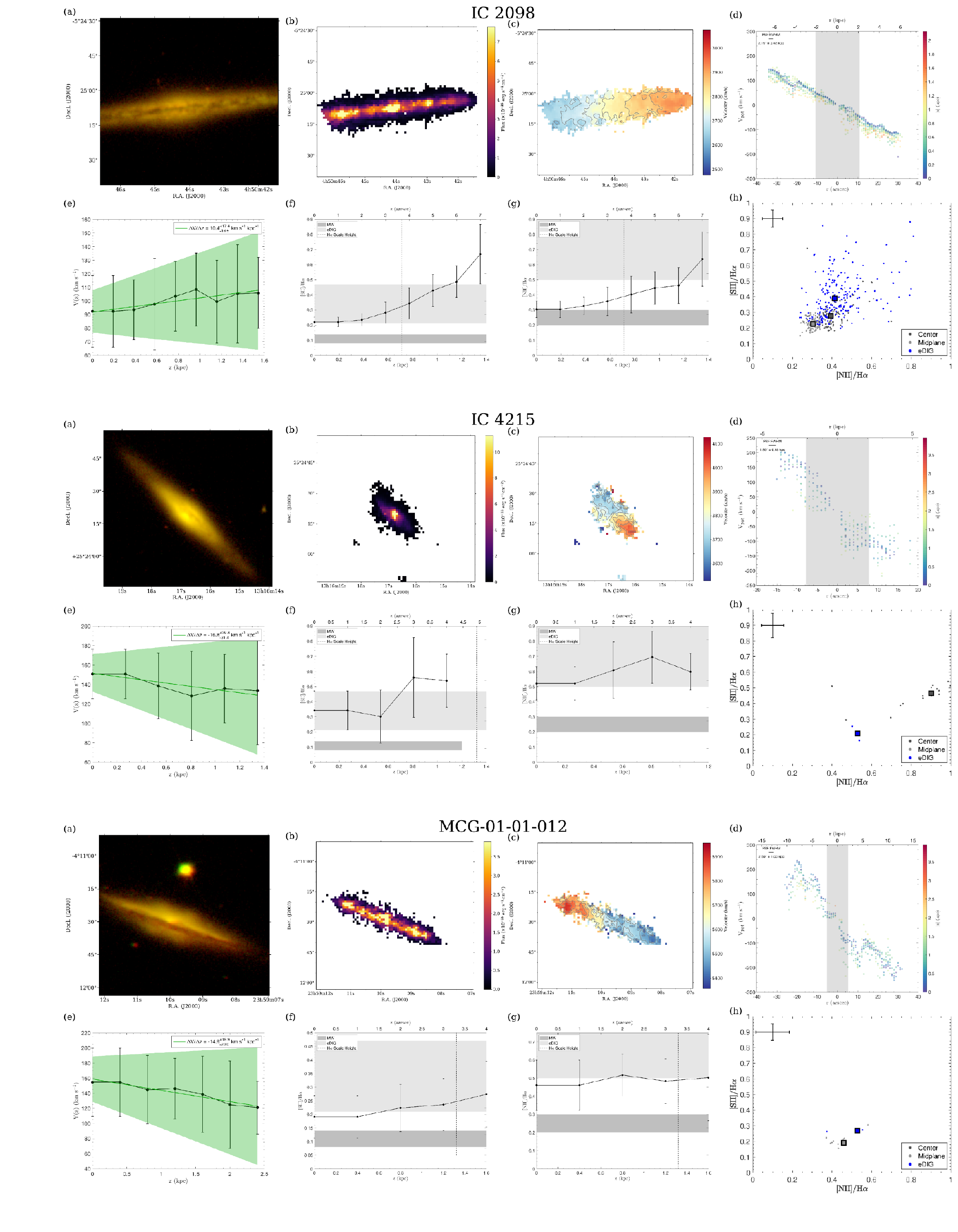}\\
\includegraphics[width=\textwidth]{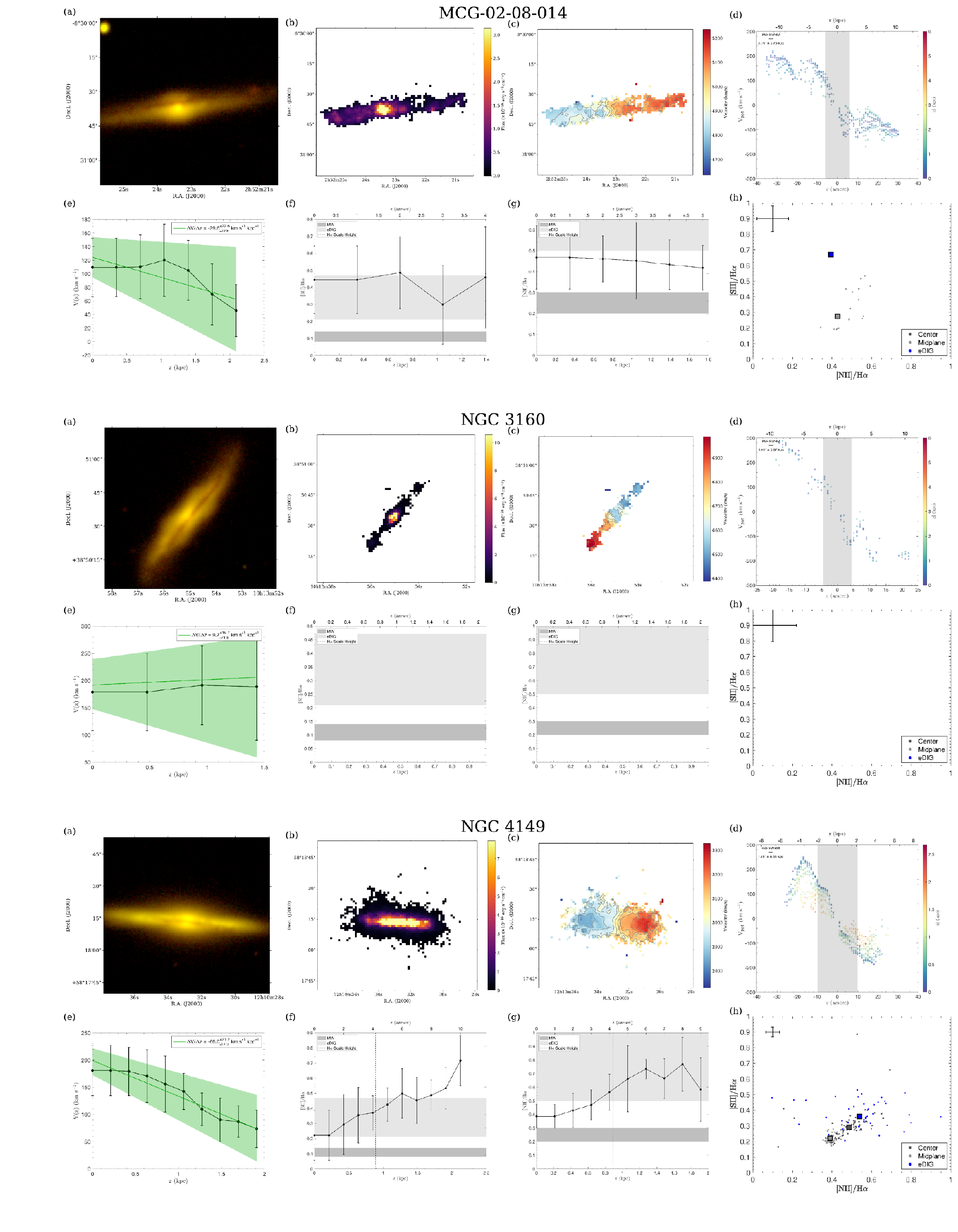}\\
\includegraphics[width=\textwidth]{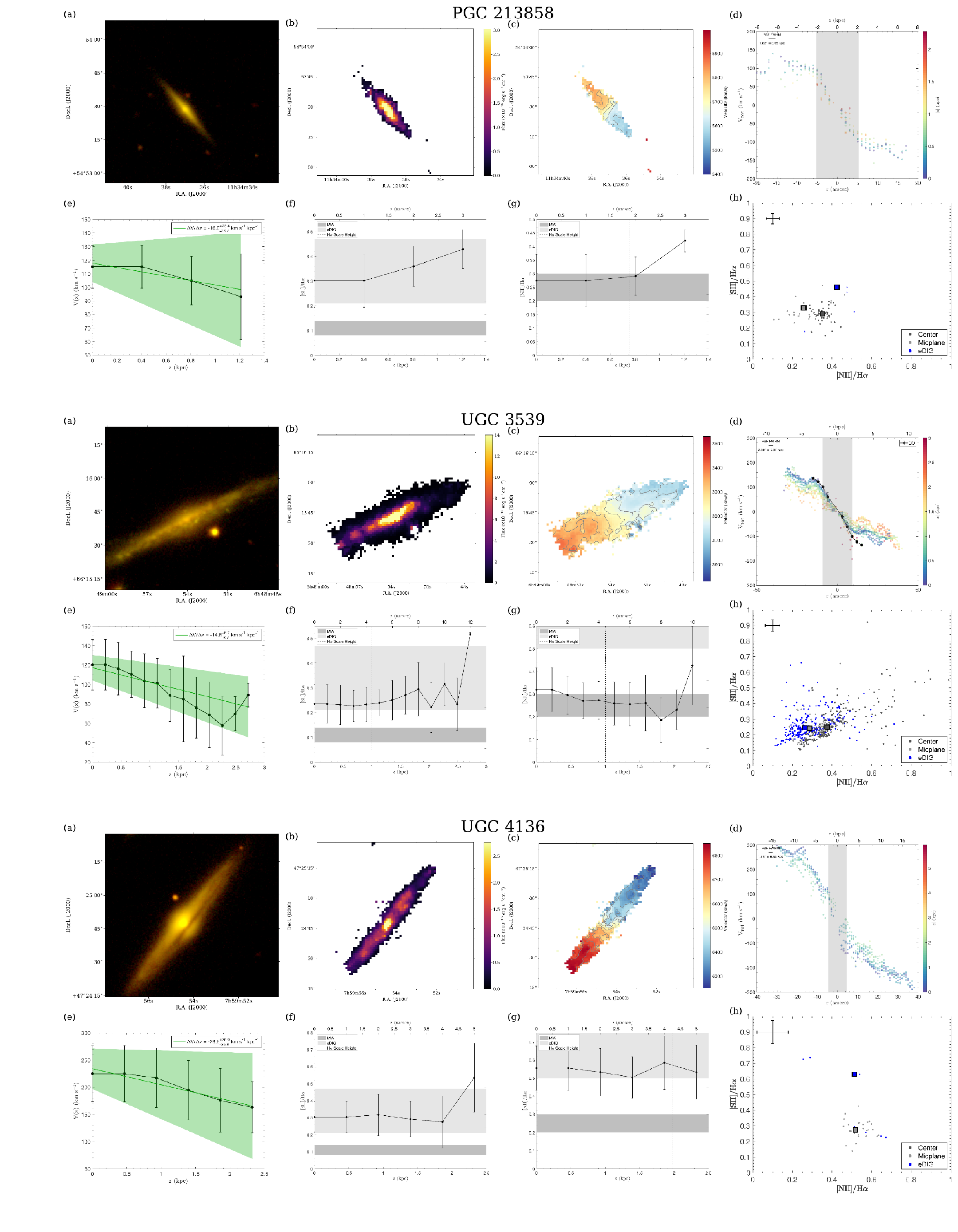}\\
\includegraphics[width=\textwidth]{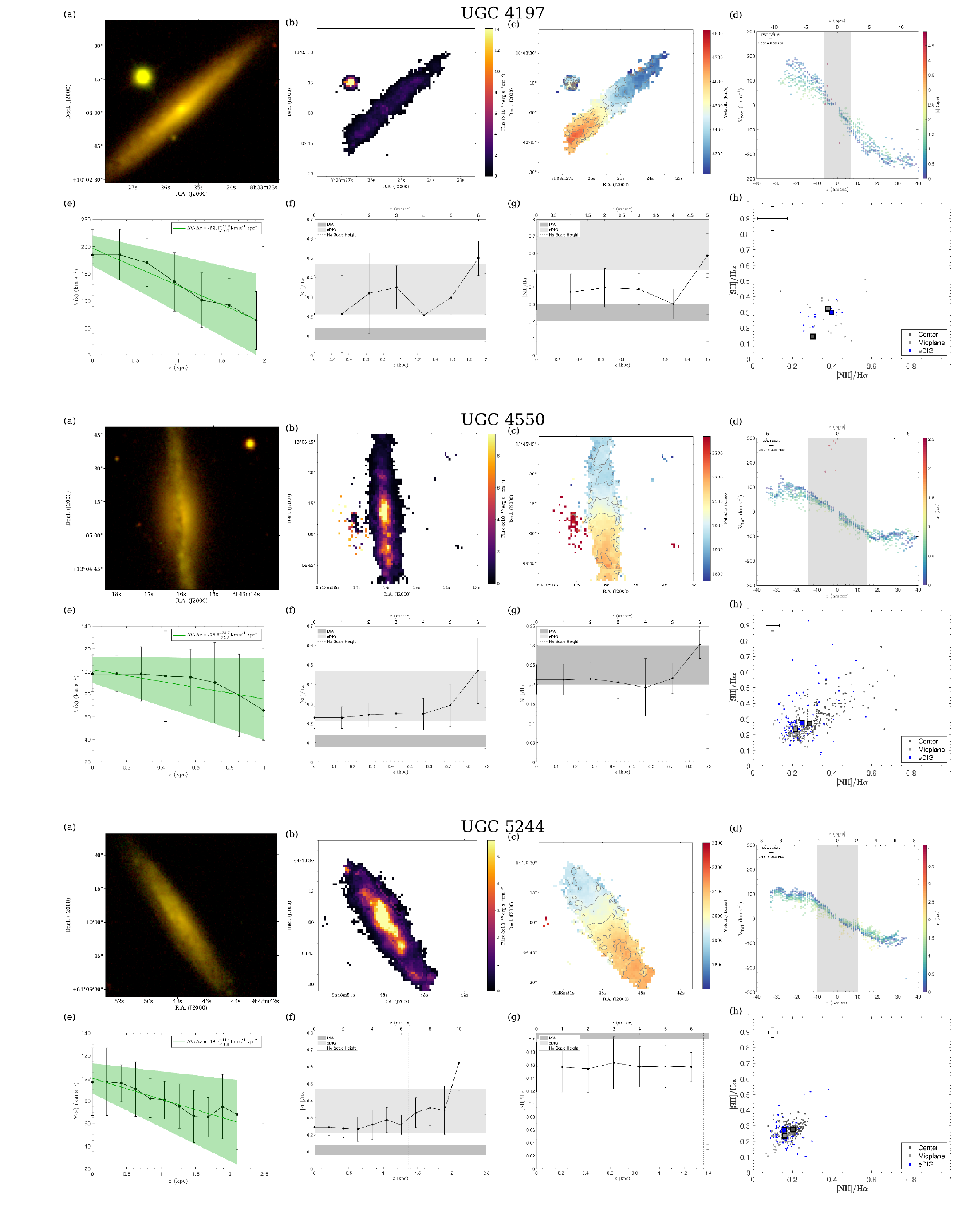}\\
\includegraphics[width=\textwidth]{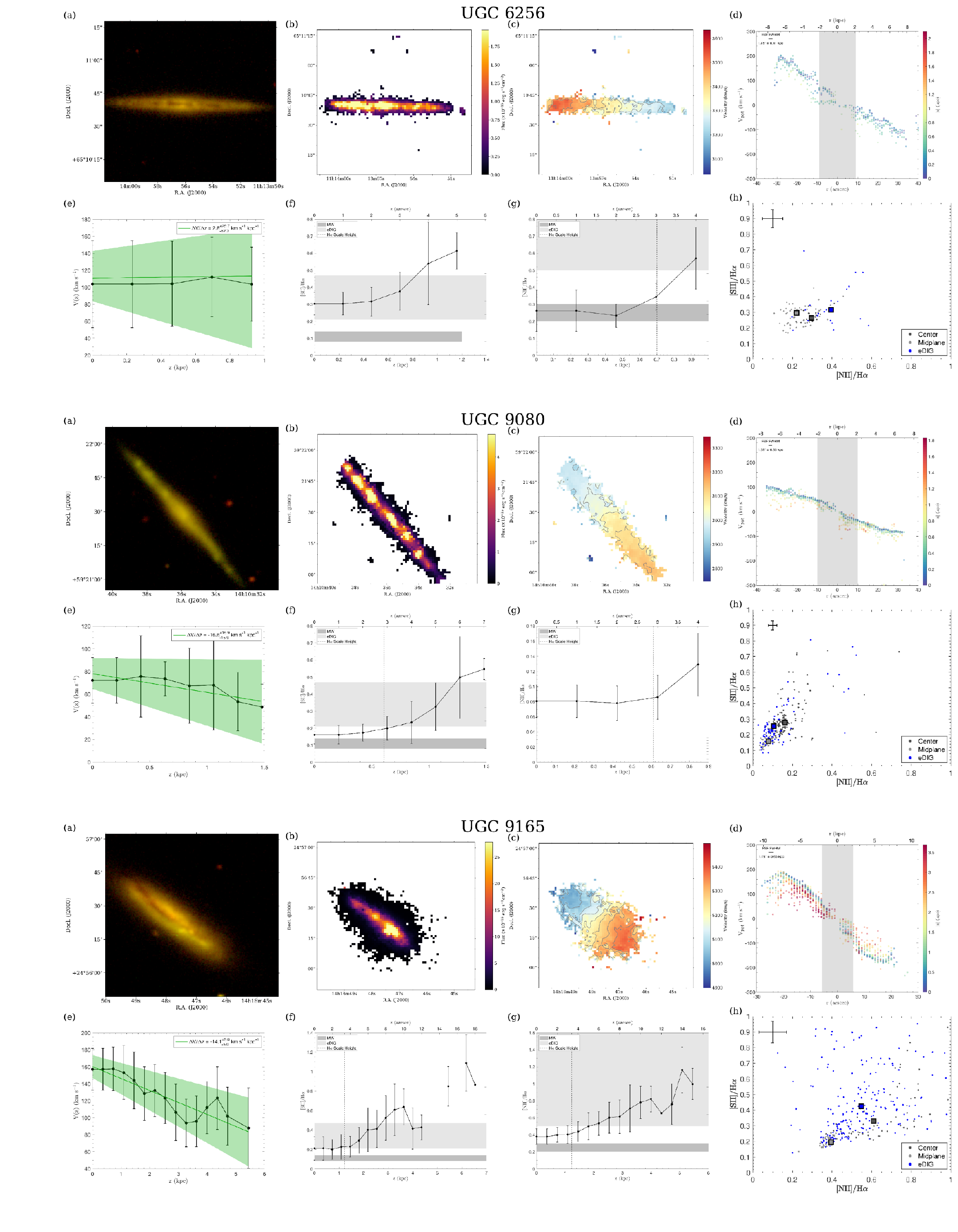}\\
\includegraphics[width=\textwidth]{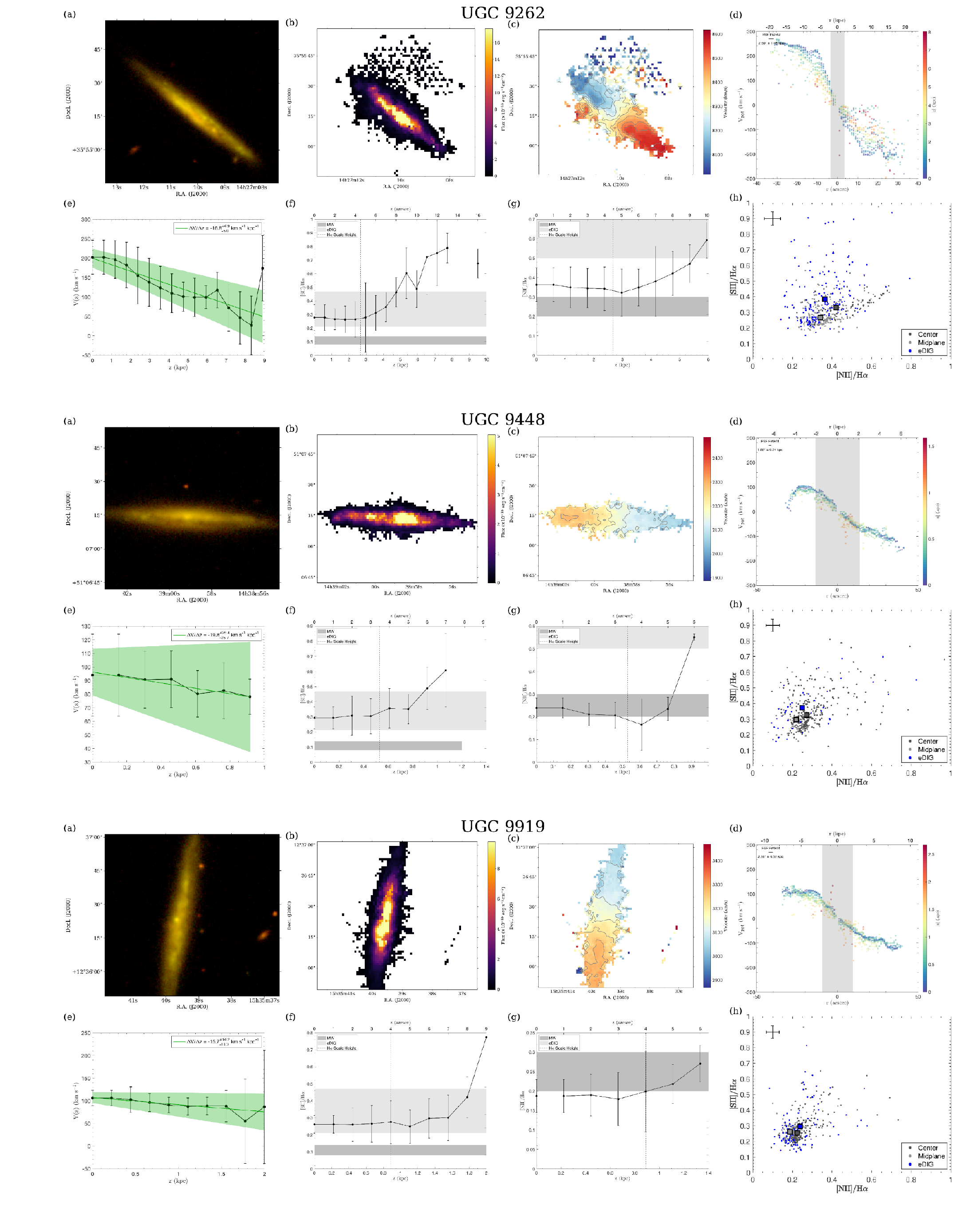}\\
\includegraphics[width=\textwidth]{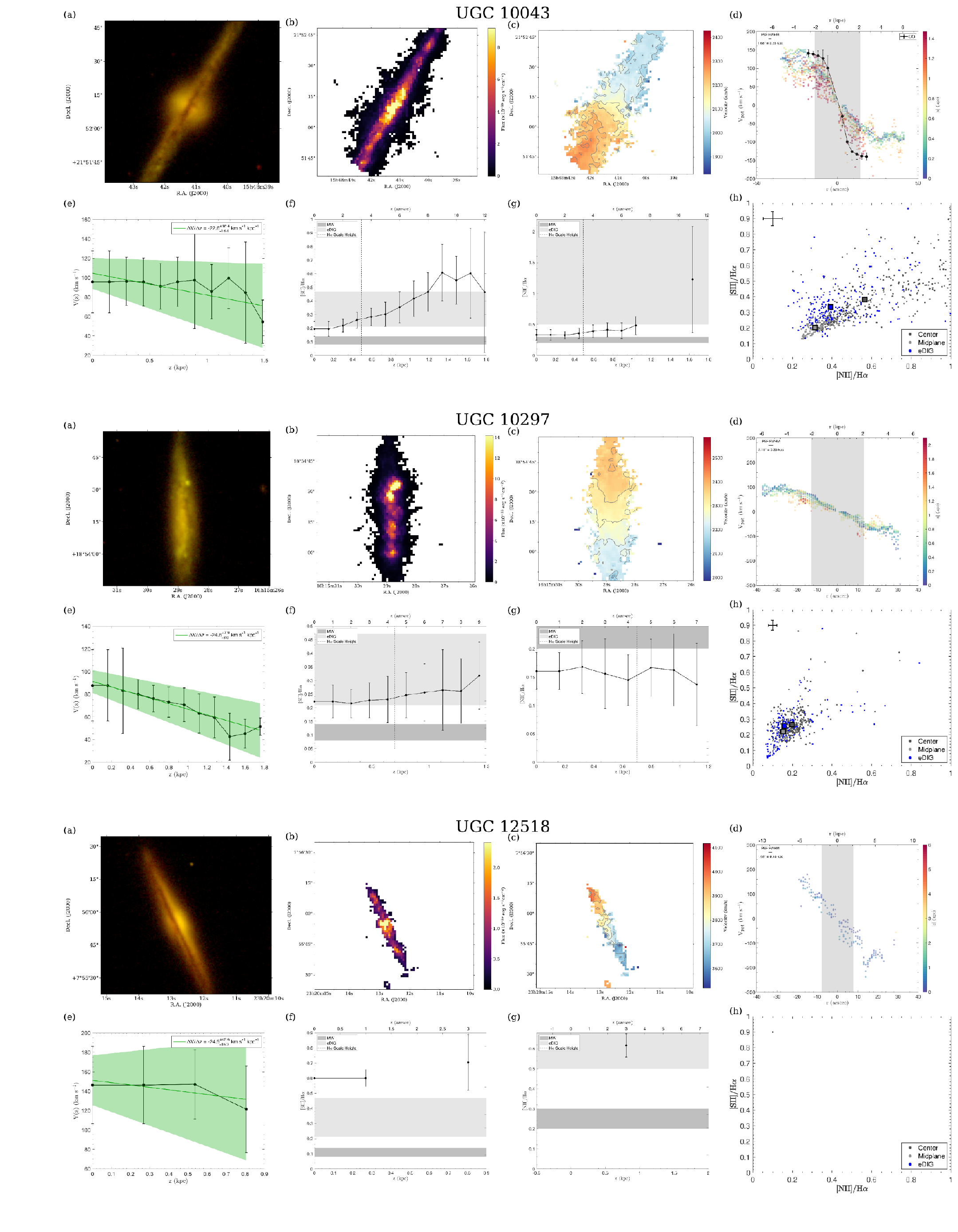}\\
\includegraphics[width=\textwidth]{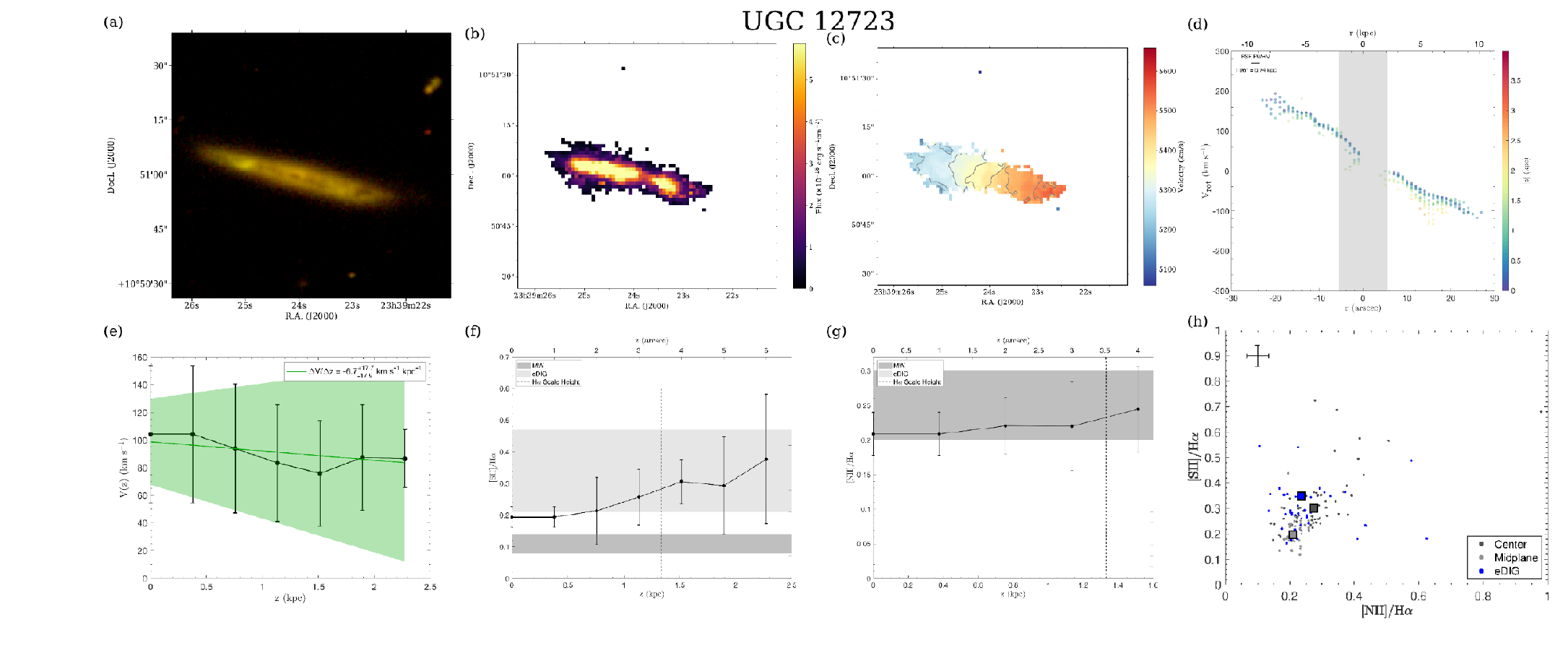}\\
\end{longtable*}
\noindent {\bf Figure \arabic{table}.} Multipanel figures for all 25 edge-on CALIFA galaxies. (a) The composite SDSS r- and g-band images. (b) The extinction-corrected \ha\ flux image. Pixels with SNR $<$ 5 are masked. (c) The \ha\ velocity map. Pixels where the flux or velocity SNR $<$ 5 are masked. Gray contours are plotted in 50 \kms\ intervals. (d) The \ha\ PV diagram in vertical slices, color-coded by height (see Section \ref{ssec:PVdiagrams}). The CO PV diagram in the midplane is shown where available (black). (e) The radially averaged \ha\ rotation velocity as a function of height above the midplane (see Section \ref{ssec:vvg}). The green line is the best fit, and the green shading shows the 68\% confidence region. The slope of the line is the vertical gradient in the rotation velocity (\DvDz). (f) The radially averaged \SII/\ha\ ratio as a function of height above the disk midplane (see Section \ref{ssec:eDIG}). Typical values observed in the midplane are $0.11\pm0.03$ \citep[dark gray shading;][]{madsen04} and $0.34\pm0.13$ in the eDIG \citep[light gray shading;][]{blanc09}. The vertical dashed line marks the median \ha\ scale height. (g) The same as (f) but for \NII/\ha. Typical values are $\sim0.25$ in the midplane \citep{madsen06} and $\gtrsim 0.5$ in the eDIG \citep{madsen04}. (h) The \SII/\ha\ versus \NII/\ha\ for pixels with EW(\ha)\,$>6$\,\AA\ in the eDIG, midplane, and center. Large squares are the medians on each group. Typical error bars are shown in the upper left corner.

\end{document}